\documentclass[a4paper,11pt]{article}
\usepackage{rotating}
\usepackage{jheppub} 
\usepackage[utf8]{inputenc}
\usepackage[T1]{fontenc} 
\usepackage{placeins} 
\usepackage{chngpage}
\usepackage{amsmath,amsfonts}
\usepackage{parskip} 
\usepackage{subcaption}
\usepackage{graphicx}
\usepackage{color}
\usepackage{xcolor}
\usepackage{relsize}
\usepackage{epstopdf}
\usepackage{hyperref}
\usepackage{mathrsfs}
\usepackage{ragged2e}
\usepackage{amssymb}
\usepackage{placeins}
\usepackage[normalem]{ ulem }
\usepackage{amsthm}
\usepackage{comment}
\usepackage{graphics}	
\usepackage{caption}
\usepackage{verbatim}
\usepackage{arydshln}
\usepackage{slashed}
\usepackage{array}
\usepackage{booktabs}
\usepackage{textcomp}
\usepackage{tcolorbox}
\usepackage[linesnumbered,lined,ruled,commentsnumbered]{algorithm2e}
\usepackage{ragged2e}
\definecolor{commentsColor}{rgb}{0.497495, 0.497587, 0.497464}
\definecolor{keywordsColor}{rgb}{0.000000, 0.000000, 0.635294}
\definecolor{stringColor}{rgb}{0.558215, 0.000000, 0.135316}
\definecolor{backgroundColor}{rgb}{0.9, 0.9, 0.9}
\usepackage{tabularx,booktabs}
\usepackage[table]{xcolor}  
\usepackage{multicol}
\usepackage{array,multirow}
\usepackage{booktabs}
\usepackage{colortbl}
\usepackage{float}
\usepackage{breqn}
\usepackage{makecell}  

\renewcommand{\to}{\rightarrow}

\newcommand{\be}{\begin{equation}}
\newcommand{\ee}{\end{equation}}
\makeatletter
\gdef\@fpheader{}
\vspace*{0cm}
\makeatother

\begin{document}

\title{Angular Coefficients from Interpretable Machine Learning with Symbolic Regression}

\author[a]{Josh Bendavid,}
\author[b]{Daniel Conde,}
\author[c]{Manuel Morales-Alvarado,}
\author[b]{Veronica Sanz,}
\author[d]{ and Maria Ubiali}
\affiliation[a]{CERN, European Organization for Nuclear Research, Geneva, Switzerland}
\affiliation[b]{Instituto de F\'isica Corpuscular (IFIC), Universidad de Valencia-CSIC, E-46980 Valencia, Spain}
\affiliation[c]{Istituto Nazionale di Fisica Nucleare (INFN), Sezione di Trieste, SISSA, Via Bonomea 265, 34136, Trieste, Italy}
\affiliation[d]{DAMTP, University of Cambridge, Wilberforce Road, Cambridge CB3 0WA, UK }

\emailAdd{josh.bendavid@cern.ch}
\emailAdd{Daniel.Conde@ific.uv.es}
\emailAdd{mmorales@sissa.it}
\emailAdd{veronica.sanz@uv.es}
\emailAdd{M.Ubiali@damtp.cam.ac.uk}

\date{\today}

\abstract{
We explore the use of symbolic regression to derive compact analytical expressions for angular observables relevant 
to electroweak boson production at the Large Hadron Collider (LHC). Focusing on the angular coefficients that govern 
the decay distributions of $W$ and $Z$ bosons, we investigate whether symbolic models can well approximate these quantities, 
typically computed via computationally costly numerical procedures, with high fidelity and interpretability. 
Using the PySR package, we first validate the approach in controlled settings, namely in angular distributions in lepton-lepton collisions in QED and 
in leading-order Drell–Yan production at the LHC. We then apply symbolic regression to extract 
closed-form expressions for the angular coefficients $A_i$ as functions of transverse momentum, 
rapidity, and invariant mass, using next-to-leading order simulations of $pp \to \ell^+\ell^-$ events. 
Our results demonstrate that symbolic regression can produce accurate and generalisable expressions that 
match Monte Carlo predictions within uncertainties, while preserving interpretability and providing insight 
into the kinematic dependence of angular observables.
}


\maketitle

\section{Introduction}
\label{sec:intro}

Accessing the polarisation of electroweak (EW) bosons at high-energy colliders is crucial to gain
insights in the electroweak-symmetry-breaking mechanism (EWSB). The $W$ and $Z$ boson are given a 
longitudinal polarisation, hence a mass, by means of the EWSB. Any deviations in the production
of longitudinal bosons in scattering processes would suggest the presence of effects beyond the Standard Model (BSM), 
implying a realisation of the EWSB that deviated from the Higgs mechanism of the Standard Model (SM).
The investigation of boson polarisation in processes at the Large Hadron Collider (LHC) is a key part of 
the analysis programme of the experimental collaborations~\cite{ATLAS:2019bsc,CMS:2020etf,CMS:2021icx,ATLAS:2022oge,LHCb:2022tbc,ATLAS:2023lsr}, 
and High-Luminosity (HL-LHC) measurements will further improve the precision of current analyses. 

The production and decay of W and Z bosons at a hadron collider can be characterized by angular coefficients, which are in turn related to the polarisation fractions~\cite{Bern:2011ie,Stirling:2012zt,Belyaev:2013nla}.
These have been measured in several experimental analyses of $W$~\cite{CMS:2011kaj,ATLAS:2012au,CMS:2020cph} and $Z$~\cite{CMS:2015cyj,ATLAS:2016rnf}  production at 
the LHC, where angular coefficients or polarized cross sections or fractions are extracted from the measured events via the angular distributions of the final state particles, 
either reconstructed explicitly in the boson rest frame, or inferred from lab frame distributions.  This angular decomposition is also a key component in the measurement of 
boson production cross sections in the full phase space~\cite{ATLAS:2023lsr} and corresponding interpretation in terms of the strong coupling constant~\cite{ATLAS:2023lhg} and in the theoretical modeling and uncertainty for measurements of the $W$ boson mass at hadron colliders~\cite{ATLAS:2024erm, LHCb:2021bjt, CMS:2024lrd, CDF:2022hxs, D0:2013jba, LHC-TeVMWWorkingGroup:2023zkn}.
%
%
While the angular coefficients can be predicted from perturbative QCD, their dependence on the boson kinematics do not have a known closed analytical formula. 

In this work, we infer analytical expressions for the angular coefficients,  that is accurate and general enough using a 
particular type of Machine Learning (ML) technique, namely Symbolic Regression (SR), that aims to discover human-interpretable symbolic 
models~\cite{Udrescu:2019mnk,cranmer2019learningsymbolicphysicsgraph}. SR describes a supervised
learning task where the model space is spanned by analytic expressions. This is typically framed as a multi-objective optimisation framework, 
jointly minimising prediction error and model complexity. In this family of algorithms, instead of fitting concrete parameters in 
some over-parametrised general model, one searches the space of simple analytic expressions for accurate and interpretable 
models\footnote{Note that alternative ML methods have been used to extract the longitudinal contributions \cite{Grossi:2023fqq} via neural networks.}.
SR is a way to combine the power of ML with the advantage of analytical 
intuition. In analogy to training a neural network one can use SR to learn a general, analytic function 
over phase space from a data set. SR gives us a way to extract
relatively simple human-readable formulas from complex simulated data sets.
In addition to the insights that a simple formula might provide, a single equation is very fast to evaluate, and can be useful for smoothing 
predictions with finite statistical precision, or as a basis for statistical or systematic uncertainties on those predictions.
SR algorithms have been shown to be able to infer well-known astrophysical and cosmological formulas, 
see for example Refs.~\cite{matchev2021analyticalmodellingexoplanettransit,lemos2022rediscoveringorbitalmechanicsmachine,Cranmer:2020wew}. 

Beyond providing compact representations of physical observables, SR can serve as a practical tool in experimental analyses. The closed-form expressions it produces can be directly integrated into fitting procedures, replacing numerical lookup tables or high-dimensional interpolation grids. This facilitates faster and more transparent parameter estimation, especially when dealing with systematic variations or complex signal models. Moreover, analytical expressions simplify error propagation and make it easier to incorporate theoretical constraints or symmetries explicitly.

Another promising application of SR is the replacement of costly Monte Carlo evaluations with fast, analytical surrogates. In simulation chains involving high-multiplicity final states, detector response, or reweighting under alternative models, SR-derived functions can reduce computational time dramatically. For instance, rather than re-running full simulations to apply new weights or systematic shifts, one could use the SR expressions to perform fast analytic reweighting of kinematic distributions, which is particularly valuable in large-scale parameter scans or real-time applications.

A distinguishing feature of SR is its interpretability. Unlike neural networks or other black-box machine learning models, SR yields explicit mathematical expressions whose structure can be inspected, validated, and directly compared to theoretical expectations. This is particularly advantageous in physics applications, where constraints from symmetries, conservation laws, or limiting behaviours (e.g., low transverse momentum limits or high energy scaling) can be imposed or checked analytically. The transparency of SR makes it well suited for tasks where physical insight is essential.

Despite its far-reaching potential, SR has only been sparsely used in high energy physics 
so far. First pioneered in Ref.~\cite{Choi:2010wa} to derive a kinematic variable that is sensitive to the mass of the 
Higgs boson in the $WW$ channel, it was subsequently used in a variety of problems, from constructing optimal observables for LHC
processes~\cite{Butter:2021rvz} to the simplification of the mathematical expression of polylogarithms~\cite{Dersy:2022bym}, and many 
other applications~\cite{Dong:2022trn,Alnuqaydan:2022ncd,AbdusSalam:2024obf,Tsoi:2023isc,Tsoi:2024pbn,Soybelman:2024mbv,Dotson:2025omi,Bahl:2025jtk,Vent:2025ddm}.
Its usage was also advocated for the study of observables sensitive to the presence of physics beyond the Standard Model (SM), 
such as those described within the framework of Effective Field Theories (EFTs). EFTs provide a systematic expansion to encode 
the effects of heavy new physics through higher-dimensional operators, which may manifest as small deviations in angular distributions~\cite{Gauld:2024glt,Li:2024iyj,Li:2025fom,Hiller:2025hpf}. 
The transparency and compactness of SR expressions make them well-suited 
to identify and characterise such deviations, especially in EFT-sensitive observables where interpretability is paramount.

Finding SR models can be a computationally intensive task~\cite{virgolin2022symbolicregressionnphard}, but the constant development of 
advanced methodologies that are able to tackle such problems in a reasonable amount of time, such as PySR~\cite{Cranmer2023InterpretableML} and
SymbolNet~\cite{Tsoi:2024ypg} is moving the field forward. 

The PySR software package~\cite{Cranmer2023InterpretableML} is an open-source library for SR, based on a multi-population evolutionary algorithm.  
The population consists of symbolic expressions, each one represented as an expression tree consisting of nodes of operator functions, constant, and input variables (or features). 
More details are given in App.~\ref{app:sr}. 
The aim of this work is to explore whether SR, implemented using PySR, 
can recover compact and accurate analytical expressions for the angular coefficients $A_i$ as functions 
of the vector boson transverse momentum ($p_T$), dilepton rapidity ($y$) and invariant mass ($m$). 
 To this end, Section~\ref{sec:benchmarks} validates the SR approach in two distinct contexts: 
 first, by recovering the well-known angular distribution from first principles in QED for lepton-lepton collisions; and second, 
 by regressing the underlying parton luminosities of the input PDF set in photon-mediated Drell-Yan production at the LHC computed at leading order. 
 Section~\ref{sec:results} presents the SR results for the angular coefficients $A_i$ as 
 functions of one, two, or all three of the variables $p_T$, $y$, and $m$. 
 We summarise our findings and list future work in Section~\ref{sec:conclusions}.
 The appendices provide technical details, including a discussion on the SR algorithm 
 and its selection criteria in App.~\ref{app:sr}, 
 the complete results for the 1D angular coefficients in App.~\ref{app:1d_eqs}, and for the 2D angular coefficients in App.~\ref{app:2d_eqs} fits. 

\section{Symbolic regression benchmarks}
\label{sec:benchmarks}
 
To assess the effectiveness of SR in recovering compact analytic expressions, 
we consider two benchmark scenarios of increasing complexity. The first involves lepton–lepton 
collisions, where the angular distributions of the final-state particles are known analytically from first principles in QED. 
This provides a controlled setting to validate the equation-recovery capabilities of the method. The second benchmark involves 
photon-mediated Drell–Yan production in proton–proton collisions, where observables depend on 
parton distribution functions (PDFs) and their combination via parton luminosities. 
In this case, SR is used to uncover interpretable approximations for 
these non-trivial functions, illustrating its potential both in data- and in simulation-driven contexts. 
A partial version of these results have been presented in Ref.~\cite{Morales-Alvarado:2024jrk}.

\subsection{Angular distribution in lepton-lepton collisions}
\label{subsec:ee}

The production of vector bosons has been extensively studied at LEP and the LHC. At an $e^+e^-$ collider such as LEP, the leading order cross section for $e^+e^-\to \mu^+\mu^-$ in the massless limit is given by 
%
\begin{equation}
    \label{eq:level1}
    \frac{d\sigma}{d\Omega} = \frac{ \alpha^2}{4s}\left(1 + \cos^2\theta\right),
\end{equation}
where $\theta$ is the scattering angle between the incoming electron and the outgoing muon, 
$d\Omega = d(\cos \theta) \, d\phi$ is the differential solid angle, $\alpha$ is the QED coupling, and $s$ is the squared centre-of-mass energy. 

We start from the simplest case given by Eq.~(\ref{eq:level1}) for leading-order $\mu^+\mu^-$ production. To train the regressor, we simulate events using \textsc{MadGraph5\_aMC@NLO}~\cite{Alwall:2014hca,Frederix:2018nkq} with a centre-of-mass energy of $\sqrt{s} = 1~\text{TeV}$ and $\alpha = 0.00755$ (the default value used internally by \textsc{MadGraph5\_aMC@NLO}). No cuts are applied to the transverse momenta, rapidities, or angular separations of the outgoing leptons. We generate distributions with varying numbers of bins, using the bin centres and corresponding cross sections (estimated from the simulation) to train the regressor. Our analysis focuses on the impact of binning on the regressor’s performance, as well as identifying the optimal selection criteria in PySR for recovering the underlying physical law.

PySR also offers a built-in denoising feature, which models the data using Gaussian processes incorporating a white noise kernel to account for statistical fluctuations. However, we do not use it in the results presented in this paper as it can lead to overcomplicated expressions without considerable gain in accuracy. 

We present the SR equations obtained using the three selection criteria in Table~\ref{tab:l1_results}, where, for readability, all expressions have been normalised by the prefactor in Eq.~(\ref{eq:level1}), after performing the azimuthal integration and restoring from natural units. For simplicity, in all cases the loss function is the mean square error (MSE).

\begin{table}[h]
\begin{center}
\resizebox{\textwidth}{!}{%
\begin{tabular}{@{}c
>{\centering\arraybackslash}m{0.3\textwidth}|
>{\centering\arraybackslash}m{0.3\textwidth}|
>{\centering\arraybackslash}m{0.3\textwidth}@{}}
\toprule
Bins & \multicolumn{1}{c}{Accuracy} & \multicolumn{1}{c}{Best} & \multicolumn{1}{c}{Score} \\
\midrule
10 &
$1.00563 + 0.99300\,x_{0}^{2} + 0.00792\,x_{0} + 0.00025\,x_{0}^{4} + 0.03187\,x_{0}^{5}$ &
$1.00563 + 0.99300\,x_{0}^{2} + 0.03187\,x_{0}^{3}$ &
$1.00563 + 0.99300\,x_{0}^{2}$ \\

\noalign{\smallskip}\hline\noalign{\smallskip}
20 &
$1.00046 + 1.02749\,x_{0}^{2} + 0.01875\,x_{0} + 0.00147\,x_{0}^{3} - 0.04450\,x_{0}^{4}$ &
$1.00419 + 0.98955\,x_{0}^{2} + 0.01837\,x_{0}$ &
$1.00419 + 0.98955\,x_{0}^{2}$ \\

\noalign{\smallskip}\hline\noalign{\smallskip}
30 &
$0.99816 + 1.05041\,x_{0}^{2} + 0.02813\,x_{0}^{3} - 0.07378\,x_{0}^{4}$ &
$1.00448 + 0.98725\,x_{0}^{2}$ &
$1.00448 + 0.98725\,x_{0}^{2}$ \\

\noalign{\smallskip}\hline\noalign{\smallskip}
50 &
$1.00075 + 1.01828\,x_{0}^{2} + 0.00098\,x_{0} + 0.02753\,x_{0}^{3} - 0.03416\,x_{0}^{4}$ &
$1.00362 + 0.98898\,x_{0}^{2}$ &
$1.00362 + 0.98898\,x_{0}^{2}$ \\

\noalign{\smallskip}\hline\noalign{\smallskip}
100 &
$1.00046 + 1.01998\,x_{0}^{2} + 0.05752\,x_{0}^{3} - 0.03588\,x_{0}^{4} - 0.03588\,x_{0}^{5}$ &
$1.00362 + 0.98926\,x_{0}^{2}$ &
$1.00362 + 0.98926\,x_{0}^{2}$ \\

\noalign{\smallskip}\hline\noalign{\smallskip}
200 &
$1.00333 + 1.00333\,x_{0}^{2} + 0.02802\,x_{0}^{3} - 0.02208\,x_{0}^{4}$ &
$1.00333 + 0.98955\,x_{0}^{2}$ &
$1.00333 + 0.98955\,x_{0}^{2}$ \\

\noalign{\smallskip}\hline\noalign{\smallskip}
500 &
$1.00046 + 1.01866\,x_{0}^{2} + 0.00615\,x_{0} + 0.01960\,x_{0}^{3} - 0.03387\,x_{0}^{4}$ &
$1.00333 + 0.98955\,x_{0}^{2}$ &
$1.00333 + 0.98955\,x_{0}^{2}$ \\

\noalign{\smallskip}\hline\noalign{\smallskip}
1000 &
$1.00333 + 0.98955\,x_{0}^{2} + 0.00715\,x_{0} + 0.01802\,x_{0}^{3}$ &
$1.00333 + 0.98955\,x_{0}^{2}$ &
$1.00333 + 0.98955\,x_{0}^{2}$ \\
\bottomrule
\end{tabular}
}
\caption{Normalised equations according to the three selection criteria for different bin sizes with $x_{0} \equiv \cos \theta$. 
The numbers that appear in these expressions have been approximated to the 5th decimal place.}
\label{tab:l1_results}
\end{center}
\end{table}

The equations selected by the \texttt{accuracy} criterion minimise the loss function but consistently fail to capture the underlying 
natural law. These equations are overly complex and tend to overfit the noise arising from the simulation.

The \texttt{score} selection criterion finds the correct equation in all cases, even in the presence of noisy data. 
This robustness is due to the way \texttt{score} balances accuracy with equation simplicity. \texttt{Score} penalizes model complexity while preserving fit quality, helping to avoid overfitting to noise.
As discussed in Sect.~\ref{app:sr}, rather than 
focusing solely on minimising the loss function like \texttt{accuracy}, \texttt{score} incorporates a complexity penalty that prevents overfitting, 
allowing it to avoid being misled by noisy data. Moreover, this criterion’s ability to generalise well, even with noisy or low-resolution data, 
highlights its strength as a reliable and effective approach. By selecting equations that are both mathematically simple and physically accurate, 
\texttt{score} ensures that the derived models not only fit the data but also reflect the underlying natural law. This makes it the optimal choice at this 
level for tasks where both interpretability and performance are critical.

Finally, the \texttt{best} criterion, while more effective than \texttt{accuracy}, manages to recover the natural law when using distributions with 
30 or more bins. For datasets with fewer bins, this method manages to substantially find the true underlying equation with a very small (around $\times 30$ smaller) additional term. In general, when exploring big combinatorial spaces to find equations that are not currently known, it is useful to use the \texttt{best} criterion as it combines accuracy and simplicity, with a slightly higher weight (compared to the \texttt{score} criterion) towards optimising accuracy in a data-driven setting. 

\begin{figure}[h!]
  \centering
  \includegraphics[width=0.8\linewidth]{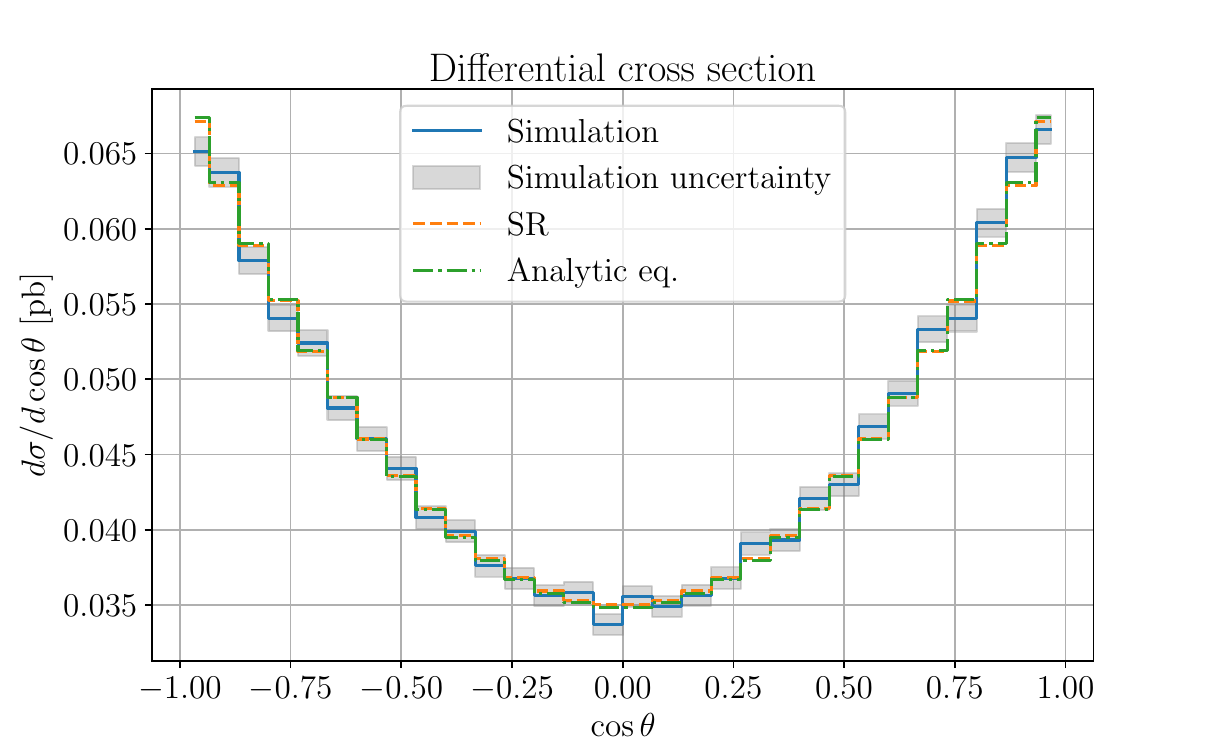}
  \caption{Angular distribution for 30 bins in $\cos\theta$.}
  \label{fig:sr_pred_l1_absolute}
\end{figure}

In Fig.~\ref{fig:sr_pred_l1_absolute} we show the angular distributions from the simulator and 
SR for $30$ bins according to the \texttt{best} selection criterion. We can clearly see the angular functional dependence 
and overall agreement between SR, trained exclusively with the simulated events, and the analytical formula from first principles.

\begin{figure}[h!]
  \centering
  \includegraphics[width=0.8\linewidth]{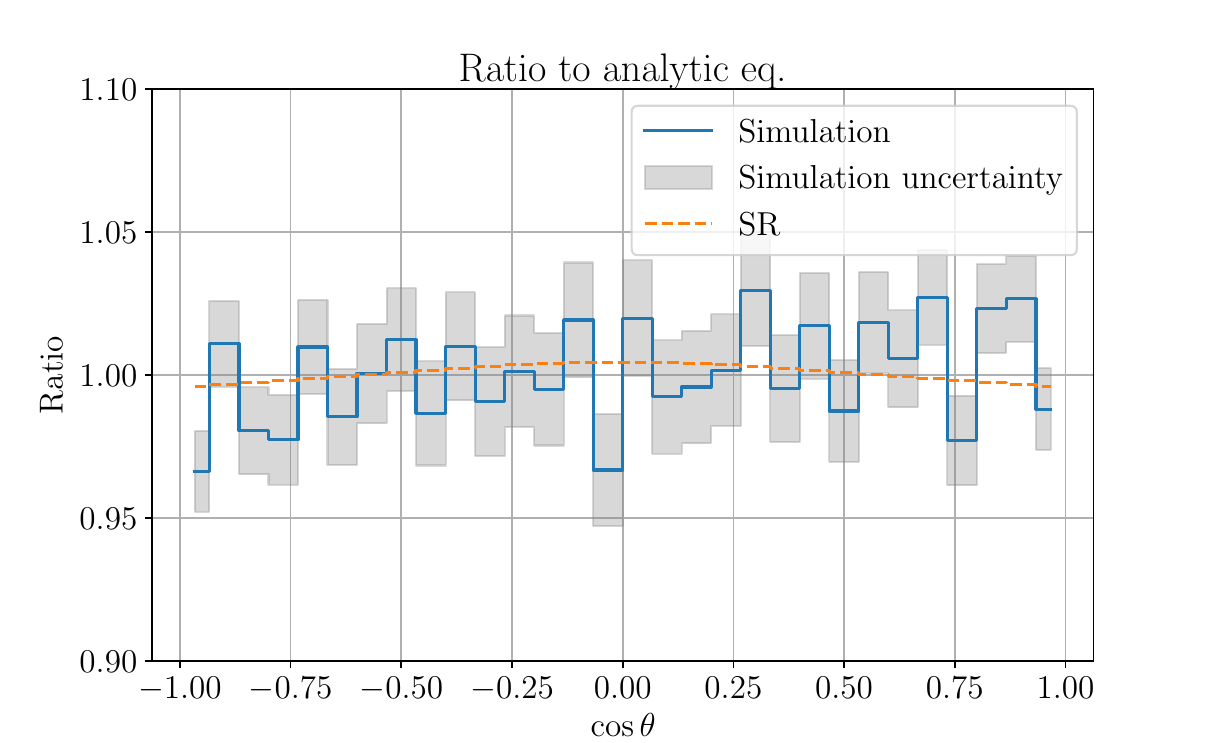}
  \caption{Angular distribution for 30 bins normalised to the analytical equation.}
  \label{fig:sr_pred_l1_normalised}
\end{figure}

For further inspection, in Fig.~\ref{fig:sr_pred_l1_normalised} we show the ratio to the analytic equation of the angular distributions 
from the simulation and from SR, from the previous plot. We can see how the SR, across the complete kinematic coverage, agrees very well 
to the percent level with the analytic formula, even when the simulation disagrees beyond uncertainties with the real underlying equation. The discrepancy between the simulation and the analytical 
formula arises from the stochastic nature of the former, where finite statistics can perturb the angular dependence that is derived from the first principles. In contrast, the SR, though trained on this imperfect data (as it is subject to statistical fluctuations), produces a closed-form approximation that inherently 
prioritises accuracy, simplicity and smoothness. The result aligns more closely with the analytic solution precisely because the SR discards noise in favor of a simpler and more physically plausible representation.

\begin{figure}[h!]
  \centering
  \includegraphics[width=0.8\linewidth]{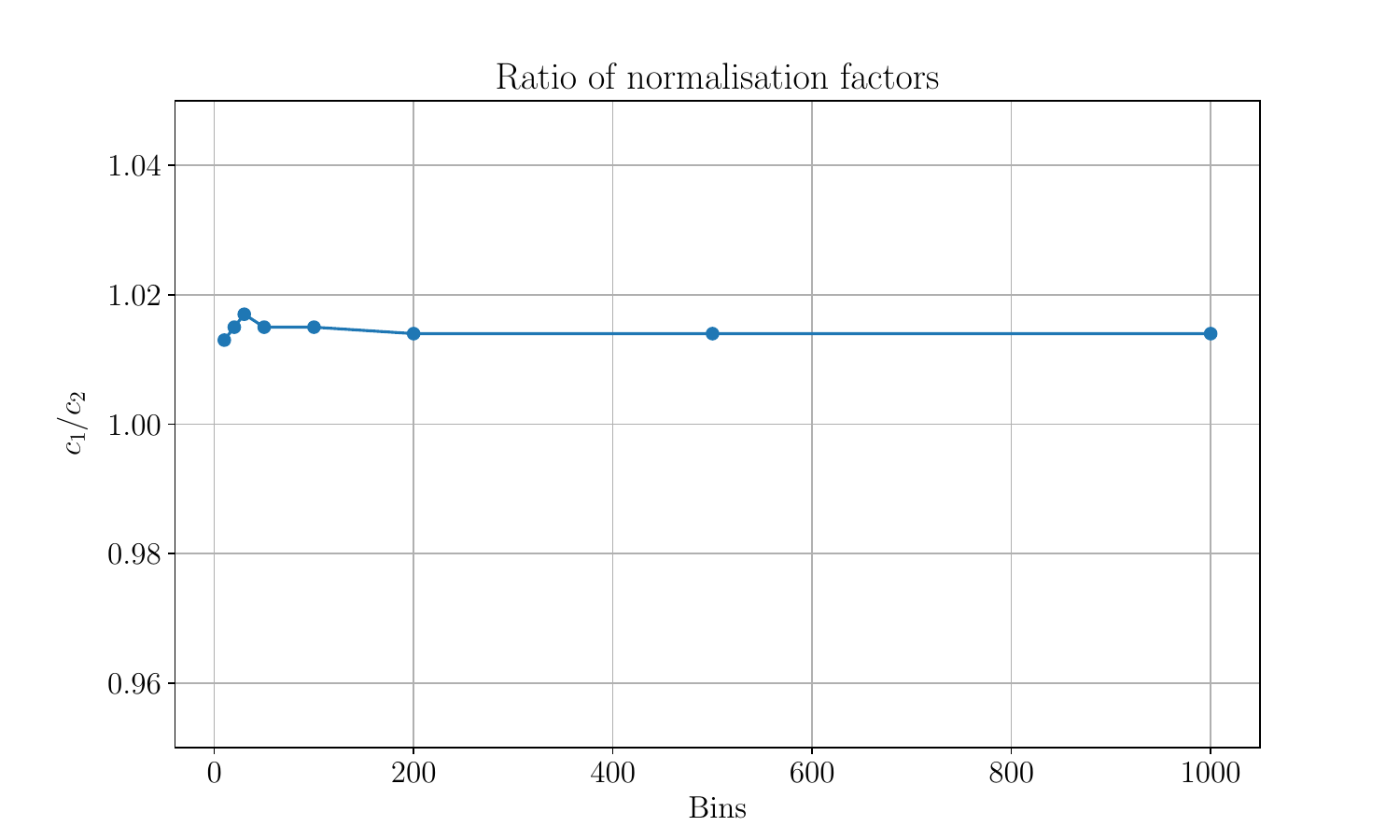}
  \caption{Relative size of normalisation coefficients for different number of bins.}
  \label{fig:sr_l1_norm}
\end{figure}

Not only the correct functional dependence in $\cos \theta$ is recovered, but also the constants to a very good accuracy. We can
assess this by parametrising the SR prediction as
\begin{equation}
    \label{eq:sr_level1}
  \text{SR} (\cos \theta) = c_1 + c_2 \cdot \cos^2 \theta,
\end{equation}
with $c_1$ and $c_2$ being normalisation factors of the prediction. Comparing Eqs.~(\ref{eq:sr_level1}) and~(\ref{eq:level1}), one finds that, from first principles, $c_1 / c_2 = 1$. 
In Fig.~\ref{fig:sr_l1_norm} we show the relative size of the normalisation factors when training the regressor on distributions with different numbers of bins.  This demonstrates that SR not only recovers the correct functional form but also captures the relative normalization with high stability.

The ratio of the normalisation factors $c_1 / c_2$ remains stable across a wide range of binning choices, staying within a 2\% deviation from the expected value of 1. This consistency demonstrates the robustness of the SR method in recovering not just the functional form but also the correct normalisation factors, regardless of the granularity of the binning. Even with very fine binning, the model manages to avoid overfitting noise and provides accurate coefficients, confirming the reliability of the SR prediction across different resolutions of the data at this level.

\subsection{Parton luminosities in photon-mediated Drell-Yan}
\label{subsec:dygamma}

Given that the angular distribution of the lepton process discussed in the previous section can be described analytically from first principles, 
it can used to validate the SR method in a controlled setting. Having established its efficacy in reproducing known results, we now turn to a 
more challenging scenario: deriving previously unknown analytic expressions for observables where a simple closed-form solution is not immediately recoverable.

In proton-proton collisions, the differential cross section is obtained by convolving the partonic cross section, calculable from perturbative QFT, 
with parton distribution functions (PDFs). PDFs cannot be computed from first principles and must instead be determined empirically from fits to data. 
There are several collaborations that provide up-to-date fits of the PDFs that are determined from global fits of data involving protons or nucleons 
in the initial state and released as public LHAPDF~\cite{Buckley:2014ana} interpolation grids in $(x,Q)$, 
see Refs.~\cite{Amoroso:2022eow,Ubiali:2024pyg} for some recent reviews. 
At leading order in QCD, the observables' dependence on PDFs is dictated by the parton luminosities~\cite{Ellis:1996mzs, Peskin:1995ev} 
\begin{equation}
{\cal L}_{ij}(m,y) = \int_{-y}^y\, d\bar{y} \,f_i\left(\frac{m}{\sqrt{s}}e^{\bar{y}},m\right)\,f_j\left(\frac{m}{\sqrt{s}}e^{-\bar{y}},m\right),
\label{eq:partonlumi}
\end{equation}
where $i,j$ are the flavour indices of the partons involved in the hard scattering process, $m$ is 
the invariant mass of the final states produced at the hard scattering process level, 
and the integration limits $y$ are given by the rapidity of the final states in the hadronic centre-of-mass frame and 
are defined in terms of $M$ and the hadronic centre-of-mass energy $\sqrt{s}$, as
\begin{equation}
    y=\log\left(\frac{\sqrt{s}}{m}\right).
\end{equation}
Once a given PDF set is selected among those available on 
LHAPDF, their analytical  expression is in principle computable from Eq.~\eqref{eq:partonlumi}, but -- depending on the parametrisation 
used by the selected PDF collaboration -- it can be extremely long and convoluted, and it depends on the interpolation 
algorithm used to transpose the PDF parametrisation and their evolution with the energy scale $m$ on the LHAPDF grid. Hence 
we can say that in this case the analytic dependence of parton luminosity is not easily derivable nor particularly handy. 

In this section, we present the first application of SR to derive compact, interpretable approximations for 
parton luminosities. 
By leveraging SR, we aim to obtain accurate analytic representations that capture their essential features.
To do that, we focus on the Drell-Yan (DY) mechanism via virtual photon, $pp \to \gamma^* \to \mu^+\mu^-$. 
We simulate the LO process at $\sqrt{s} = 1$ TeV using the CT10 NLO PDF set~\cite{Gao:2013xoa}, implementing just a cut on the dilepton invariant mass $m>1$ GeV, being fully inclusive in the remaining kinematics like the dilepton rapidity, distance, etc. 
In this case $i$ and $j$ in Eq.~\eqref{eq:partonlumi} are the indices of the active flavour quarks and antiquarks respectively.
The double differential cross section is given by 
\begin{equation}
\label{eq:l2_pdf}
\frac{d^2\sigma}{dm dy} = \frac{8 \pi \alpha^2}{9 m s}   
\sum_{q} Q_{q}^2 \left[ {\cal L}_{q\bar q}(m,y) + {\cal L}_{\bar q q}(m, y)\right]
\equiv \frac{8 \pi \alpha^2}{9 m s} F(m, y),
\end{equation}
where sum runs over the $q=u, d, s, c$ quark flavours. 
We regress the function $F(m, y)$ with SR, as we are exclusively interested in this quantity as the other pieces of the distribution are known. 
Comparing the SR results to Eq.~(\ref{eq:l2_pdf}) provides a closed analytical expression to parametrise the 
parton luminosities. 

\begin{table}[h]
\begin{center}

\resizebox{\textwidth}{!}{%
\begin{tabular}{@{}c>{\centering\arraybackslash}m{0.5\textwidth}>{\centering\arraybackslash}m{0.2\textwidth}@{}}
\toprule
Complexity & \multicolumn{1}{c}{Equation} & \multicolumn{1}{c}{Score} \\
\midrule
$3$ & $\frac{9.16 \cdot 10^{4}}{m}$ & $0.359$\\

\rowcolor{lightgray} 
$33$ & 
\begin{minipage}{0.5\textwidth} 
\vspace{-0.7em}
\[
\frac{2.86 \cdot 10^{5} \left(0.0461 \cdot 1.15^{y^2}\right)^{- 0.0250 \cdot 1.15^{y^2}}}{m^{2}}
\]
\end{minipage} 
& $0.387$ \\
$35$ & 
\begin{minipage}{0.5\textwidth} 
\vspace{-0.5em}
\[
\frac{2.86 \cdot 10^{5} \left(0.0461 \cdot 1.15^{y^2}\right)^{- 0.0250 \cdot 1.15^{y^2}}}{m^{2} + 0.117}
\]
\end{minipage} 
& $0.00967$ \\
\bottomrule
\end{tabular}
}
\caption{Selection of \texttt{best} SR expressions $F(M, y)$ with their complexities and scores. Constants are approximated for display purposes.}
\label{tab:selected_equations}
\end{center}
\end{table}

\begin{figure}[ht!]
  \centering
  \begin{minipage}[b]{0.45\linewidth}
    \includegraphics[width=\linewidth]{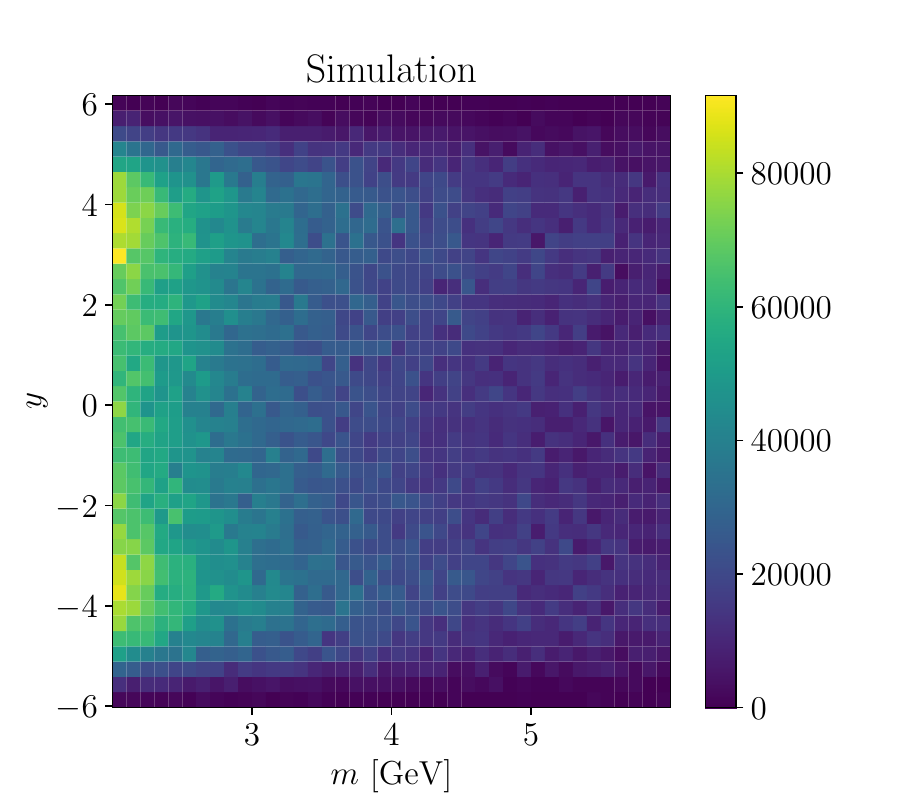}
    \caption{Parton luminosity values obtained from the reweighted simulation of the 
    LO Drell-Yan double-differential cross section Eq.~\eqref{eq:l2_pdf}.}
    \label{fig:l2_mg}
  \end{minipage}
  \hfill 
  \begin{minipage}[b]{0.45\linewidth}
    \includegraphics[width=\linewidth]{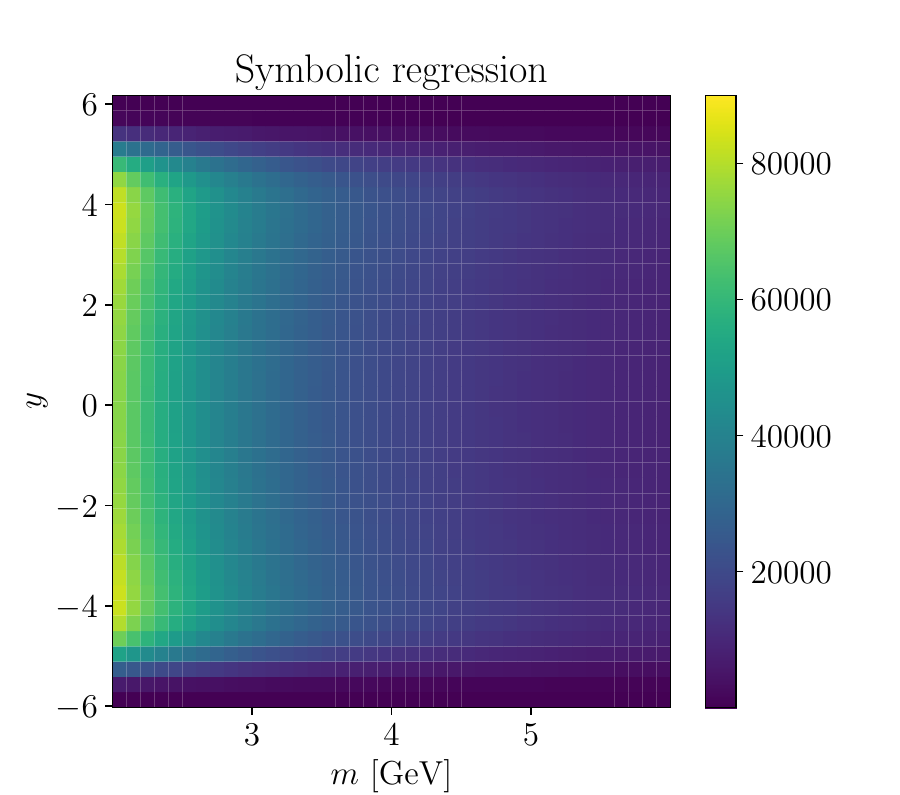}
    \caption{Parton luminosity values obtained with the symbolic regression model corresponding to a complexity of 33, see Table~\ref{tab:selected_equations}.}
    \label{fig:l2_sr}
    \label{fig:l2_lhapdf}
  \end{minipage}
  
  
  \begin{minipage}[b]{0.5\linewidth}
    \centering 
    \includegraphics[width=\linewidth]{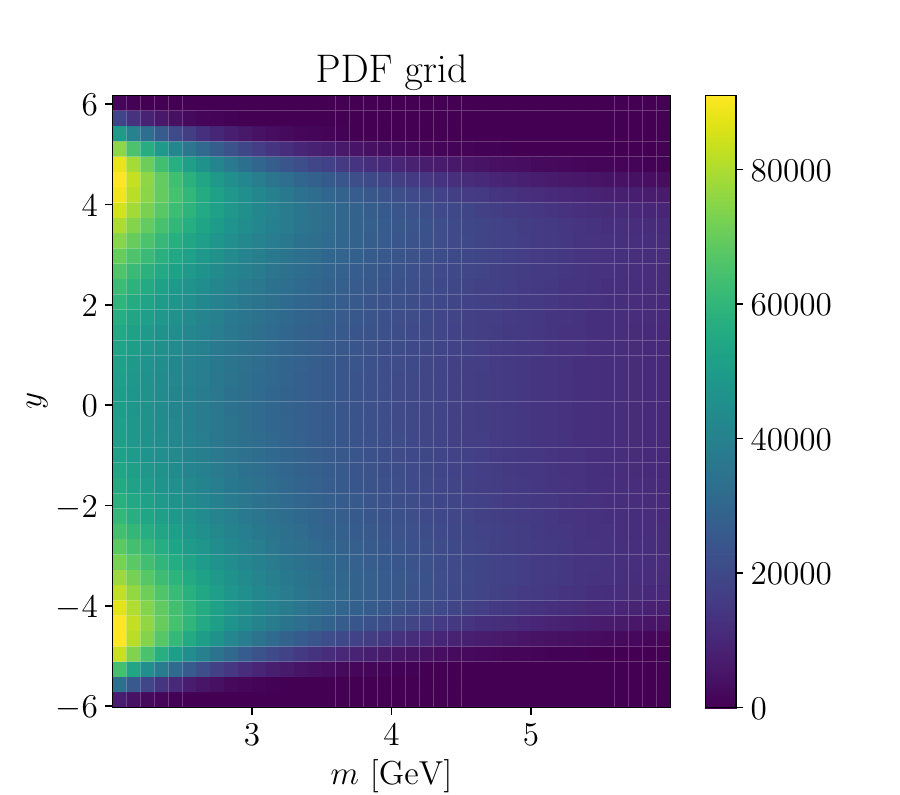}
    \caption{Benchmark for parton luminosity values obtained directly with numerical evaluation of Eq.~\eqref{eq:partonlumi} using the 
    CT10 NLO~\cite{Gao:2013xoa} central member from the LHAPDF grid~\cite{Buckley:2014ana}.}
  \end{minipage}
\end{figure}

The parton luminosities are computed numerically by reweighting the double differential distribution with 
the prefactor $\frac{8 \pi \alpha^2}{9 m s}$ as per Eq.~(\ref{eq:l2_pdf}), and they are presented in Fig.~\ref{fig:l2_mg}. 
Applying SR to this distribution generates a so-called, in the PySR terminology, hall of fame set of the best candidate models spanning a range of complexities, 
with representative examples summarised in Table~\ref{tab:selected_equations}. To make the notation more transparent in the equations that we obtain, 
dimensionful variables (like the invariant mass $m$) are understood to be divided by their corresponding dimension (GeV in the case of $m$) in order to make the 
equation dimensionally consistent. 
Among the candidates listed in the table, the model with the highest score, highlighted in gray in Table~\ref{tab:selected_equations}, 
has a complexity of $33$ and is shown in Fig.~\ref{fig:l2_sr}. As a baseline for our benchmark, Fig.~\ref{fig:l2_lhapdf} displays 
the corresponding integral computed directly by using Eq.~\eqref{eq:partonlumi} and the central member of the CT10NLO set 
from the LHAPDF library~\cite{Buckley:2014ana}.
From Figs.~\ref{fig:l2_mg}--\ref{fig:l2_sr}, we observe that SR produces a smooth functional approximation that agrees well with the 
direct computation of Eq.~\eqref{eq:partonlumi} displayed in Fig.~\ref{fig:l2_lhapdf}. The SR solution successfully extrapolates 
into unphysical kinematic regions by appropriately suppressing the parton luminosities (though not vanishing completely), 
while simultaneously reducing statistical fluctuations in the high invariant mass regime. 

The SR equations shown in Table~\ref{tab:selected_equations} demonstrate that at low complexity ($3$), 
the parton luminosities can be well-described by simple 
inverse power laws in the invariant mass of the final state $m$. As the complexity increases and resolution in $y$ is achieved, the SR automatically discovers 
the required $y \to - y$ symmetry dictated by fundamental particle kinematics.
Notably, while the highest-complexity solution ($35$) achieves greater accuracy by construction, it receives a lower overall score, illustrating the inherent 
trade-off between model complexity and interpretability in SR.

To compare the simulation (Fig.~\ref{fig:l2_mg}) and the SR results (Fig.~\ref{fig:l2_sr}) to the PDF grid baseline (Fig.~\ref{fig:l2_lhapdf}), in Table~\ref{tab:metrics_comparison} we use the following fit quality metrics: coefficient of determination $R^2$, and the weighted mean absolute percent error (wMAPE), defined as
\begin{equation*}
\text{wMAPE} = 
\frac{\sum_{i=1}^{n} |y_i - \hat{y}_i|}
     {\sum_{i=1}^{n} |y_i|},
\qquad
R^2 = 1 - 
\frac{\sum_{i=1}^{n} (y_i - \hat{y}_i)^2}
     {\sum_{i=1}^{n} (y_i - \bar{y})^2},
\label{eq:metrics}
\end{equation*}
where $n$ is the number of data points, $y_i$ the actual value, $\hat{y}_i$ the predicted value, and $\bar{y}$ the mean of $y_i$. The wMAPE provides a scale-independent measure of average deviation that remains well-defined even in the presence of bins with vanishing or near-zero values. Unlike the standard mean absolute percentage error (MAPE), which diverges when $y_i \to 0$, the wMAPE weights errors by the magnitude of the actual values, effectively normalising the total absolute error by the total signal. This makes it particularly robust for datasets with highly inhomogeneous bin populations or values spanning several orders of magnitude, as in this specific case. 

\begin{table}[htb]
\centering
\begin{tabular}{@{}lcc@{}}
\toprule
Metric & Reweighted simulation & Symbolic regression \\
\midrule
wMAPE & 18.21 \% & 16.35 \% \\
$R^2$ & 0.8898 & 0.9030 \\
\bottomrule
\end{tabular}
\caption{Comparison of metrics between the reweighted Monte Carlo simulation and SR result relative to the benchmark value of parton luminosities computed directly from Eq.~\eqref{eq:partonlumi}.}
\label{tab:metrics_comparison}
\end{table}

Compared to the reweighted simulation, SR achieves a lower wMAPE and a higher $R^2$, indicating improved fit quality and stronger correlation with the baseline reference. These results demonstrate the capability of SR to model the data with better precision and accuracy, now also incorporating the effects of non-perturbative QCD objects such as the PDFs and their associated parton luminosities.

\section{Angular coefficients from symbolic regression}
\label{sec:results}

The aim of the previous section was to validate the results obtained with SR. 
We first checked the case of lepton collisions and  recovered the equations from first principles in QED, 
and in the photon mediated proton collisions we recovered the parton luminosities associated to the underlying PDF set used in 
the simulation and, as a byproduct, we obtained a simple closed-form analytic equation to describe them. 
Now, we focus on the angular coefficients in the Drell-Yan process at ${\cal O}(\alpha_s)$, including real QCD corrections (jet emission). 
More specifically, we look at the tree-level process $p \ p \to l^+ l^- j$, with $l = e, \mu$. We do this at $\sqrt{s} = 8$ TeV, 
and consider distributions that are differential in the transverse momentum $p_T$ of the $Z$ vector boson, 
in the rapidity of the lepton pair $y$ and in their invariant mass $m$. 
In Sect.~\ref{subsec:expr} we review the analytical expression of the Drell-Yan cross section in terms of 
angular coefficients, then in Sect.~\ref{subsec:anal} we describe the analysis details. 
Subsequently in Sect.~\ref{subsec:1D} we present the one-dimensional (1D) 
SR equations for the angular coefficients as a function of $p_T$, $y$ and $m$ independently (the latter two variables 
just for $A_4$, which exhibits interesting structure and is used experimentally). 
In Sect.~\ref{subsec:2D} we present SR equations for the angular coefficients as a function of $(p_T, |y|)$ (and in $(m, |y|)$ for $A_4$). Finally, 
in Sect.~\ref{subsec:3D} we present the fully triple-differential 
SR expression for the angular coefficients depending on $(p_T, |y|, m)$.

\subsection{Angular coefficients}
\label{subsec:expr}

Precision electroweak measurements at the LHC often rely on the decomposition of the lepton angular distributions in the
Collins-Soper frame~\cite{Collins:1977iv} into nine spherical harmonic polynomials $P_i$, multiplied by 
angular coefficients $A_i$~\cite{Mirkes:1992hu,Mirkes:1994dp,Mirkes:1994eb,Mirkes:1994nr}. 
For lepton pair production, the full five-dimensional differential cross section 
describing the kinematics of the two Born-level leptons can be written as:
\begin{equation}
    \frac{d^5\sigma}{dp_T\, dy \, dm \,d\cos\theta \,d\phi} = \frac{3}{16\pi}\frac{d^3\sigma^{U+L}}{dp_T\, dy\, dm}
    \left[(1+\cos^2\theta) + \sum_{i=0}^7 P_i(\theta,\phi)\,A_i \right],
    \label{eq:full_coeff}
\end{equation}
where $p_T$ is the transverse momentum of the $Z$ boson, $y$ the rapidity of the lepton-antilepton pair, 
$m$ the invariant mass of the dileptons, while $\theta$ and $\phi$ are, 
respectively, the polar and azimuthal angle of the lepton in the Collins-Soper frame, which can be calculated in terms of the  
kinematic variables in the laboratory frame. 
While the fully differential Drell-Yan cross section is naturally expressed in terms of the laboratory-frame 
variables $p_T$, $y$, and $m$, the angular variables are conveniently described in the Collins-Soper frame.

The Collins-Soper frame is defined as the rest frame of the dilepton system, with the $z$-axis taken along the bisector of 
the incoming beam directions. The positive $z$-direction is chosen to align with the $z$-direction of the lepton pair in the 
laboratory frame. In this frame, the angle $\theta$ (sometimes called $\theta^*$ in the literature) is defined as the angle 
between the $z$-axis and the momentum of the negatively charged lepton. The $x$-axis lies in the plane spanned by the incoming beams 
and is orthogonal to the $z$-axis, while the $y$-axis is fixed by requiring a right-handed Cartesian coordinate system. 
The angle $\phi$ (sometimes called $\phi^*$ in the literature) is defined as the angle between the plane of the incoming hadrons and the outgoing negative lepton.

Explicitly, we can express the Collins-Soper angular variables in terms of the laboratory kinematic as 
\begin{align}
\cos\theta &= \frac{2(l^+ \bar{l}^- - l^- \bar{l}^+)}{Q \sqrt{Q^2 + \vec{Q}_T^2}}, \\
\tan\phi &= \frac{\sqrt{Q^2 + \vec{Q}_T^2}}{Q} \cdot \frac{\vec{\Delta}_T \cdot \hat{R}_T}{\vec{\Delta}_T \cdot \vec{Q}_T},
\end{align}
where
\begin{equation}
l^{\pm} = \frac{1}{\sqrt{2}} \left( p_l^E \pm p_l^z \right),
\quad
l, \bar{l} = \{e^-, \mu^-\}, \{e^+, \mu^+\},
\tag{2.2}
\end{equation}
with $Q_{\mu}$ the dilepton momentum, $Q = m$ the invariant mass of the dilepton system, $\Delta_{\mu} = l_{\mu} - \bar{l}_\mu$,  $\vec{\Delta}_{T}$ 
and $\vec{Q}_T$ being respective spatial transverse components, and $\hat{R}_T$ being a transverse unit vector in the direction of 
$\vec{P}_{\text{proton}} \times \vec{Q}$. Additionally, the spherical harmonics in Eq.~\eqref{eq:full_coeff} are given by
\begin{align}
    P_0(\theta,\phi)&=\frac{1}{2}(1-3\cos^2\theta),\notag\\
    P_1(\theta,\phi)&=\sin2\theta\,\cos\phi,\notag\\
    P_2(\theta,\phi)&=\frac{1}{2}\sin^2\theta\cos2\phi,\notag\\
    P_3(\theta,\phi)&=\sin\theta\cos\phi,\notag\\
    P_4(\theta,\phi)&=\cos\theta,\notag\\
    P_5(\theta,\phi)&=\sin^2\theta\sin2\phi,\notag\\
    P_6(\theta,\phi)&=\sin2\theta\sin\phi,\notag\\
    P_7(\theta,\phi)&=\sin\theta\sin\phi.\notag
\end{align}
In this way, the dependence on $p_T$, $y$ and $m$ is entirely contained in the unpolarised cross section $\sigma^{U+L}$ and in the $A_i$ angular coefficients. 
The angular coefficients $A_i$ are functions of $(p_T,y,m)$ and can be 
extracted by evaluating weighted averages over angular distributions obtained from Monte Carlo simulations 
at a given perturbative order~\cite{Bern:2011ie}.  Specifically, each coefficient can be calculated in terms of the expectation value of an angular function
\begin{align}
    \label{eq:coefs}
    A_0&= 4-10\langle\cos^2\theta\rangle,\\
    A_1&=\langle 5\sin2\theta\,\cos\phi\rangle\notag,\\
    A_2&=\langle 10\sin^2\theta\cos2\phi\rangle\notag,\\
    A_3&=\langle 4\sin\theta\cos\phi\rangle\notag,\\
    A_4&=\langle 4\cos\theta\rangle \notag,\\
    A_5&=\langle\sin\theta\sin\phi\rangle\notag,\\
    A_6&=\langle 5\sin2\theta\sin\phi\rangle\notag,\\
    A_7&=\langle 5\sin^2\theta\sin2\phi\rangle\notag,
\end{align}
where $\langle \dots \rangle$ denotes taking the normalised weighted average over the Collins-Soper angular variables $\theta$, $\phi$ and is defined as

\begin{equation}\label{eq:av}
\langle f(\theta, \phi) \rangle \equiv 
\frac{\int_{-1}^{1} \mathrm{d} \cos\theta \int_{0}^{2\pi} \mathrm{d}\phi \, \mathrm{d}\sigma(\theta, \phi) \, f(\theta, \phi)}
{\int_{-1}^{1} \mathrm{d} \cos\theta \int_{0}^{2\pi} \mathrm{d}\phi \, \mathrm{d}\sigma(\theta, \phi)}.
\end{equation}

At leading order (LO) or in matrix-element plus parton-shower (ME+PS) simulations, the coefficients $A_5$, $A_6$, and $A_7$ vanish. 
This is because the associated angular functions are odd under parity and “naive” time-reversal transformations, both of which invert 
the azimuthal angle: $\phi \to -\phi$. These symmetries ensure that the integrals of such functions cancel out in the absence of absorptive phases. 
However, at next-to-leading order (NLO) in QCD, these coefficients can receive small contributions from the absorptive parts of 
one-loop amplitudes~\cite{Hagiwara:1984hi,Gauld:2017tww,Lyubovitskij:2025oig}. However, their transverse momentum distributions remain strongly suppressed at the LHC due to cancellations 
between the forward and backward rapidity regions.

In contrast, the coefficients $A_0$ to $A_4$ are expected to exhibit nontrivial structure, and we focus our analysis on these five. 
Among them, $A_0$ and $A_2$ are particularly interesting due to the so-called Lam–Tung relation~\cite{Lam:1978zr,Lam:1978pu,Lam:1980uc}, $A_0 = A_2$, which holds 
up to ${\cal O}(\alpha_s)$ in the Drell–Yan process. This relation arises from the spin-1 nature of the intermediate boson and the helicity 
structure of the quark–antiquark annihilation, and is formally analogous to the Callan–Gross relation in deep inelastic scattering~\cite{Callan:1969uq}. 
Notably, Lam and Tung showed that the relation remains exact at $\mathcal{O}(\alpha_s)$ in QCD, even though the individual structure functions receive 
large radiative corrections. This makes $A_0 = A_2$ a robust signature of the underlying partonic dynamics and a unique test of the QCD-improved parton model. 
Deviations from this relation at high transverse momentum can signal the onset of higher-twist effects, transverse-momentum dependence, 
or contributions beyond LO, and are thus of particular experimental and theoretical 
interest~\cite{Gehrmann-DeRidder:2015wbt,Gehrmann-DeRidder:2016jns,Gauld:2024glt,Piloneta:2024aac,Li:2024iyj,Li:2025fom,Arroyo-Castro:2025slx}.

%

\subsection{Analysis details}
\label{subsec:anal}

The simulation used as an input for the SR task were generated using \texttt{MadGraph5}
\texttt{aMC@NLO}~\cite{Alwall:2014hca} at leading order accuracy for the process $pp \to \ell^+ \ell^- j$. 
The simulation is done at fixed order and does not include any parton shower. It employes 
dynamic renormalisation and factorisation scales, which sets the scale to the transverse mass of the dilepton system. 
The reference scale is set to \( \mu = m_Z \). The input parton distribution function is the \texttt{NNPDF40\_nlo\_as\_01180} PDF set~\cite{NNPDF:2021njg}. 
Only the events with invariant mass of the dilepton system in the range \( 81~\mathrm{GeV} < m < 110~\mathrm{GeV} \) are kept, following 
the experimental selection cuts in Ref.~\cite{ATLAS:2023lsr}. A minimum jet transverse momentum of \( p_T^j > 0.01~\mathrm{GeV} \) was imposed to 
reduce the computation time. No other cuts were imposed on jets or leptons. Events were analysed at parton level.

In the case of the $p_T$ dependence, a total of 6 million events were generated, divided into five exclusive 
bins\footnote{1 million events were produced in each bin with the exception of B2, where 2 million events 
were used to access slightly higher statistical sensitivity towards the upper bound of the bin.} of the 
transverse momentum of the leading jet $p_T^j$:

\begin{itemize}
  \item B1: $0.01~\text{GeV} < p_T^j \leq 10~\text{GeV}$,
  \item B2: $10~\text{GeV} < p_T^j \leq 30~\text{GeV}$,
  \item B3: $30~\text{GeV} < p_T^j \leq 50~\text{GeV}$,
  \item B4: $50~\text{GeV} < p_T^j \leq 80~\text{GeV}$,
  \item B5: $p_T^j > 80~\text{GeV}$.
\end{itemize}

This binning strategy allows for good coverage of the relevant $p_T$ ranges for Drell–Yan production at the LHC, 
while ensuring sufficient event statistics in each region to perform reliable angular coefficient extraction.

In each bin of kinematic variables (e.g., $p_T$, $y$, or  $m$), the angular coefficients \( A_i \) are 
extracted from the differential cross section in Eq.~\eqref{eq:full_coeff}. 
Each coefficient \( A_i \) is extracted from the data by using the projection integral of the form given 
in Eq.~\eqref{eq:coefs} and ~\eqref{eq:av}. In practice, this is computed using a weighted average over 
simulated events, which is a discretisation of Eq.~\eqref{eq:av}
\begin{equation}
\label{eq:av_disc}
\langle f \rangle = \frac{\sum_i w_i f_i}{\sum_i w_i},
\end{equation}
where \( f_i \) is the angular function evaluated for event \( i \), and \( w_i \) is the event weight. 
The weights account for the cross section and can be defined for each event $i$ as
\begin{equation*}
w_i = \frac{\sigma}{N_{\text{events}}},
\end{equation*}
with \( \sigma \) the integrated cross section and \( N_{\text{events}} \) the total number of events for the simulation run from which event \( i \) originates. Note that the events are generated in mutually exclusive bins of $p_T^j$, so within each $p_T$ bin the events can be treated as effectively unweighted. However, when combining samples across bins in other variables, the proper event weights are used following Eq.~\eqref{eq:av_disc}. 
The variance of the estimator $\langle f \rangle$, assuming the values \( f_i \) are independent and identically distributed with finite variance, is given by
\begin{equation}
\mathrm{Var}(\langle f \rangle) = \frac{1}{N} \left( \frac{\sum_i w_i^2 (f_i - \bar{f})^2}{\left(\sum_i w_i\right)^2} \right),
\end{equation}
where \( \bar{f} \) is the weighted mean $\bar{f} = \frac{\sum_i w_i f_i}{\sum_i w_i}$.
%
%
This procedure yields an uncertainty estimate consistent with the statistical fluctuations expected in a binned analysis of finite-size 
Monte Carlo samples, while preserving the normalisation to the correct cross section per bin. The effects of systematic uncertainties are not included and are left for future work.

\subsection{1D angular coefficients}
\label{subsec:1D}

In this section, we present analytical expressions obtained via SR for the angular coefficients as functions of $p_T$, $y$, and $m$, independently; 
that is, we compute $A_i(p_T)$ for $i = 0,1,2,3,4$, and $A_4(y)$, and $A_4(m)$. The 1D dependence on $y$ and $m$ of the other angular coefficients can 
be easily obtained in a analogous way, but here we focus on $A_4$. 
The input data provided to PySR were taken from the simulation, with the angular coefficients calculated using Eq.~\eqref{eq:coefs}. 
The operator set available for the SR task consists of the binary operators $+, -, \times, \div$ for all individual 
angular coefficients $A_0, \dots, A_4$. To provide the possibility of slightly higher expressivity, in case of the Lam-Tung relation~\cite{Lam:1978pu,Lam:1978zr,Lam:1980uc} 
we use $+, -, \times, \text{^}$ (the power operator). 
In all cases, we let the algorithm run for 1000 iterations, and we verify that the quality of the results is stable upon increasing the number of iterations. Uncertainty estimations for the output of 1D angular coefficients are investigated in the appendix \ref{app:SR_uncertainties}.

Naturally, while a larger set of operators (and of potentially higher complexities) could be incorporated to expand the combinatorial space of possible equations, 
and the tuning of many hyperparameters could be explored -- such as number of iterations, early stopping, nested constraints, custom functions, etc -- 
the settings we have chosen are already sufficient to accurately capture the shape of the observed distributions with simple analytical expressions. 

\begin{figure}[ht!]
  \centering

  \begin{minipage}[b]{0.48\linewidth}
    \includegraphics[width=\linewidth]{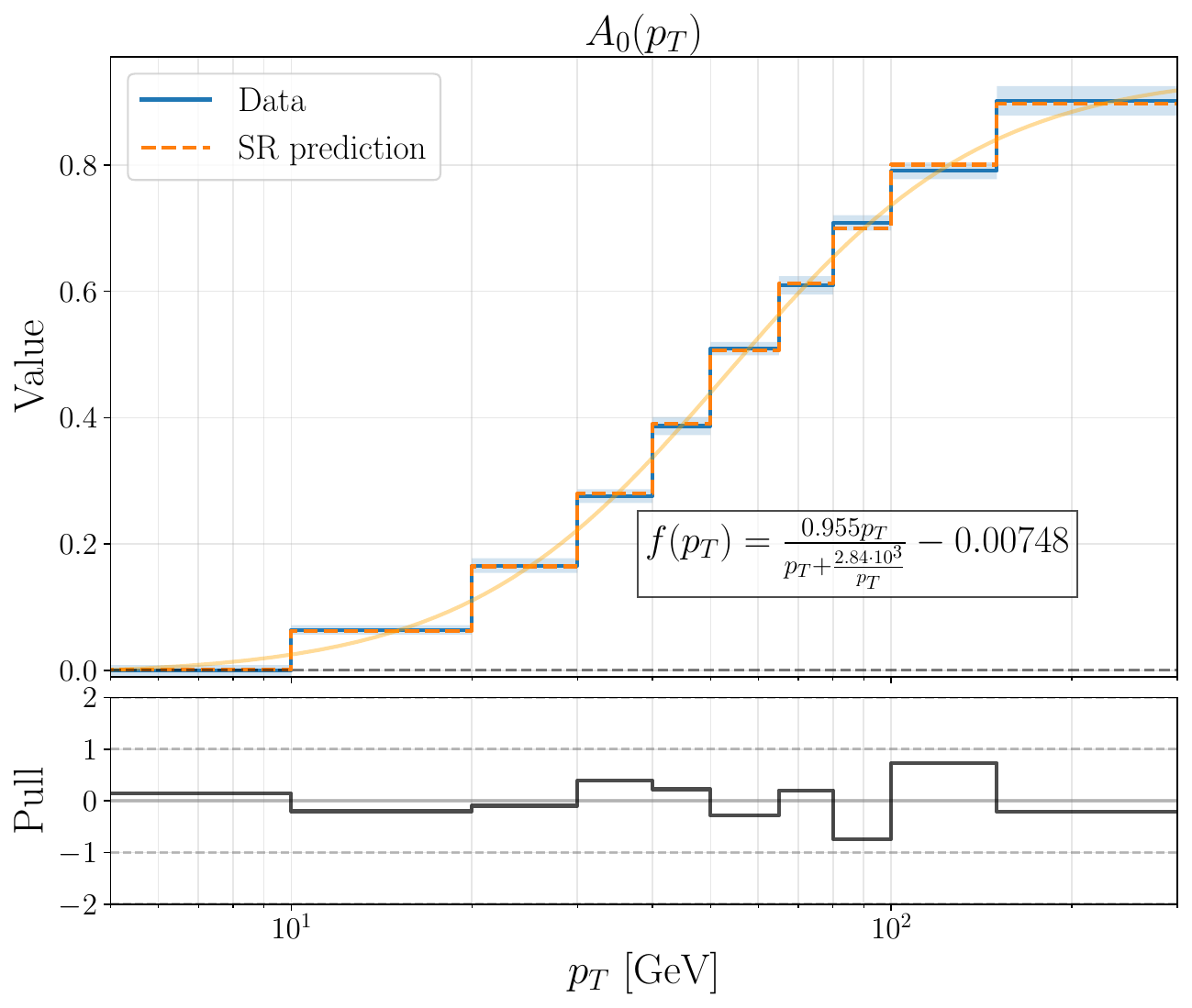}
    \caption{$A_0 (p_T)$ simulation vs. SR.}
    \label{fig:a0_1d}
  \end{minipage}
  \begin{minipage}[b]{0.48\linewidth}
    \includegraphics[width=\linewidth]{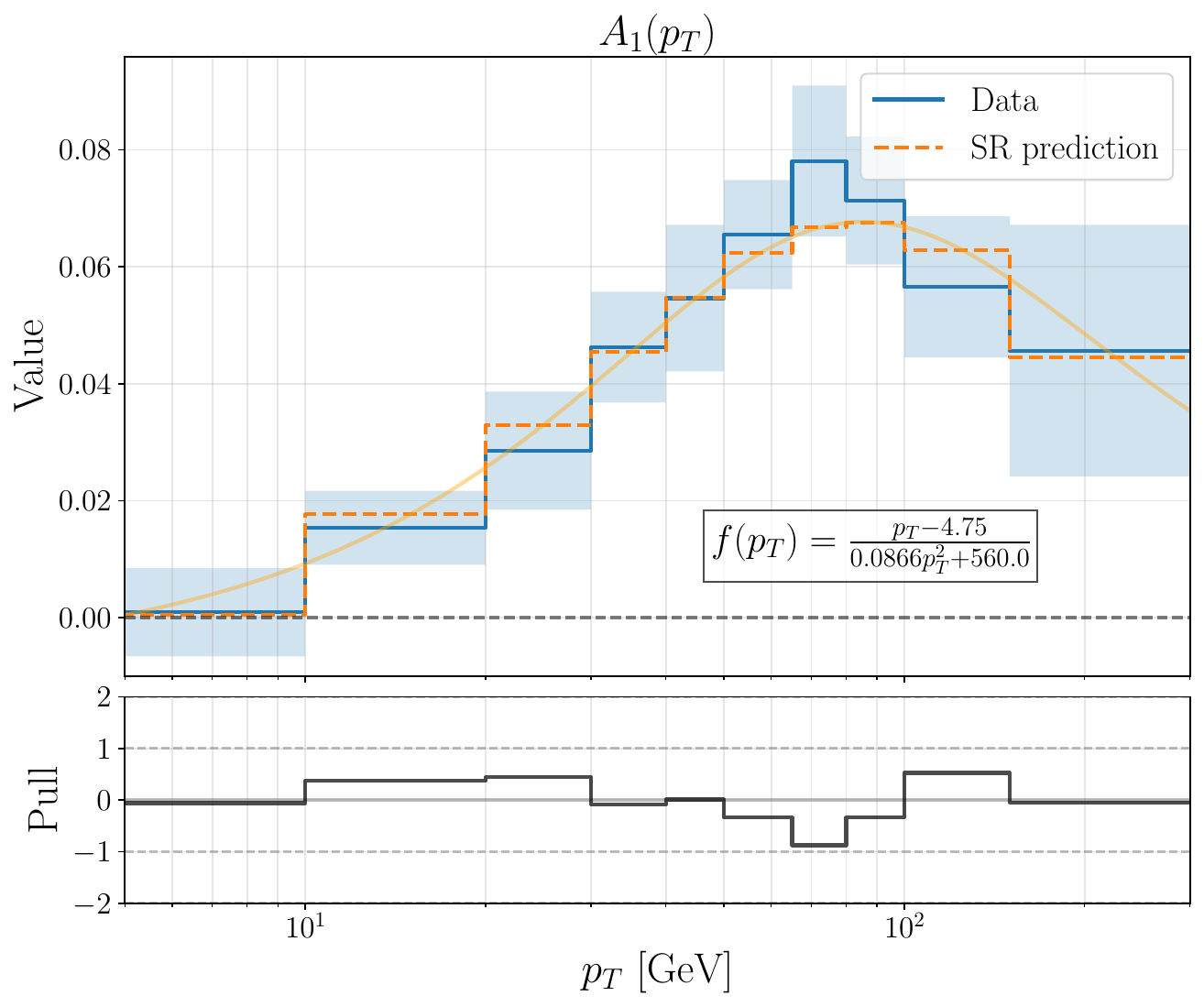}
    \caption{Same as Fig.~\ref{fig:a0_1d} for $A_1$.}
    \label{fig:a1_1d}
  \end{minipage}

  \begin{minipage}[b]{0.48\linewidth}
    \includegraphics[width=\linewidth]{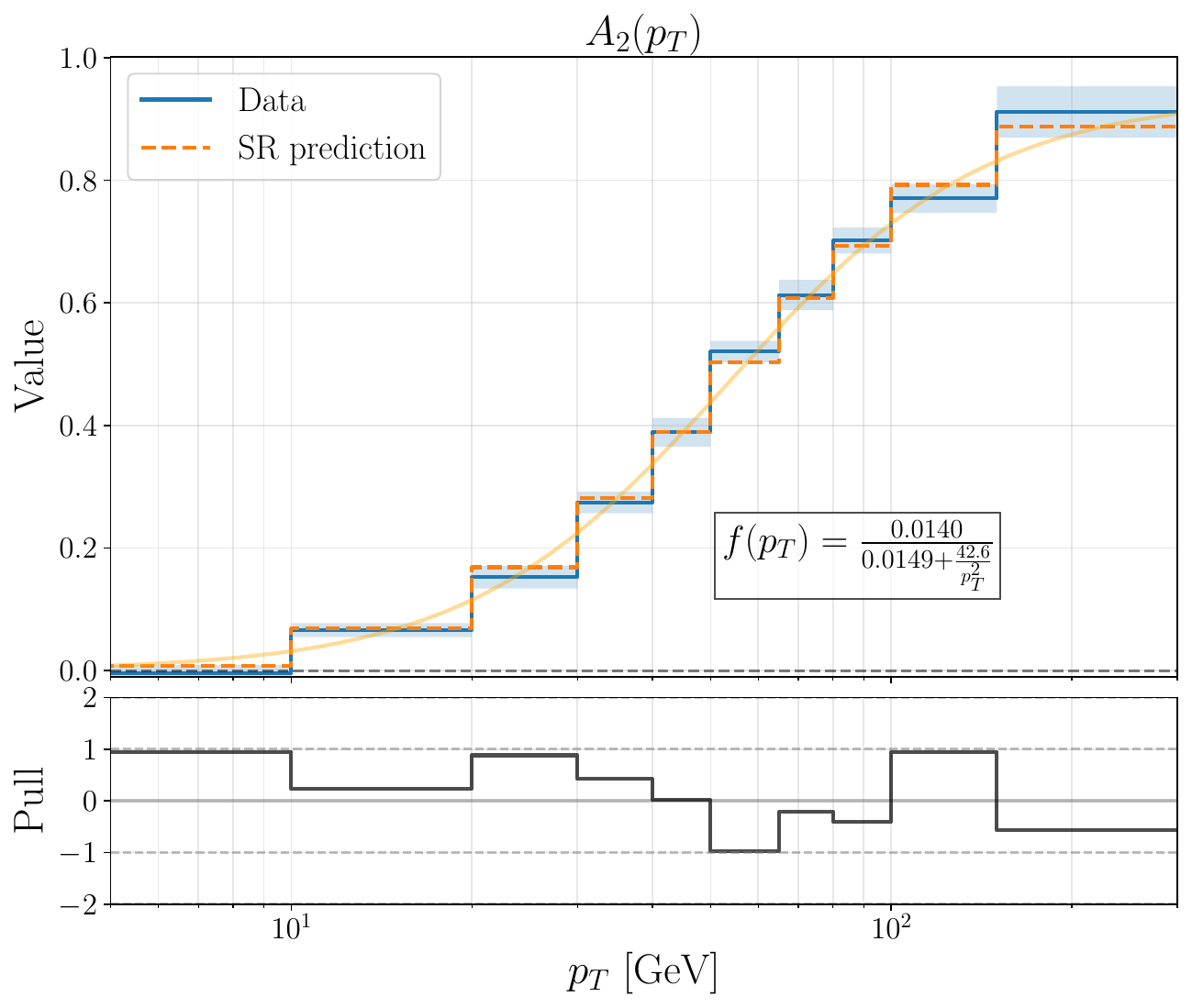}
    \caption{Same as Fig.~\ref{fig:a0_1d} for $A_2$.}
    \label{fig:a2_1d}
  \end{minipage}
  \begin{minipage}[b]{0.48\linewidth}
    \includegraphics[width=\linewidth]{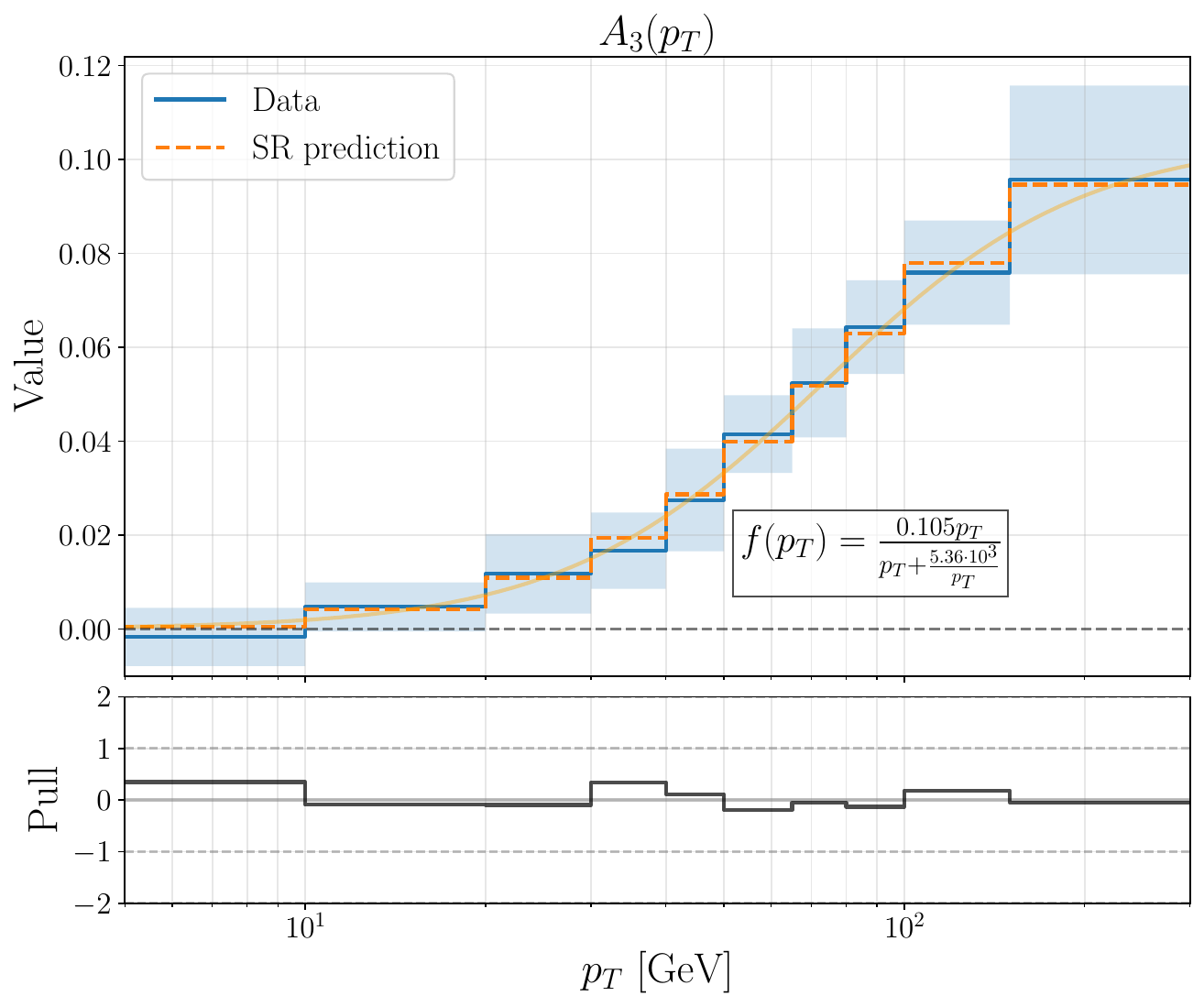}
    \caption{Same as Fig.~\ref{fig:a0_1d} for $A_3$.}
    \label{fig:a3_1d}
  \end{minipage}

  \vspace{1em}

  \begin{minipage}[b]{0.48\linewidth}
    \centering
    \includegraphics[width=\linewidth]{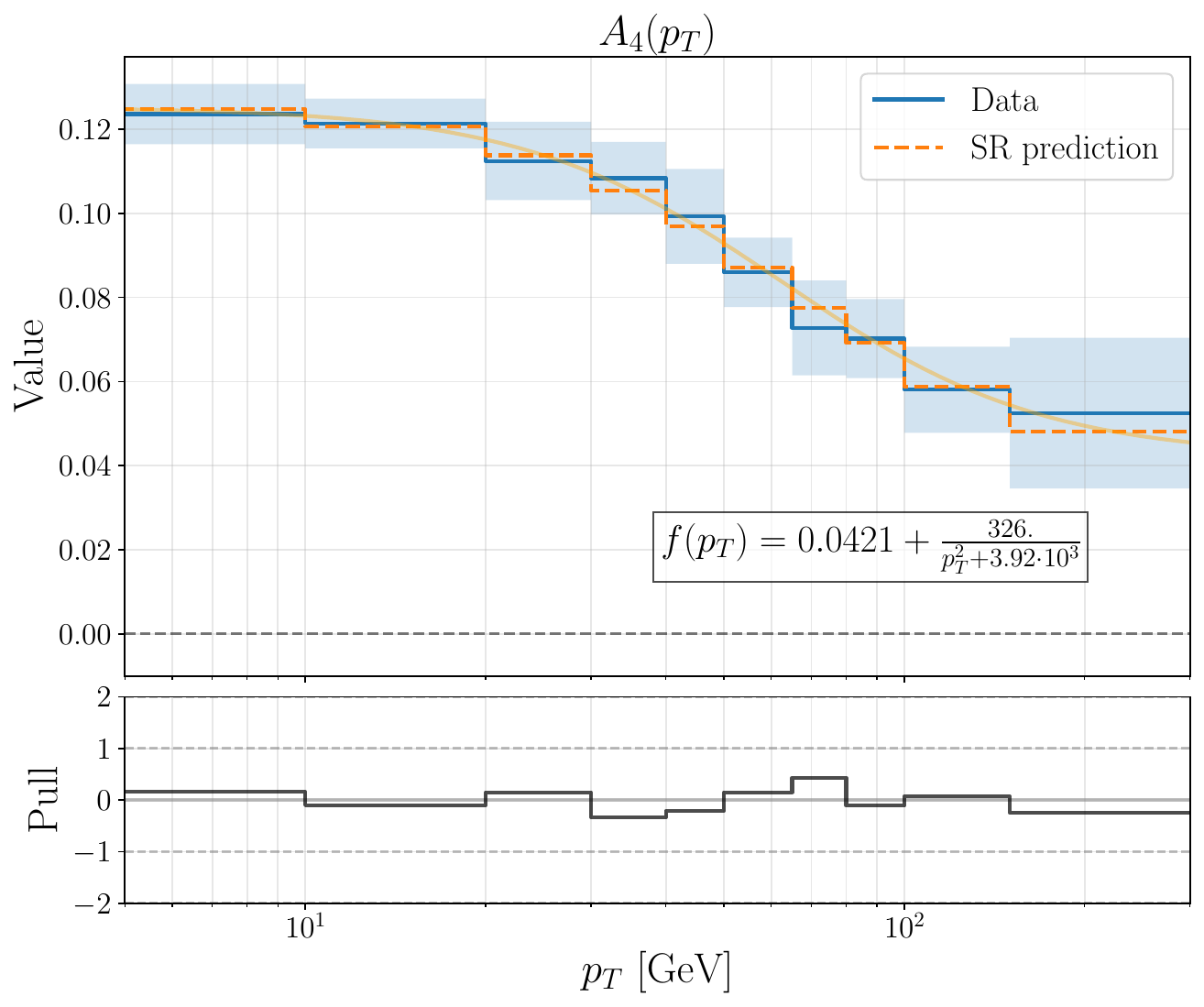}
    \caption{Same as Fig.~\ref{fig:a0_1d} for $A_4$.}
    \label{fig:a4_1d}
  \end{minipage}
  \begin{minipage}[b]{0.48\linewidth}
    \includegraphics[width=\linewidth]{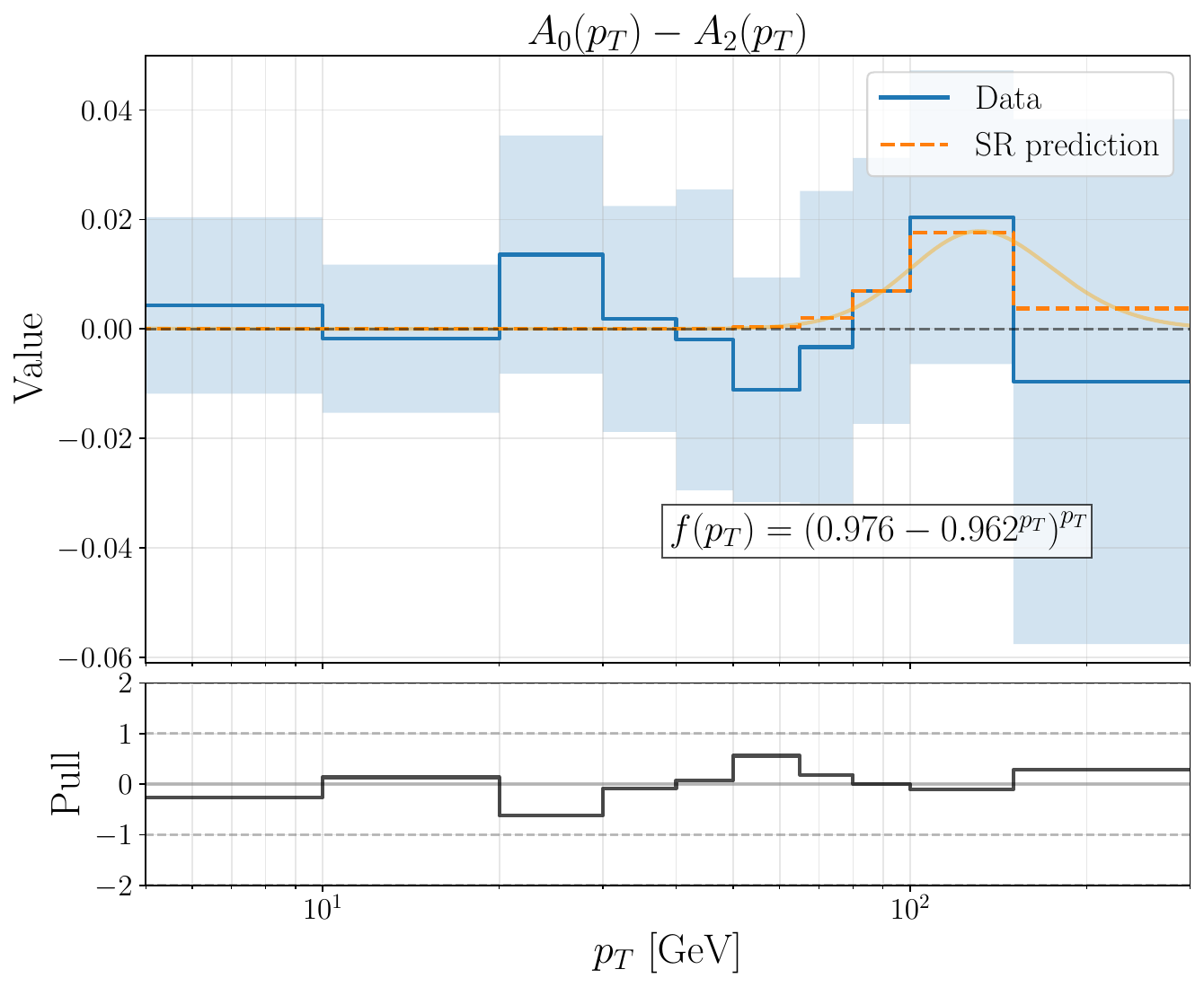}
    \caption{Same as Fig.~\ref{fig:a0_1d} for $A_0 - A_2$.}
    \label{fig:a0ma2_1d}
  \end{minipage}
\end{figure}

In Figs. ~\ref{fig:a0_1d}, ~\ref{fig:a1_1d}, ~\ref{fig:a2_1d}, ~\ref{fig:a3_1d}, ~\ref{fig:a4_1d} and~\ref{fig:a0ma2_1d} we show the SR analytic expressions for respective 
angular coefficients $A_0, \dots, A_4$ as a function of the $p_T$ of the EW boson, as well as checking the Lam-Tung relation. In the figures, we show the values of the angular 
coefficients as functions of $p_T$, computed from the simulation, alongside the corresponding SR expressions. The lower panel shows the pull, defined as $(\text{SR prediction} - \text{simulation}) / \text{uncertainty}$, which measures the deviation between SR and full MC simulation in units of the corresponding statistical uncertainty. Across all angular coefficients, we find that SR provides a very good description of the simulated behaviour across different $p_T$ regimes. The symbolic expressions reproduce the simulation within 
uncertainties over the complete kinematic range. 

The SR regressor was trained only on the bin centres and their corresponding angular coefficient values, using a 
relatively small number of training points to search within a highly combinatorial function space. Despite being trained 
on a limited set of discrete values (bin centres), SR produces smooth and continuous functional expressions that effectively 
interpolate across the full kinematic domain. These expressions can capture the underlying trends of the angular coefficients 
with greater continuity and resolution than standard binned distributions, which inherently lose information due to their discretised nature.

The expressions shown in Figs.~\ref{fig:a0_1d}–~\ref{fig:a4_1d} are composed solely of rational functions. As such, their behaviour 
in extreme kinematic regimes can diverge if no data are available to constrain the functional form in those regions. For this reason, 
extrapolating the SR expressions beyond the domain covered (by the bin centres) by the training data should be avoided. 
If new data become available in such regimes, retraining the model is recommended to ensure reliable predictions. 
Caution is also required when interpolating within the domain, due to the potential appearance of poles introduced by 
the division operator $\div$, which can lead to artificial singularities in otherwise smooth regions.

In the previous figures, we show only the single equation selected according to the \texttt{best} criterion. 
However, as discussed in earlier sections, PySR does not produce just one solution, but rather a hall of fame, 
a family of candidate equations that achieve varying levels of accuracy, complexity, and score in describing the data. 
To illustrate this point, in Table~\ref{table:a0_1d} we present the hall of fame for the $A_0$ coefficient 
and highlight in gray the best symbolic expression. 
As the selection criterion indicates, the chosen expression is neither the most complex nor the one with the lowest loss 
(i.e., highest accuracy). This reflects the essential trade-off between precision and interpretability. Similarly, the 
selected equation is not the one with the highest score (which would correspond the \texttt{score} selection criterion), but 
rather the one with the highest score among those whose loss remains close to that of the most accurate model, to be precise
the highest score among those whose loss \(L\) is within 1.5 times the minimum achievable loss \(L_{\text{min}}\):
\begin{equation}
\label{eq:cond}
L \leq 1.5 \times L_{\text{min}}.
\end{equation}
For a detailed explanation of the selection criteria, see Appendix~\ref{app:sr}. 

\begin{table}[h]
\footnotesize
\setlength{\tabcolsep}{5pt}
\renewcommand{\arraystretch}{1.5}
\begin{center}
\begin{tabular}{@{}c c c c@{}}
\toprule
Equation & Complexity & Loss & Score \\
\midrule
$f(p_T) = 0.313$ & 1 & 0.0795 & 0.0 \\
$f(p_T) = 0.00640\,p_T$ & 3 & 0.0129 & 0.908 \\
$f(p_T) = \frac{p_T}{p_T + 67.5}$ & 5 & 0.00962 & 0.148 \\
\makecell{$f(p_T) = \frac{p_T}{p_T + \frac{3.30 \cdot 10^3}{p_T}}$} & 7 & $1.48\cdot10^{-4}$ & 2.09 \\
\makecell{$f(p_T) = \frac{p_T}{1.05\,p_T + \frac{3.08 \cdot 10^3}{p_T}}$} & 9 & $3.56\cdot10^{-5}$ & 0.712 \\
\hline
\hline
\rowcolor{lightgray} 
\makecell{$f(p_T) = \frac{0.955\,p_T}{p_T + \frac{2.84 \cdot 10^3}{p_T}} - 0.00748$} & 11 & $1.89\cdot10^{-5}$ & 0.315 \\
\makecell{$f(p_T) = \frac{0.955\,p_T}{p_T + \frac{2.84 \cdot 10^3}{p_T}} - 0.00748$} & 13 & $1.89\cdot10^{-5}$ & $1.99\cdot10^{-5}$ \\
\makecell{$f(p_T) = \frac{0.953\,p_T}{p_T + \frac{2.84 \cdot 10^3}{p_T}}$\\$- 0.00634 - \frac{0.0102}{p_T}$} & 15 & $1.87\cdot10^{-5}$ & 0.00562 \\
\makecell{$f(p_T) = \frac{0.953\,p_T}{p_T - 0.0158 + \frac{2.84 \cdot 10^3}{p_T}}$\\$- 0.00634 - \frac{0.0102}{p_T}$} & 17 & $1.87\cdot10^{-5}$ & $1.23\cdot10^{-4}$ \\
\makecell{$f(p_T) = \frac{0.953\,p_T}{p_T - 0.0158 + \frac{2.84 \cdot 10^3}{p_T - 0.0424}}$\\$- 0.00634 - \frac{0.0102}{p_T}$} & 19 & $1.87\cdot10^{-5}$ & $8.02\cdot10^{-4}$ \\
\bottomrule
\end{tabular}
\end{center}
\caption{SR hall of fame for the $A_0$ coefficient. The symbolic expression with the best score is highlighted in gray, and the horizontal line corresponds to 
Eq.~\eqref{eq:cond}. The best score is the highest score among those below the horizontal line.}
\label{table:a0_1d}
\end{table}

Note that the equation shown in Fig.~\ref{fig:a0_1d} corresponds to the one with complexity 11 in 
Table~\ref{table:a0_1d}\footnote{Notice that different expression trees can simplify to the same algebraic equations, 
like the one with complexity 13. This is a feature of PySR and it arises from its simplify-optimise loop, which allows 
for the exploration of more expressions while optimising the size of the search space.}.

Similar considerations apply to the other angular coefficients $A_1$ through $A_4$, whose complete halls of fame are provided in Tables~\ref{table:1d_a1}–\ref{table:1d_a4} in Appendix~\ref{app:1d_eqs}, altogether with the $A_0 - A_2$ difference in Table~\ref{table:a0_minus_a2_1d}. It is useful to have a family of equations that describe the data, and not just one, to be able to adapt the description of the angular coefficient to the analyses that the user may wish to implement: one could work with the equation with the lowest loss if the accuracy is the determining factor, or with simpler models and higher score if computational cost or complexity are a constraint. It is interesting to note that several equations in the $A_0$ hall of fame exhibit a similar scaling to the one that can be obtained in the small $p_T$ limit from first principle calculations in quantum field theory\footnote{Defining $r \equiv p_T/M$, it can be shown that in the vanishing $p_T$ limit $A_0^{p_T \to 0} \approx r^2 / (1+r^2)$.}. Given that our analysis has finite resolution at small $p_T$ because the distribution has to be binned, the agreement naturally cannot be exact (and the analytical expression can only be used in a small part of the kinematic coverage, while the SR expression spans the whole range). 

After discussing the dependence of the angular coefficients on transverse momentum \( p_T \), we turn to their one-dimensional dependence on rapidity and invariant mass, \( y \) and \( m \). In this case, for brevity we limit ourselves to the $A_4$ angular coefficient, since this is the only one which is non-zero at tree-level and is therefore the most relevant when integrating over $p_T$ (although the coefficient $A_0, \dots, A_3$ could be regressed in the same way and present no inherent extra difficulty). 

\begin{figure}[ht!]
    \centering
    \includegraphics[width=0.49\linewidth]{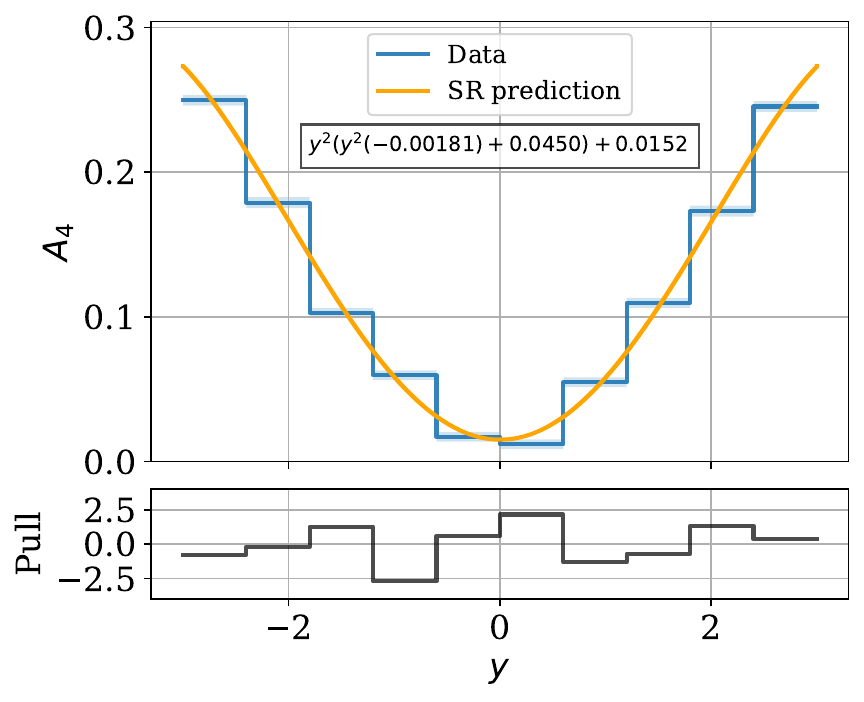}
    \includegraphics[width=0.49\linewidth]{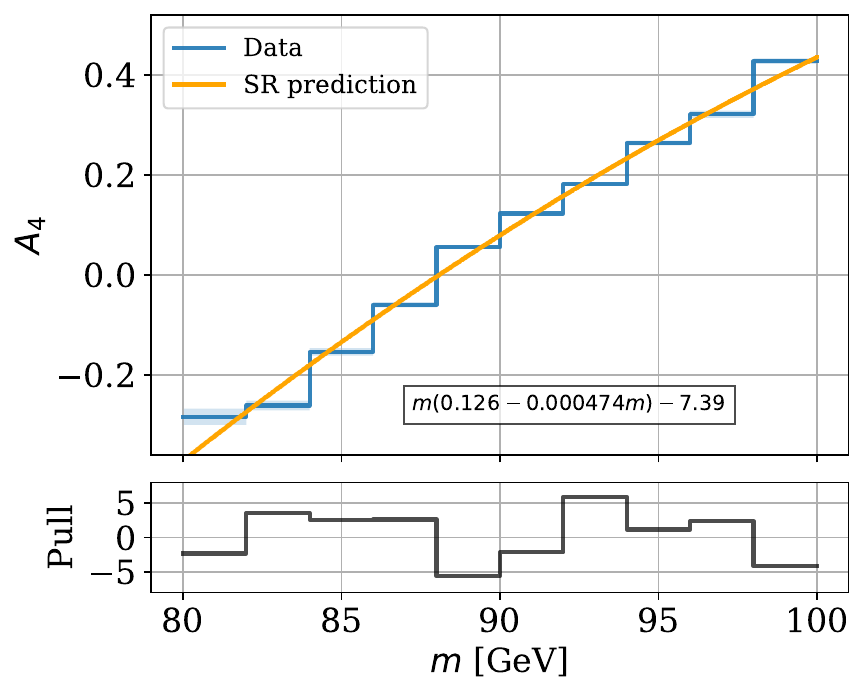}
    \caption{$A_4$ as a function of dilepton rapidity (left) and invariant mass (right).}
    \label{fig:A41D}
\end{figure}

In Fig.~\ref{fig:A41D}, we show the simulation and SR result for $A_4$. The left plot shows $A_4(y)$, which is described by the regressor as a simple polynomial function of $y$. It is interesting to note that the SR expression recovers the $y \leftrightarrow -y$ symmetry which comes from fundamental particle kinematics (note that the regression could also have been performed over the absolute value of the rapidity $|y|$, in which case the symmetry is imposed by construction). The right plot shows $A_4(m)$, which can be described by a simple linear function in $m$ with a small quadratic correction. The halls of fame for both $A_4$ distributions are given respectively in Tables~\ref{table:a4_1d_y} and~\ref{table:a4_1d_m} in App.~\ref{app:1d_eqs}.



In the next section, we revisit the same events but analyze them using two-dimensional distributions in \( (p_T, |y|) \) or \( (m, |y|) \). In these double-differential settings, PySR also succeeds in identifying meaningful functional dependencies, as the inclusion of extra kinematic dependence provides additional information that reveals the underlying structure of the 1D distributions we have shown. As an illustration of this behaviour, Appendix~\ref{app:1d_eqs} shows how the symbolic fit to \( A_0(y) \) changes when applying different \( p_T \) cuts, leading to different shapes and functional forms. 

\subsection{2D angular coefficients}
\label{subsec:2D}
In this section, we introduce an additional kinematic variable in the description of the angular coefficients and perform 2D SR in the $(p_T, |y|)$ plane. 
The central values and uncertainties of the simulation are shown in Figs.~\ref{fig:Ai_2d_cv_unc_1} and~\ref{fig:Ai_2d_cv_unc_2} in Appendix~\ref{app:2d_eqs}. 
In the case of $A_4$, we also study the distribution on the $(m, |y|)$ plane. For these double-differential distributions we use the binary operators $+, -, \times, \text{^}$.

\begin{figure}[ht!]
  \centering

  \begin{minipage}[b]{0.9\linewidth}
    \includegraphics[width=\linewidth]{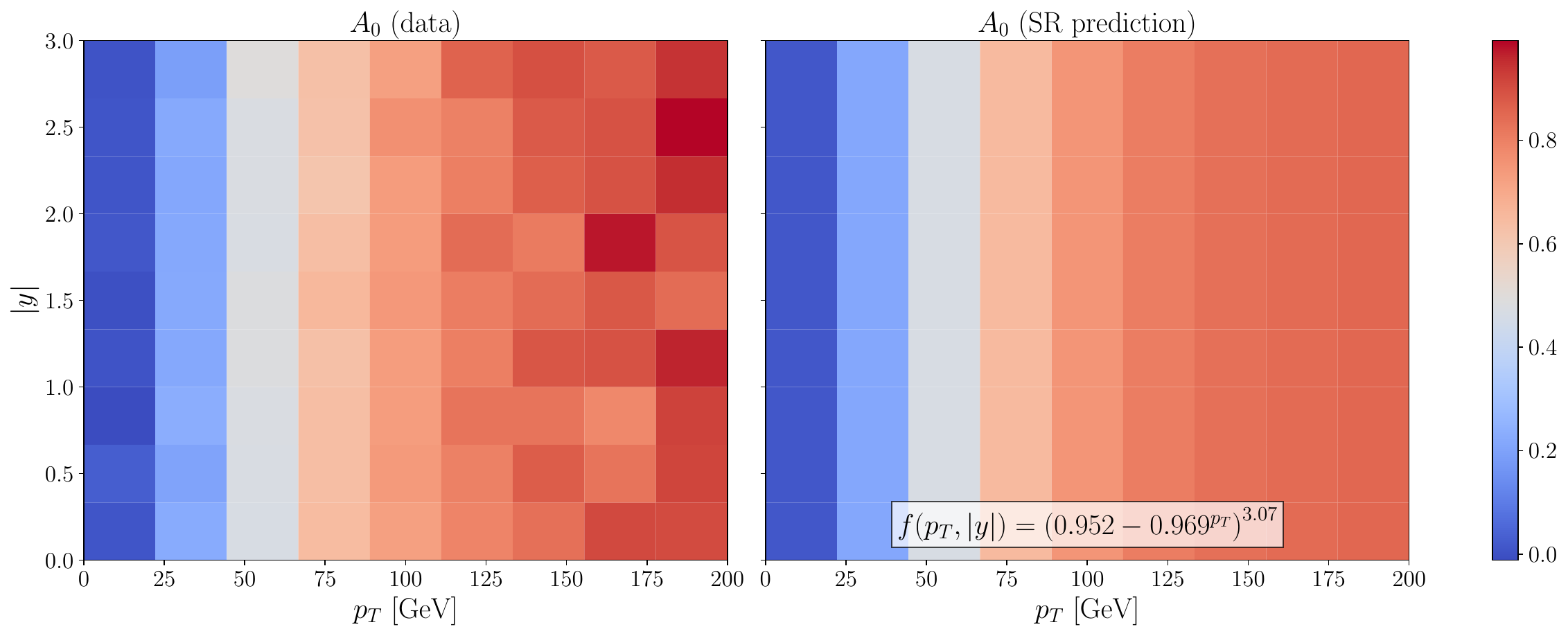}
    \label{fig:a0_2d}
  \end{minipage}

  \begin{minipage}[b]{0.9\linewidth}
    \includegraphics[width=\linewidth]{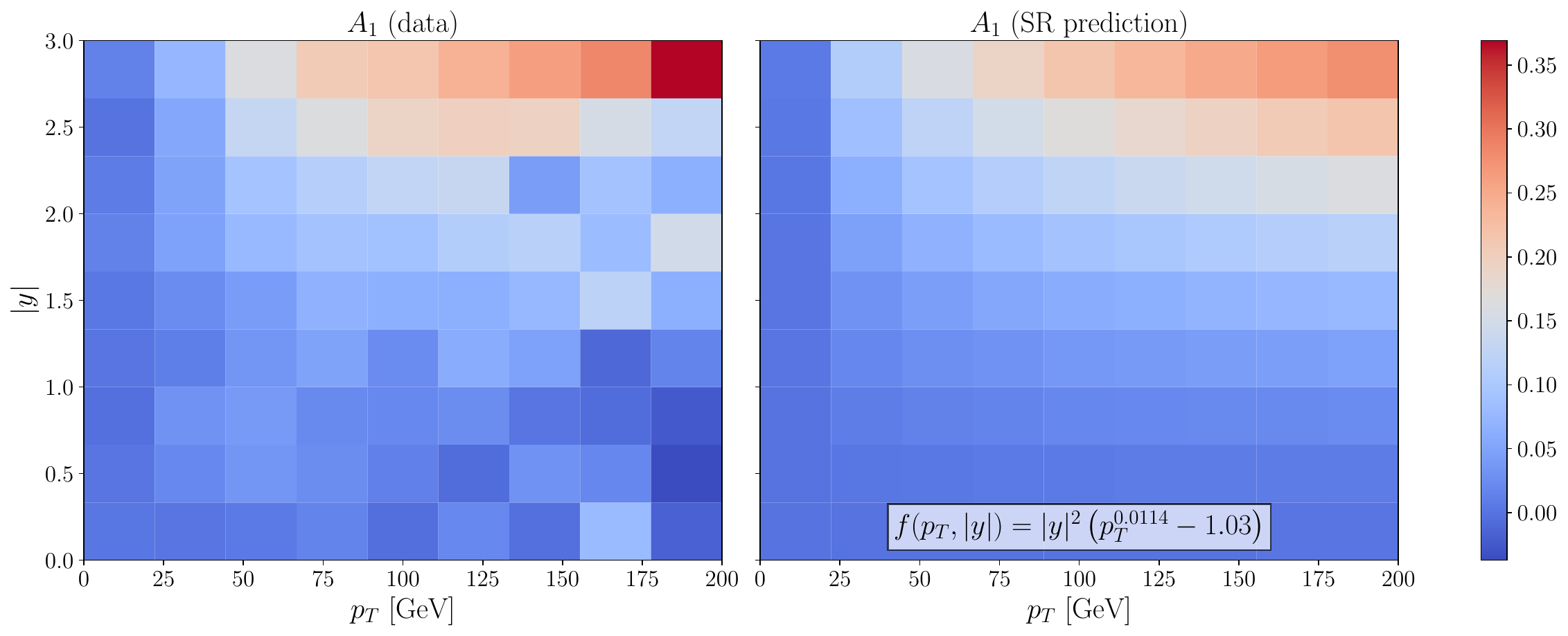}
    \label{fig:a1_2d}
  \end{minipage}

  \begin{minipage}[b]{0.9\linewidth}
    \includegraphics[width=\linewidth]{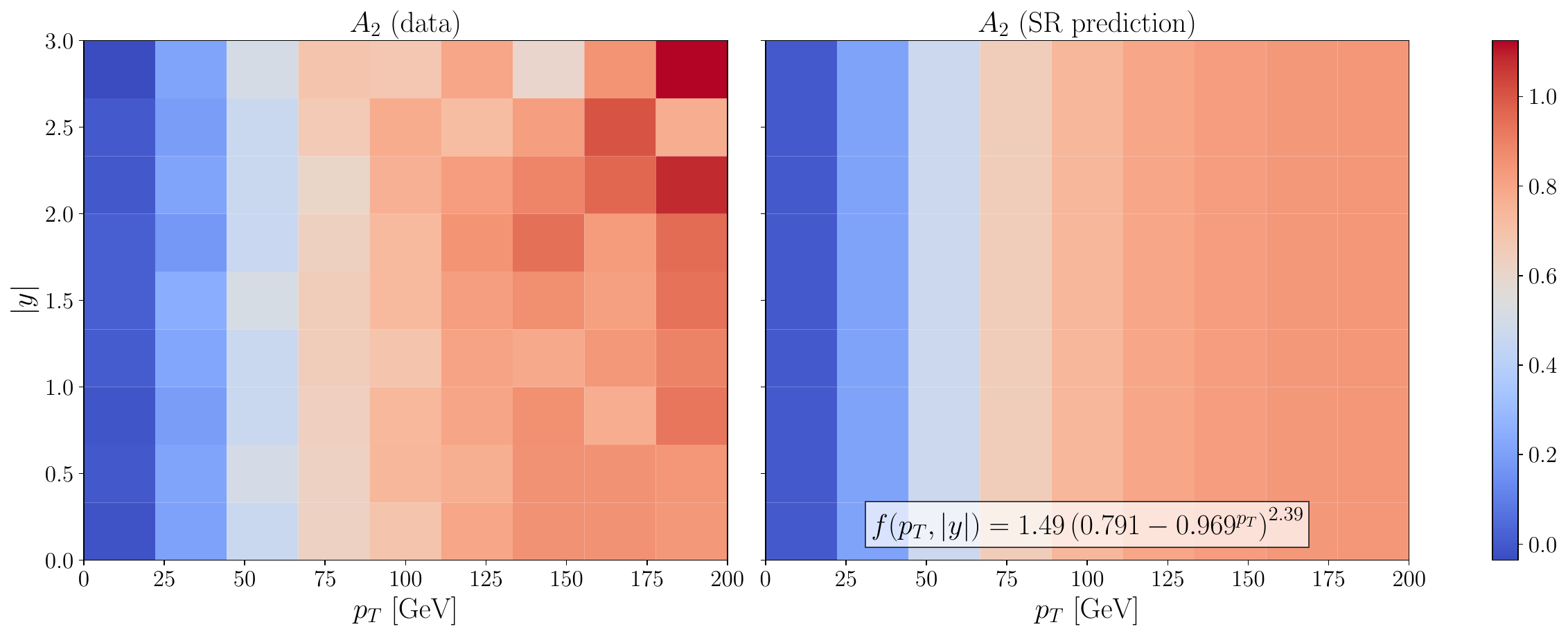}
    \label{fig:a2_2d}
  \end{minipage}

  \caption{2D kinematic distributions for $A_0$, $A_1$, and $A_2$ in $(p_T, |y|)$: data from the simulation (left) and SR result (right).}
  \label{fig:ai_2d_1}
\end{figure}

\begin{figure}[ht!]
  \centering

  \begin{minipage}[b]{0.9\linewidth}
    \includegraphics[width=\linewidth]{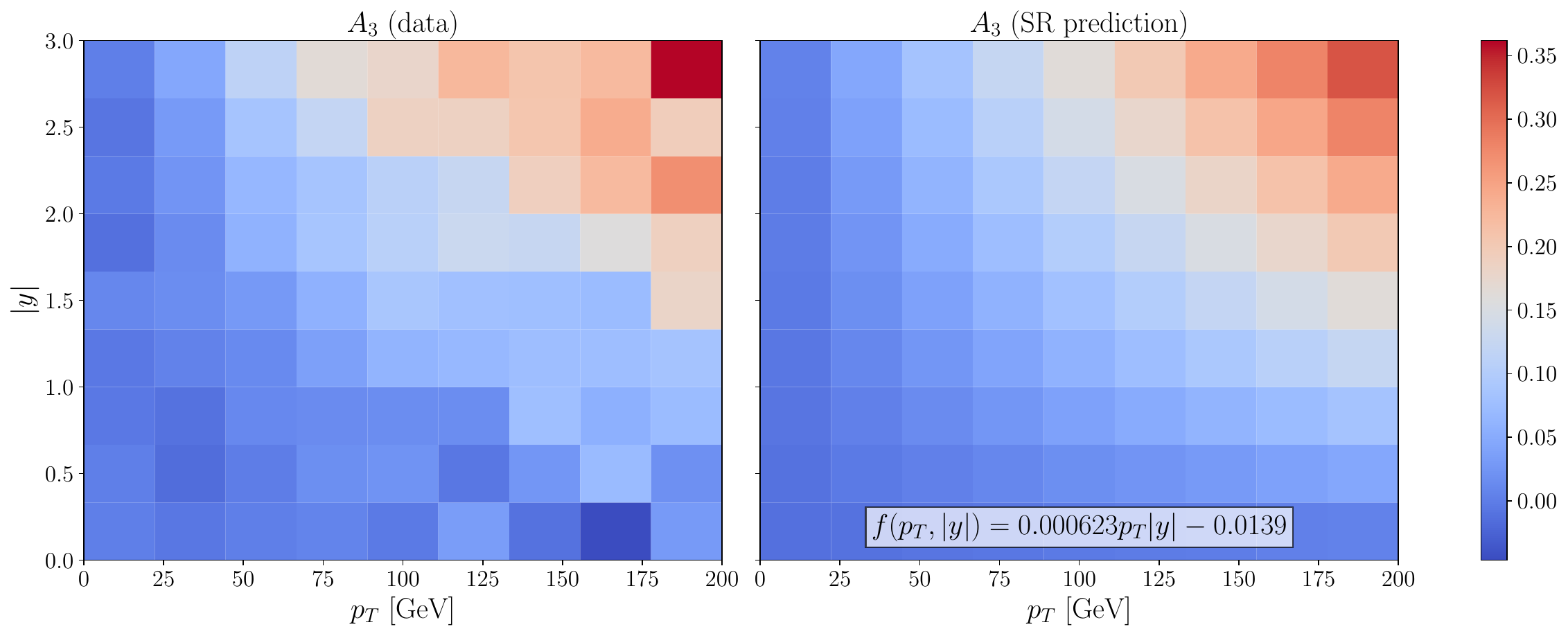}
    \label{fig:a3_2d}
  \end{minipage}

  \begin{minipage}[b]{0.9\linewidth}
    \includegraphics[width=\linewidth]{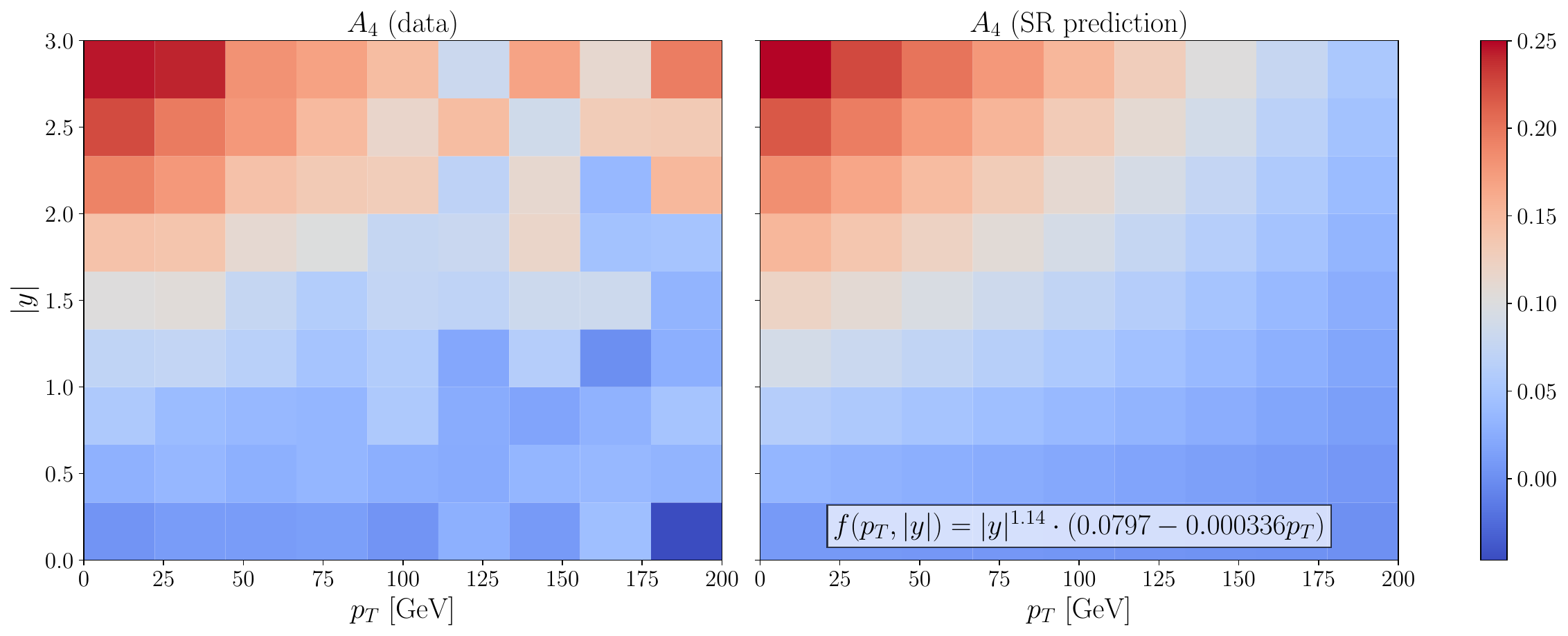}
    \label{fig:a4_2d}
  \end{minipage}

  \caption{2D kinematic distributions for $A_3$ and $A_4$ in $(p_T, |y|)$: data from the simulation (left) and SR result (right).}
  \label{fig:ai_2d_2}
\end{figure}

Figures~\ref{fig:ai_2d_1} and~\ref{fig:ai_2d_2} display the results of the Monte Carlo simulation alongside the corresponding SR expressions for the angular coefficients $A_0$ through $A_4$ in the 2D kinematic space $(p_T, |y|)$. Compared to the 1D case, the inclusion of rapidity as a second variable allows the regression to potentially capture more distinctive and characteristic patterns for each coefficient. 

\begin{table}[h!]
\scriptsize
\setlength{\tabcolsep}{4pt}
\begin{center}
\begin{tabular}{@{}c c c c@{}}
\toprule
Equation & Complexity & Loss & Score \\
\midrule
$f(p_T, |y|) = 0.386$ & 1 & 0.0869 & 0.0 \\
$f(p_T, |y|) = 0.00697 p_T$ & 3 & 0.00763 & 1.22 \\
$f(p_T, |y|) = p_T^{0.177} - 1.56$ & 5 & 0.00225 & 0.611 \\
\hline
\hline
\rowcolor{lightgray} 
$f(p_T, |y|) = \left(0.952 - 0.969^{p_T}\right)^{3.07}$ & 7 & 0.000306 & 0.998 \\
$f(p_T, |y|) = \left(0.952 - 0.969^{p_T}\right)^{3.07} - 0.00134$ & 9 & 0.000305 & $1.38 \cdot 10^{-3}$ \\
$f(p_T, |y|) = \left(0.929 - \left(0.723^{p_T} p_T\right)^{0.0826}\right)^{1.56}$ & 11 & 0.000247 & 0.104 \\
$f(p_T, |y|) = \left(0.930 - \left(0.727^{p_T} \left(p_T - 2.29\right)\right)^{0.0839}\right)^{1.56}$ & 13 & 0.000244 & $7.67 \cdot 10^{-3}$ \\
$f(p_T, |y|) = \left(0.930 - \left(p_T \left(0.727 - 0.693^{p_T}\right)^{p_T}\right)^{0.0839}\right)^{1.56}$ & 15 & 0.000240 & $7.27 \cdot 10^{-3}$ \\
$f(p_T, |y|) = \left(0.930 - \left(\left(0.727 - 0.721^{p_T}\right)^{p_T} \left(p_T + |y|\right)\right)^{0.0839}\right)^{1.56}$ & 17 & 0.000239 & $3.45 \cdot 10^{-3}$ \\
$f(p_T, |y|) = \left(\left(0.930 - \left(\left(0.727 - 0.721^{p_T}\right)^{p_T} \left(p_T + |y|\right)\right)^{0.0839}\right)^{1.56}\right)^{0.997}$ & 19 & 0.000238 & $1.05 \cdot 10^{-3}$ \\
\bottomrule
\end{tabular}
\end{center}
\caption{Same as Table~\ref{table:a0_1d} for the $A_0$ coefficient in 2D $(p_T, |y|)$.}
\label{table:a0_2d}
\end{table}

In all cases, SR results yield a smoother and differentiable representation compared to the Monte Carlo sample, which is affected by statistical fluctuations in the tail of the phase space. This enhanced smoothness is particularly useful for rebinning observables flexibly and for analytically probing the scaling behaviour of angular coefficients in specific kinematic limits (e.g., low $p_T$), where theoretical predictions from first principles are available. Figures~\ref{fig:ai_2d_1} and~\ref{fig:ai_2d_2} display the \texttt{best} expression selected by PySR.

Table~\ref{table:a0_2d} presents the hall of fame for the $A_0$ coefficient. We can see that how, in general, the transverse momentum $p_T$ is the maximally informative kinematic variable, and the rapidity $|y|$ only appears at very high complexities in the hall of fame. The selected expression for $A_0$ is of complexity 7 and reads
\begin{equation}
A_0(p_T, y) = \left(0.952 - 0.969^{p_T}\right)^{3.07},
\end{equation}
where, as we previously mentioned in the text, the transverse momentum \( p_T \) is expressed in units of GeV as to reach dimensional consistency. The equation features simple dependencies on powers of $p_T$. Note that since the range of validity of the symbolic expression is guaranteed only in the intrapolation region between bin centres used to train the regressor the base of the outer power function is always non-negative and $A_0(p_T, |y|)$ is always well defined. While empirical in nature, the equation provides a useful tool for exploring how angular coefficients evolve across different kinematic regimes.

The complete halls of fame for the 2D distributions in $(p_T, |y|)$ for the remaining angular coefficients $A_1, \dots, A_4$ are shown in Tables~\ref{table:a1_2d},~\ref{table:a2_2d},~\ref{table:a3_2d},~\ref{table:a4_2d} in Appendix~\ref{app:2d_eqs}.

We can also use SR to find simple closed-form equations to describe the 2D distribution of the angular coefficients in $(m, |y|)$. The procedure can be performed on every angular coefficient to obtain $A_i(m, |y|)$, $i = 0, \dots, 4$ but in this study we simply limit ourselves to $A_4(m, |y|)$, which is of great interest in experimental measurements of, for example, Drell-Yan forward-backward asymmetries and the weak mixing angle~\cite{CMS:2024ony, ATLAS:2015ihy, LHCb:2024ygc}.

\begin{figure}[ht!]
    \centering
    \includegraphics[width=\linewidth]{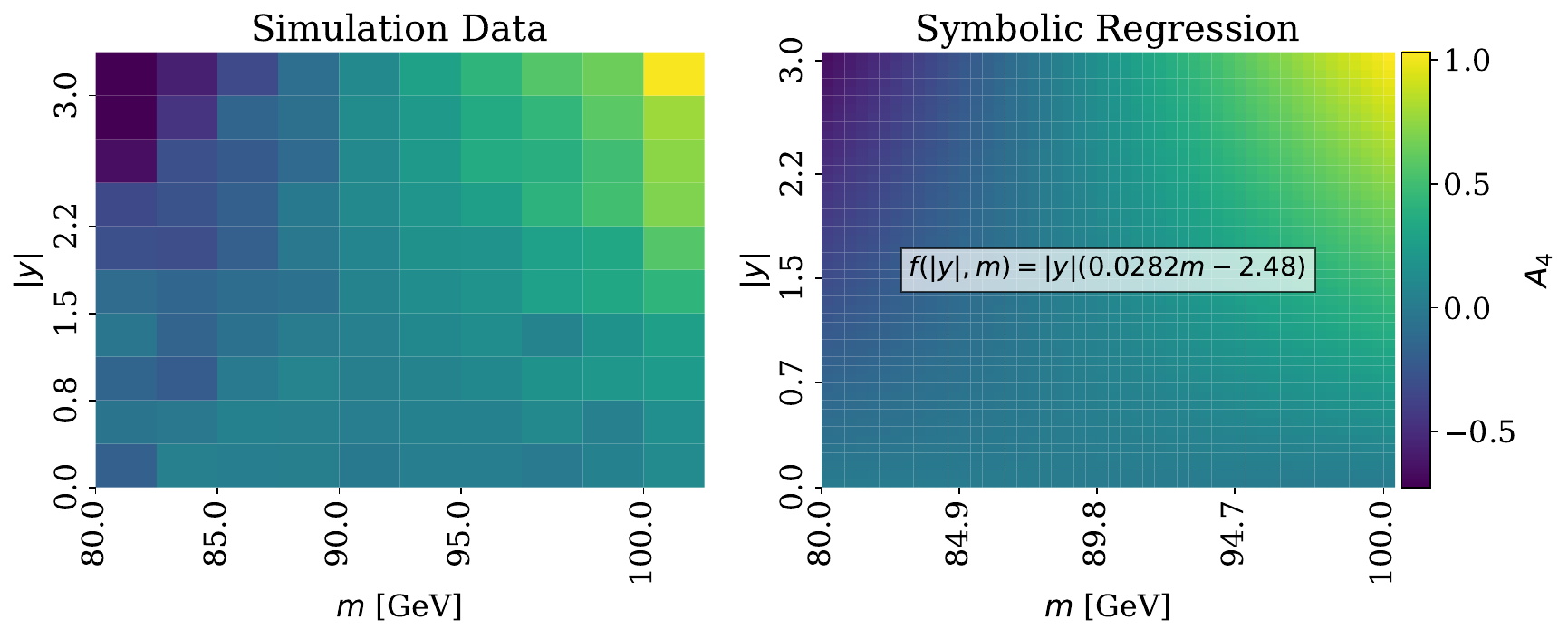}
    \caption{$A_4$ as a function of invariant mass and rapidity.}
    \label{fig:A42D}
\end{figure}

In Fig.~\ref{fig:A42D} we show $A_4(m, |y|)$ from the simulation data (central values) and from the SR result. In this case we show the SR result in higher resolution in $m$ and $|y|$ to graphically display the good interpolation capabilities of the SR equation. We see an accurate and smooth description of the 2D distribution by the regressor across the full phase space. When comparing to the 1D invariant mass distribution in Fig.~\ref{fig:A41D}, in the 2D case we obtain a linear dependence on $m$ and no quadratic contributions which, as can be seen, are highly suppressed with respect to the linear term in the 1D case. In this way, the regressor finds it more informative and simpler to just parametrise the 2D distribution as a product of two linear functions, one in $m$ and the other in $|y|$\footnote{In terms of the rapidity dependence, while in the 1D case the regressor found the $y \to -y$ symmetry, in the 2D case the training was performed directly on $|y|$, so the symmetry is imposed by construction.}. The 2D hall of fame for $A_4(m, |y|)$ is shown in Table~\ref{table:a4_2d_m_y}.

\begin{table}[h]
\footnotesize
\setlength{\tabcolsep}{5pt}
\renewcommand{\arraystretch}{1.5}
\begin{center}
\begin{tabular}{@{}c c c c@{}}
\toprule
Equation & Complexity & Loss & Score \\
\midrule
$f(m, |y|) = 0.1471$ & 1 & $0.02951$ & $0.00000$ \\
$f(m, |y|) = 0.097909\, \left| y \right|$ & 3 & $0.02182$ & $0.15090$ \\
$f(m, |y|) = 0.041079\, m - 3.6042$ & 5 & $0.01529$ & $0.17770$ \\
\hline
\hline
\rowcolor{lightgray} 
$f(m, |y|) = \left| y \right| \left(0.028228\, m - 2.4791\right)$ & 7 & $0.00116$ & $1.29100$ \\
$f(m, |y|) = \left| y \right|^{1.1313} \left(0.025495\, m - 2.2393\right)$ & 9 & $0.00107$ & $0.04041$ \\
$f(m, |y|) = \left| y \right|^{1.2063} \left(0.023983\, m - 2.1119\right) + 0.013482$ & 11 & $0.00104$ & $0.01424$ \\
$f(m, |y|) = \left| y \right|^{1.183} \left(0.024469\, \left(0.68814^{\left| y \right|} + m\right) - 2.1597\right)$ & 13 & $0.00103$ & $0.00148$ \\
\bottomrule
\end{tabular}
\end{center}
\caption{Same as Table~\ref{table:a0_1d} for the $A_4$ coefficient in 2D $(m, |y|)$.}
\label{table:a4_2d_m_y}
\end{table}

\subsection{Towards 3D angular coefficients and $m$ dependence}

We also explore the complete, 3D fully-differential dependence of the angular coefficients on $(p_T, |y|, m)$. 
Before directly delving into the SR task, we checked whether there is a statistically significant correlation between 
the angular coefficients and the invariant mass by using a permutation test on the dependence of the two variables, 
respectively; we show the results of these tests in Table~\ref{table:permutation_tests}. 
According to the $p$-values for $m$, the $A_3$ and $A_4$ coefficients do exhibit a significant dependence on the invariant mass $m$;  
we can expect the SR expression for them to include the invariant mass. The values in the permutation test were weighted by the Monte Carlo 
uncertainty of the angular coefficient value in each bin: this generally meant that the values of the angular coefficient calculated 
far from the pole of the invariant mass at $m = m_Z$ had much less weight than the values calculated close to the pole. 
The same weighting scheme was used in the subsequent SR search for the expressions.


\begin{table}[h]
\small
\setlength{\tabcolsep}{7pt}  
\begin{center}
\begin{tabular}{@{}c c@{}}
\toprule
Angular coefficient & p-value for $m$ \\
\midrule
\multicolumn{1}{c|}{$A_0$} &  0.949 \\
\multicolumn{1}{c|}{$A_1$} &  0.971 \\
\multicolumn{1}{c|}{$A_2$} &  0.977 \\
\multicolumn{1}{c|}{$A_3$} &  0.0   \\
\multicolumn{1}{c|}{$A_4$} &  0.0   \\
\bottomrule
\end{tabular}
\end{center}
\caption{Permutation tests p-values.}
\label{table:permutation_tests}
\end{table}


The results of the SR search for each of the angular coefficients are shown in Table~\ref{table:3d_pySR} (where, as usual, all dimensionful 
variables should be understood as appropriately adimensionalised). These were chosen according to the \texttt{best} criterion (described in 
App.~\ref{app:sr}). Consistent with the permutation tests, the SR result for $A_3$ and $A_4$ depends on all three kinematic variables in quite simple ways.
The loss value for these expressions is generally higher than the one for the expressions searched with less dimensionality 
(see App.~\ref{app:1d_eqs} and~\ref{app:2d_eqs} for more details). 

\begin{table}[h]
\small
\setlength{\tabcolsep}{6pt}  
\renewcommand{\arraystretch}{1.6}  
\begin{center}
\begin{tabular}{@{}c c c c c >{\raggedright\arraybackslash}p{7cm}@{}}
\toprule
Coeff. & Complexity & Loss & Score & Expression \\
\midrule
\multicolumn{1}{c|}{$A_0$} &   8  &  0.00229 & 0.221 & $f(p_T, |y|,m) = \text{exp}\left(-\frac{71.7}{p_{T}^{1.28}}\right)^{1.63}$ \\
\multicolumn{1}{c|}{$A_1$} &   8  &  0.00160 & 0.0569 & $f(p_T, |y|,m) = \text{exp}\left(-\frac{8.99}{(p_{T}^{y})^{0.149}}\right)$ \\
\multicolumn{1}{c|}{$A_2$} &   7  &  0.00606 & 0.369 & $f(p_T, |y|,m) = 1.06 + (-1.17)\cdot 0.987^{p_{T}}$ \\
\multicolumn{1}{c|}{$A_3$} &   9  &  0.00146 & 0.421 & $f(p_T, |y|,m) = p_{T} \ |y| \ (0.000167\,m - 0.0147)$ \\
\multicolumn{1}{c|}{$A_4$} &  13  &  0.00141 & 0.213 & $f(p_T, |y|,m) = |y| \ (0.990^{p_{T}} + 0.326)(0.0236\,m - 2.07)$ \\
\bottomrule
\end{tabular}
\end{center}
\caption{The results for the PySR search for each of the angular coefficients for the full 3D features $(p_{T}, |y|, m)$. 
These were chosen according to the \texttt{best} criterion out of hall of fame expressions that weren't undefined within the limits of the $(p_{T}, |y|, m)$ space.}
\label{table:3d_pySR}
\end{table}

\begin{figure}[ht!]
    \centering
    \includegraphics[width=\linewidth]{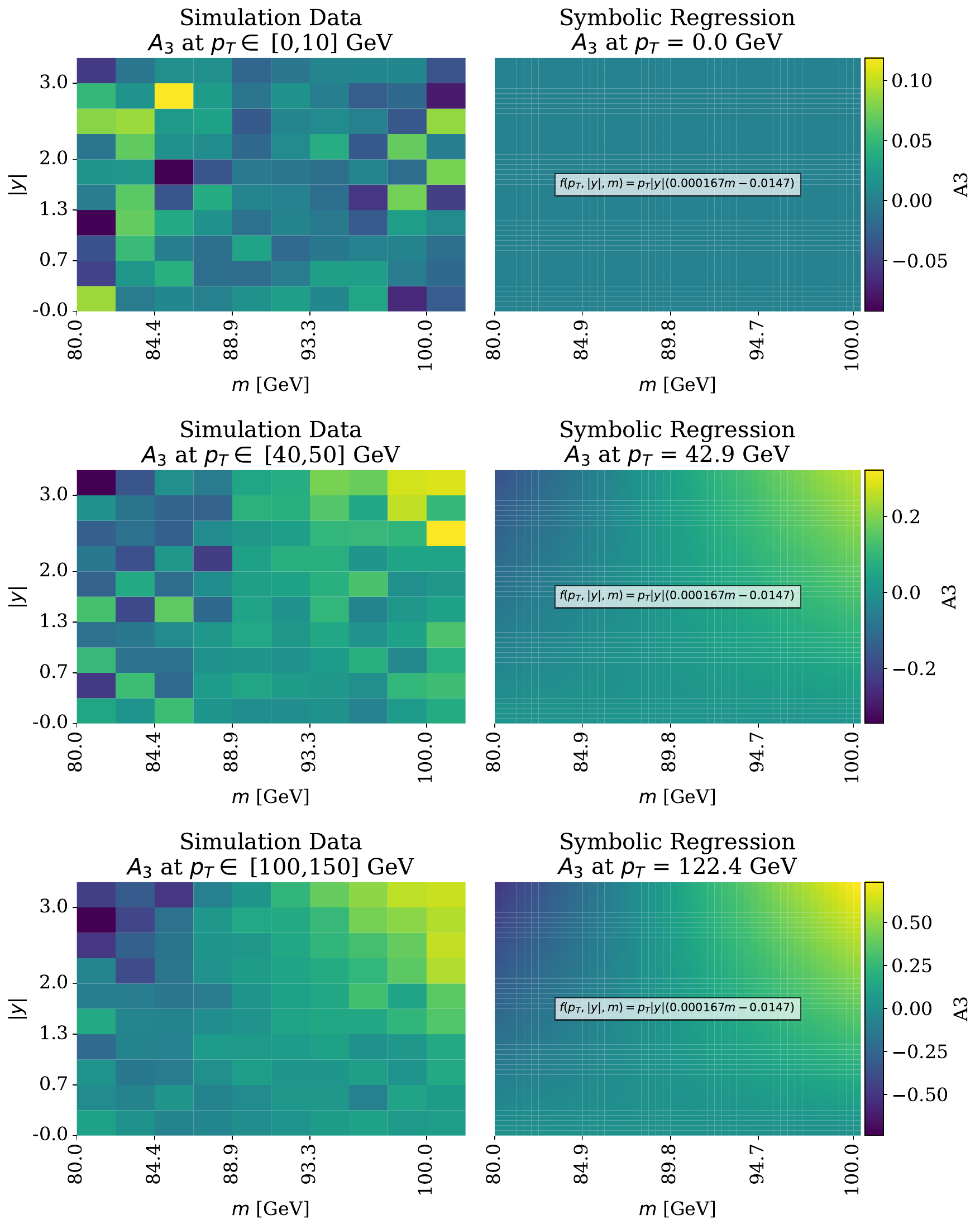}
    \caption{$A_3(p_T,|y|,m)$ shown as heatmaps in rapidity and invariant mass for three different slices in transverse momentum.}
    \label{fig:A3_3D}
\end{figure}

\begin{figure}[ht!]
    \centering
    \includegraphics[width=\linewidth]{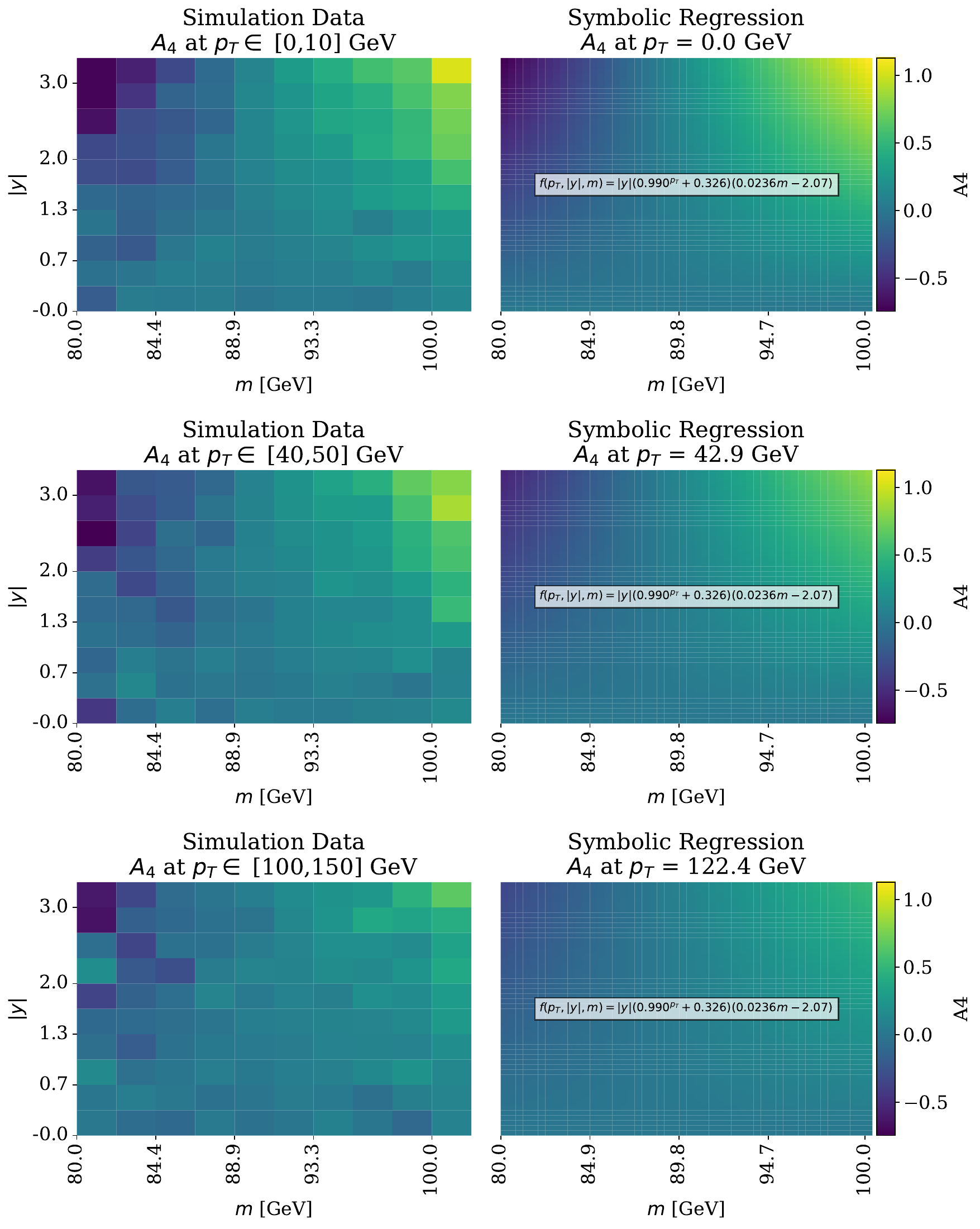}
    \caption{$A_4(p_T,|y|,m)$ shown as heatmaps in rapidity and invariant mass for three different slices in transverse momentum.}
    \label{fig:A4_3D}
\end{figure}

We verify in the expressions from Table \ref{table:3d_pySR} that the behavior in the limit $p_T \rightarrow 0$ is as expected, 
with all angular coefficients (including $A_3$) but $A_4$ vanishing. We show plots of $A_3(p_T,|y|,m)$ and $A_4(p_T,|y|,m)$ 
for different slices in $p_T$ in Figs. \ref{fig:A3_3D} and \ref{fig:A4_3D}, respectively. Due to the high cross section of events 
in the first bin of $p_T \in [0,10]$ GeV, we can compare 
\begin{equation*}
    A_4(0,|y|,m) = |y|(1.326)(0.0236m-2.07), 
\end{equation*}
to the 2D expression we obtained in the previous section
\begin{equation*}
    A_4(|y|,m) = |y|(0.0282m-2.48),
\end{equation*}
shown in Fig.~\ref{fig:A42D}. Despite one being found via regressions of different dimensionality, 
they are remarkably similar. For higher values of $p_T$ for these coefficients, we see an overall increase of the value $A_3$ 
(consistent with Fig.~\ref{fig:a3_1d}) and a waning of $A_4$. This last effect can be corroborated by the 2D SR result in Fig.~\ref{fig:ai_2d_2}, 
where we see a decreasing $A_4$ for increasing $p_T$. These results highlight the consistency of the results obtained using SR. 

\label{subsec:3D}

\section{Conclusions}
\label{sec:conclusions}

In this paper we have explored the use of SR as a tool to uncover interpretable, closed-form expressions for the angular coefficients $A_i$ in Drell–Yan-like processes at the LHC. 
Using our Monte Carlo simulations, we trained SR models to approximate the functional dependence of the angular coefficients on relevant kinematic variables 
such as $Z$ boson transverse momentum \( p_T \), the dilepton rapidity \( y \), and invariant mass \( m \).

Our results show that SR is capable of producing compact, differentiable expressions that not only reproduce the simulation data within uncertainties, but also exhibit key 
physical symmetries such as the expected \( y \to -y \). In the 1D case, we find good agreement between SR predictions and the Monte Carlo data, while in variables 
where the coefficients are weakly dependent (such as \( y \) or \( m \)), the SR tends to overfit noise, a behavior that we quantified and cross-checked with permutation tests. 
We observed that performing the symbolic regression in the two-dimensional kinematic space $(p_T, y)$ significantly enhances the ability of SR to disentangle the behavior of each 
angular coefficient, revealing distinct and physically meaningful structures that are less evident in one-dimensional fits.

Beyond reproducing known trends, the expressions learned by SR offer the possibility of compact analytic parametrisations that can speed up experimental analyses,  
serve as input to phenomenological fits, as well as support comparisons with first-principles predictions (e.g., low-\( p_T \) limits or Lam–Tung relations). 
The methodology that we present here is flexible and data-driven, but care must be taken when extrapolating beyond the kinematic domain covered by the training 
data -- particularly in rational expressions where poles may arise. The family of equations we have obtained can balance accuracy, simplicity, and generalisability, 
as illustrated through the hall of fame of solutions for each coefficient. While some representative results are displayed in Sect.~\ref{sec:results}, 
full results for all coefficients are presented in detail in the appendices.

This work demonstrates the potential of SR as a novel and interpretable approach to characterise angular observables in collider physics. Unlike black-box models, 
SR produces closed-form expressions that provide physical insight and can be directly compared with theoretical predictions. 
Future work may extend this methodology to include higher-order perturbative corrections or be applied directly to experimental data, 
where the ability to capture and interpret subtle effects, such as parton showers, hadronisation and other non-perturbative effects, is crucial. 
Addressing these limitations and integrating SR into the full simulation pipeline remains a promising direction to be further investigated in future studies.



\section*{Acknowledgements}
We are very grateful to Miles Cranmer for help on PySR and to Francesco Merlotti for useful discussions.
The work of DC and VS is supported by the Spanish grants  PID2023-148162NB-C21, CNS-2022-135688, and 
CEX2023-001292-S (MCIU/AEI/10.13039/501100011033).
M.U. is supported by the European Research Council under the European Union’s Horizon
2020 research and innovation Programme (grant agreement n.950246), and partially by the
STFC consolidated grant ST/T000694/1 and ST/X000664/1.

\appendix
\section{Symbolic regression}
\label{app:sr}

Symbolic regression (SR) is a supervised machine learning technique that aims to discover a closed-form analytical expression that maps a set of input variables to an output. Unlike traditional regression methods, which assume a fixed functional form (e.g., linear or polynomial), SR searches over a wide space of mathematical expressions to find the most accurate and interpretable model.

It presents an alternative to other regression techniques: it does not fix the functional form that the mapping has to have (as in linear regression, for example), and aims to find a good mapping that is not overly complex (like a neural network with hundreds or thousands of trainable parameters and non-trivial activation functions).
Compared to other machine learning techniques, SR offers the crucial advantage of yielding explicit mathematical equations that describe the relationships between variables. This transparency facilitates interpretability, supports hypothesis generation, and can even aid in identifying new patterns or approximate physical laws. Those qualities are particularly valuable in scientific applications.

\begin{figure}[ht]
  \centering

  \begin{subfigure}[t]{0.32\linewidth}
    \centering
    \includegraphics[width=\linewidth]{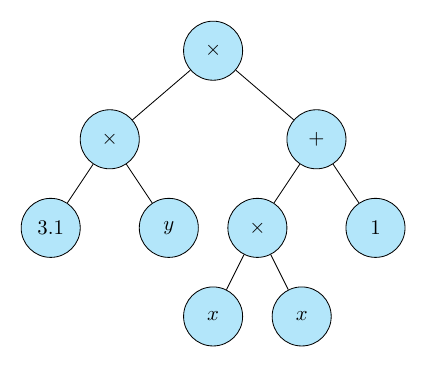}
  \end{subfigure}
  \hfill
  \begin{subfigure}[t]{0.32\linewidth}
    \centering
    \includegraphics[width=\linewidth]{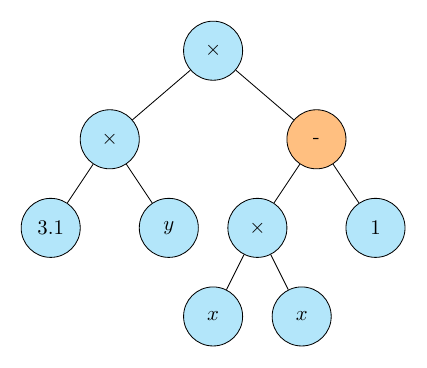}
  \end{subfigure}
  \hfill
  \begin{subfigure}[t]{0.32\linewidth}
    \centering
    \includegraphics[width=\linewidth]{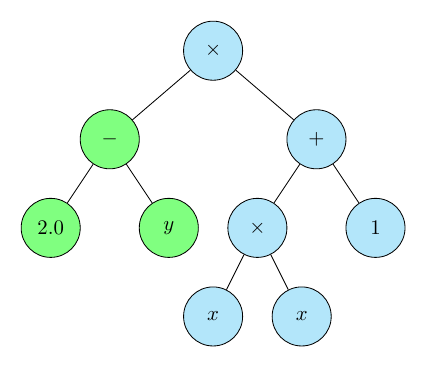}
  \end{subfigure}

  \caption{Left: example tree. Centre: mutation. Right: crossover.}
  \label{fig:tree}
\end{figure}

For this study, we make use of the PySR package \cite{Cranmer2023InterpretableML}, which makes use of a multi-population evolutionary algorithm that evaluates symbolic expressions (equations) represented by expressions trees. In Fig. \ref{fig:tree} (left) we show an example of an expression tree.

An expression tree is a structured representation of a mathematical formula, where nodes can consist of operators ($+$ or $\times$, for example), constants, or input features. For example, the equation that the example tree represents is
\begin{equation*}
    3.1 y \cdot (x^2 + 1),
\end{equation*}
where $x$ and $y$ could represent features, and $\times$ and $+$ represent the binary operations of multiplication and sum, respectively. 

During the PySR optimisation process, expression trees are evolved to discover better equations (according to a given selection criterion). This is achieved through standard operations in evolutionary algorithms, such as mutation: the random alteration of sub-expressions, as illustrated in Fig.~\ref{fig:tree} (centre), where a `$+$` operation is replaced by a `$-$`, and crossover: the recombination of parts from different trees, as shown in Fig.~\ref{fig:tree} (right), where the leftmost nodes have been inherited from another tree. These transformations drive the search toward increasingly accurate and compact models.

There are three selection criteria in PySR:

\paragraph{\texttt{Accuracy}:} This metric minimizes the mean squared error between the predicted values $\hat{Y}_i$ and the actual values $Y_i$
\begin{equation}
L = \frac{1}{n} \sum_{i=1}^{n} w_i (Y_i - \hat{Y}_i)^2.
\end{equation}

Where $w_i$ is a measure of the statistical weight of the values $\hat{Y}_i$.  In our case, these values are the calculation of an angular coefficient based on a sample of values in a particular bin of $p_{T}$, $y$ or $m$, and we can estimate the uncertainty of this value with

\begin{equation}
    w_i = \frac{\sqrt{N_i}}{s_i},
\end{equation}

where $N_i$ is the number of values in the corresponding bin, and $s_i$ is a measure of dispersion of the values used for the calculations of the angular coefficients. The weights $w_i$ encode the confidence in each data point and are computed based on the statistics of the MC sample in each bin. In particular, $w_i = \sqrt{N_i} / s_i$, where $N_i$ is the number of events in the bin, and $s_i$ is the standard deviation associated with the angular coefficient calculation in that bin.

\paragraph{\texttt{Score}:} This metric maximizes the rate of change of the logarithm of the loss function \(L\) with respect to the complexity \(c\) of the tree
\begin{equation}
-\frac{\partial \log(L)}{\partial c}.
\end{equation}
This formulation encourages the selection of equations that achieve a rapid reduction in loss with minimal increase in complexity. In practice, this favors simpler models that already explain much of the variance in the data—prioritizing compactness and interpretability over marginal gains in accuracy. Such compact expressions are easier to analyze, validate, and interpret in physical terms, making this criterion particularly useful in scientific contexts.

\paragraph{\texttt{Best}:} This metric selects the tree with the highest \texttt{Score} among those whose loss \(L\) is within 1.5 times the minimum achievable loss \(L_{\text{min}}\):
\begin{equation}
L \leq 1.5 \times L_{\text{min}}.
\end{equation}
The \texttt{Best} criterion thus implements a practical trade-off between interpretability and fidelity. Rather than choosing the model with the absolute lowest error—often more complex and harder to interpret—it selects an expression that is nearly as accurate but significantly simpler. This balance is crucial in physics applications, where overly complex models may obscure the underlying mechanisms of the process under study.

These metrics guide the evolutionary algorithm in selecting and optimizing the fittest expression trees, balancing accuracy, complexity, and overall interpretability.

In terms of computational cost, the PySR runs were carried out locally on a standard workstation (macOS, 6-core Intel CPU, no GPU acceleration). Each optimisation typically converged within $\mathcal{O}(10)$ minutes for the datasets of this study. As a result, the method remains computationally inexpensive and easily applicable to other problems of similar size and complexity.

\section{Estimation of symbolic regression uncertainties}
\label{app:SR_uncertainties}

SR provides analytical expressions for the functional dependence of the data, but it does not inherently supply an estimate of the uncertainty associated with its predictions. Because SR relies on a stochastic evolutionary search algorithm, each independent training run can be regarded as a draw from a distribution of possible equations and corresponding predictions. Under the assumption that repeated runs sample this space in an unbiased way, and that the distribution of outputs is approximately normal, multiple PySR trainings can be used as an ensemble method to estimate the predictive uncertainty of the SR model.

\begin{figure}[h!]
  \centering
  \includegraphics[width=0.60\linewidth]{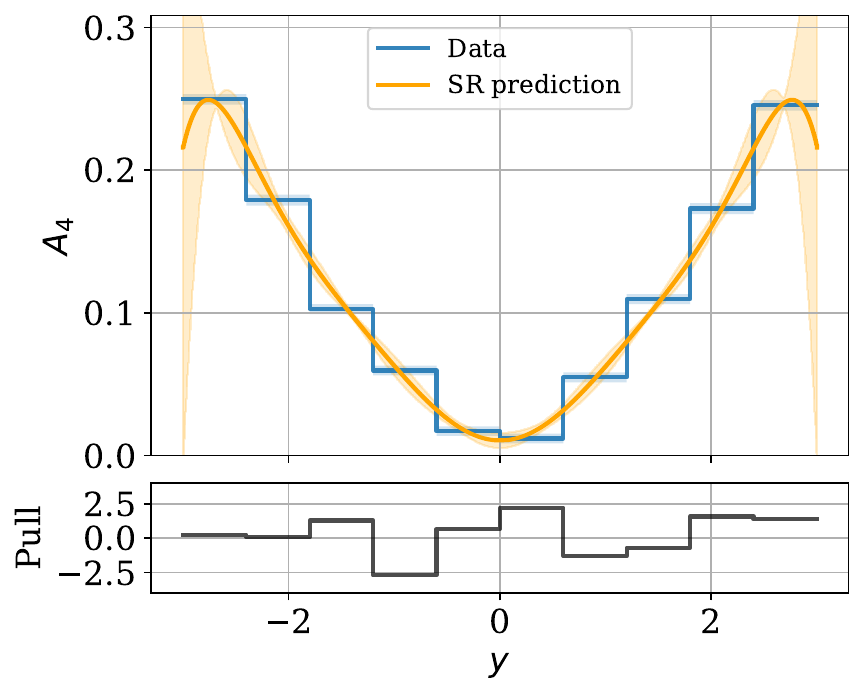}
  \caption{$A_4$ values with respect to the dilepton rapidity, with uncertainty band estimated from repeated PySR experiments. The pull is calculated with the central prediction.}
  \label{fig:A4_uncertainty}
\end{figure}

Here we apply this ensemble procedure to the coefficient $A_4(y)$, as shown in Fig.~\ref{fig:A4_uncertainty}. Five independent PySR runs are performed, yielding a set of {\tt best} expressions which are summarised in Table~\ref{table:A4_uncertainty}. The central prediction displayed in the figure is obtained as the average of the predictions from these five models, with the associated uncertainty band corresponding to one standard deviation in the ensemble. The spread among these predictions increases near the boundaries of the phase space, highlighting the limited constraints provided by the training data in the extrapolation regions. By contrast, the predictions agree closely at the bin centres used for training, reflecting the consistency in the predictive values within the data-supported domain. Because the best expressions differ in analytical form, assigning parameter-level uncertainties to any single equation is not meaningful within this framework.

\begin{table}[h!]
\centering
\begin{tabular}{r r p{10cm}}
\hline
\textbf{Loss} & \textbf{Score} & \textbf{Analytical Form} \\
\hline
$1.39\times10^{-5}$ & 0.155 & $f(y) = y\,(y - 0.0571)\,(y\,((y + 0.152)\,0.000349\,y^{2} - 0.00612\,y) + 0.0584) + 0.00965$ \\
$1.12\times10^{-5}$ & 0.291 & $f(y) = y^{2}\,(y^{2}\,(y^{2}\,(y^{2}\,(-0.000493) + 0.00694) - 0.0315) + 0.0846) + 0.00783$ \\
$1.85\times10^{-5}$ & 0.0955 & $f(y) = y^{2}\,(y^{2}\,(y^{2}\,0.000326 + 0.00571) + 0.0568) + 0.00851$ \\
$2.69\times10^{-5}$ & 0.0632 & $f(y) = y^{2}\,(y\,(y\,(-0.00204) - 0.000203) + 0.0468) + 0.0138$ \\
$2.39\times10^{-5}$ & 0.190 & $f(y) = y^{2}\,(y\,(-0.00694\,y + 0.000433\,y^{3}) + 0.0600) + 0.00938$ \\
\hline
\end{tabular}
\caption{Results from five PySR experiments showing loss, score, and analytical expression.}
\label{table:A4_uncertainty}
\end{table}

Finally, we investigated whether resampling the input data according to their reported uncertainties would contribute additional dispersion to the SR predictions. In practice, this resampling had little effect compared to the variation produced simply by repeating the SR training itself, reinforcing that the dominant source of predictive spread originates from the stochastic SR process rather than from fluctuations in the input data.

Uncertainty quantification in SR is an open problem, and a comprehensive treatment would require a dedicated study that lies beyond the scope of this work. The analysis presented here is therefore intended only as a qualitative indication of the uncertainty size associated with SR predictions across the interpolation and extrapolation regions of the kinematic coverage.

\section{1D SR equations}
\label{app:1d_eqs}

In this appendix, we present the explicit expressions obtained via SR for each of the angular coefficients in terms of a single variable, 
either $p_T$ or dilepton rapidity $y$. The aim of this appendix is to emphasise that the SR finds a family of equations that can describe the data at different accuracies, 
complexities and scores. The choice of equation depends on the intended application and can be tailored to prioritize accuracy, simplicity, interpretability, or other relevant criteria.

\subsection{Dependence on $p_T$}
Tables \ref{table:1d_a1},  \ref{table:1d_a2},  \ref{table:1d_a3},  \ref{table:1d_a4} show the SR hall of fame in $p_T$ of the $A_1$, $A_2$, $A_3$, and $A_4$ angular coefficients, respectively.

\begin{table}[h]
\footnotesize
\setlength{\tabcolsep}{5pt}
\renewcommand{\arraystretch}{1.5}
\begin{center}
\begin{tabular}{@{}c c c c@{}}
\toprule
Equation & Complexity & Loss & Score \\
\midrule
$f(p_T) = 0.0367$ & 1 & 0.000670 & 0.0 \\
$f(p_T) = 0.000632\,p_T$ & 3 & 0.000522 & 0.124 \\
$f(p_T) = 0.0568 - \frac{0.325}{p_T}$ & 5 & 0.000220 & 0.432 \\
\makecell{$f(p_T) = p_T \cdot \left(0.00129 - 5.15 \cdot 10^{-6}\,p_T\right)$} & 7 & $7.56\cdot10^{-5}$ & 0.534 \\
$f(p_T) = 0.0727 + \frac{3.18 - p_T}{p_T^2}$ & 9 & $6.53\cdot10^{-5}$ & 0.0729 \\
$f(p_T) = \frac{p_T}{0.0767\,p_T^2 + 689.}$ & 11 & $3.50\cdot10^{-5}$ & 0.312 \\
\hline
\hline
\rowcolor{lightgray} $f(p_T) = \frac{p_T - 4.75}{0.0866\,p_T^2 + 560.}$ & 13 & $1.54\cdot10^{-5}$ & 0.409 \\
\makecell{$f(p_T) = \frac{p_T - 4.75}{0.0866\,p_T (p_T - 4.75) + 571.}$} & 15 & $1.37\cdot10^{-5}$ & 0.0613 \\
\makecell{$f(p_T) = \frac{p_T - 5.11}{- p_T \cdot (1.65 - 0.0923\,p_T) + 592.}$} & 17 & $1.13\cdot10^{-5}$ & 0.0949 \\
\makecell{$f(p_T) = \frac{p_T - 5.11}{- 1.05\,p_T \cdot (1.65 - 0.0923\,p_T) + 592.}$} & 19 & $1.07\cdot10^{-5}$ & 0.0253 \\
\bottomrule
\end{tabular}
\end{center}
\caption{SR hall of fame for the $A_1$ coefficient. The symbolic expression with the best score is highlighted in gray, and the horizontal line corresponds to 
Eq.~\eqref{eq:cond}. The best score is the highest score among those below the horizontal line.}
\label{table:1d_a1}
\end{table}

\begin{table}[h]
\footnotesize
\setlength{\tabcolsep}{5pt}
\renewcommand{\arraystretch}{1.5}
\begin{center}
\begin{tabular}{@{}c c c c@{}}
\toprule
Equation & Complexity & Loss & Score \\
\midrule
$f(p_T) = 0.284$ & 1 & 0.0749 & 0.0 \\
$f(p_T) = 0.00646\,p_T$ & 3 & 0.0113 & 0.945 \\
$f(p_T) = 0.00608\,p_T + 0.0314$ & 5 & 0.0108 & 0.0223 \\
\makecell{$f(p_T) = p_T \cdot \left(0.00935 - 2.31 \cdot 10^{-5}\,p_T\right)$} & 7 & $2.96\cdot10^{-3}$ & 0.648 \\
\makecell{$f(p_T) = p_T \cdot \left(0.0114 - 3.21 \cdot 10^{-5}\,p_T\right) - 0.0801$} & 9 & $9.94\cdot10^{-4}$ & 0.546 \\
\hline
\hline
\rowcolor{lightgray} 
$f(p_T) = \frac{0.0140}{0.0149 + \frac{42.6}{p_T^2}}$ & 11 & $1.39\cdot10^{-4}$ & 0.983 \\
\makecell{$f(p_T) = \frac{0.0220}{0.0239 + \frac{60.1}{p_T(p_T - 3.50)}}$} & 13 & $9.85\cdot10^{-5}$ & 0.173 \\
\makecell{$f(p_T) = \frac{0.0128\,p_T}{0.0135\,p_T + \frac{36.7}{p_T}} - 0.0117$} & 15 & $9.48\cdot10^{-5}$ & 0.0188 \\
\makecell{$f(p_T) = \frac{0.0128\,p_T}{0.0135\,p_T + \frac{0.302}{0.00822\,p_T - 0.00317}} - 0.0102$} & 17 & $9.43\cdot10^{-5}$ & 0.00253 \\
\makecell{$f(p_T) = \frac{0.00885\,p_T}{0.00953\,p_T - 0.0233 + \frac{26.2}{p_T}} - 0.0114$} & 19 & $9.36\cdot10^{-5}$ & 0.00406 \\
\bottomrule
\end{tabular}
\end{center}
\caption{Same as Table~\ref{table:1d_a1} for the $A_2$ coefficient.}
\label{table:1d_a2}
\end{table}

\begin{table}[h]
\footnotesize
\setlength{\tabcolsep}{5pt}
\renewcommand{\arraystretch}{1.5}
\begin{center}
\begin{tabular}{@{}c c c c@{}}
\toprule
Equation & Complexity & Loss & Score \\
\midrule
$f(p_T) = 0.0229$ & 1 & 0.000665 & 0.0 \\
$f(p_T) = 0.000585\,p_T$ & 3 & $5.44\cdot10^{-5}$ & 1.25 \\
$f(p_T) = \frac{p_T}{p_T + 1.58 \cdot 10^3}$ & 5 & $4.45\cdot10^{-5}$ & 0.101 \\
$f(p_T) = p_T \cdot \left(0.000732 - 1.19 \cdot 10^{-6}\,p_T\right)$ & 7 & $3.33\cdot10^{-5}$ & 0.145 \\
\hline
\hline
\rowcolor{lightgray} 
$f(p_T) = \frac{0.105\,p_T}{p_T + \frac{5.36 \cdot 10^3}{p_T}}$ & 9 & $2.52\cdot10^{-6}$ & 1.29 \\
\makecell{$f(p_T) = \frac{0.104\,p_T}{p_T + \frac{5.08 \cdot 10^3}{p_T}} - 0.000941$} & 11 & $2.18\cdot10^{-6}$ & 0.0714 \\
\makecell{$f(p_T) = \frac{0.104\,p_T}{p_T + \frac{5.26 \cdot 10^3}{p_T}} - \frac{0.00877}{p_T}$} & 13 & $1.87\cdot10^{-6}$ & 0.0763 \\
\makecell{$f(p_T) = \frac{0.105\,p_T}{p_T + \frac{5.36 \cdot 10^3}{p_T}} - \frac{0.0482}{p_T^2}$} & 15 & $1.74\cdot10^{-6}$ & 0.0360 \\
\makecell{$f(p_T) = \frac{0.105\,p_T}{p_T + \frac{5.36 \cdot 10^3}{p_T}} - \frac{0.265}{p_T^3}$} & 17 & $1.70\cdot10^{-6}$ & 0.0136 \\
\makecell{$f(p_T) = \frac{0.105\,p_T}{p_T + \frac{5.36 \cdot 10^3}{p_T}} - \frac{1.27}{p_T^4}$} & 19 & $1.69\cdot10^{-6}$ & 0.00350 \\
\bottomrule
\end{tabular}
\end{center}
\caption{Same as Table~\ref{table:1d_a1} for the $A_3$ coefficient.}
\label{table:1d_a3}
\end{table}

\begin{table}[h]
\footnotesize
\setlength{\tabcolsep}{5pt}
\renewcommand{\arraystretch}{1.5}
\begin{center}
\begin{tabular}{@{}c c c c@{}}
\toprule
Equation & Complexity & Loss & Score \\
\midrule
$f(p_T) = 0.100$ & 1 & 0.000543 & 0.0 \\
$f(p_T) = \frac{14.3}{p_T + 106.0}$ & 5 & $2.11\cdot10^{-5}$ & 0.812 \\
$f(p_T) = -0.0147 + \frac{19.2}{p_T + 128.9}$ & 7 & $2.03\cdot10^{-5}$ & 0.0185 \\
\hline
\hline
\rowcolor{lightgray} 
$f(p_T) = 0.0421 + \frac{325.8}{p_T^2 + 3.92 \cdot 10^3}$ & 9 & $3.93\cdot10^{-6}$ & 0.823 \\
\makecell{$f(p_T) = 0.0432 + \frac{299.3}{p_T (p_T - 2.38) + 3.66 \cdot 10^3}$} & 11 & $3.66\cdot10^{-6}$ & 0.0354 \\
\makecell{$f(p_T) = \frac{p_T + 218.0}{p_T^2 + 2.65 \cdot 10^3} + 0.0404$} & 13 & $3.41\cdot10^{-6}$ & 0.0357 \\
\makecell{$f(p_T) = 0.0434 + \frac{292.4}{1.16\,p_T^2 - 11.7\,p_T + 3.66 \cdot 10^3}$} & 15 & $3.25\cdot10^{-6}$ & 0.0235 \\
\makecell{$f(p_T) = 0.0434 + \frac{292.4}{1.16\,p_T^2 - 12.2\,p_T + 3.68 \cdot 10^3}$} & 17 & $3.16\cdot10^{-6}$ & 0.0146 \\
\makecell{$f(p_T) = 0.0434 + \frac{292.4}{1.16\,p_T^2 - 13.4\,p_T + 3.70 \cdot 10^3}$} & 19 & $3.11\cdot10^{-6}$ & 0.00765 \\
\bottomrule
\end{tabular}
\end{center}
\caption{Same as Table~\ref{table:1d_a1} for the $A_4$ coefficient.}
\label{table:1d_a4}
\end{table}

\begin{table}[h]
\footnotesize
\setlength{\tabcolsep}{5pt}
\renewcommand{\arraystretch}{1.5}
\begin{center}
\begin{tabular}{@{}c c c c@{}}
\toprule
Equation & Complexity & Loss & Score \\
\midrule
$f(p_T) = 0.00191$ & 1 & $6.40\cdot10^{-5}$ & 0.0 \\
$f(p_T) = 0.310^{p_T} + 0.00141$ & 5 & $6.28\cdot10^{-5}$ & 0.00469 \\
\hline
\hline
\rowcolor{lightgray} 
$f(p_T) = \left(0.976 - 0.962^{p_T}\right)^{p_T}$ & 7 & $4.17\cdot10^{-5}$ & 0.205 \\
$f(p_T) = 2.74 \left(0.971 - 0.964^{p_T}\right)^{p_T}$ & 9 & $4.07\cdot10^{-5}$ & 0.0126 \\
\makecell{$f(p_T) = \left(- 0.964^{p_T} + p_T^{-0.980} + 0.970\right)^{p_T}$} & 11 & $3.83\cdot10^{-5}$ & 0.0292 \\
\makecell{$f(p_T) = \left(- 1.07 \cdot 0.965^{p_T} + \frac{1.07}{p_T^{0.859}} + 0.965\right)^{p_T}$} & 13 & $3.71\cdot10^{-5}$ & 0.0159 \\
\makecell{$f(p_T) = \left(- 0.965^{p_T} + \left(p_T + 0.947\right)^{-0.859} + 0.965\right)^{p_T}$\\$- 0.000576$} & 15 & $3.67\cdot10^{-5}$ & 0.00610 \\
\makecell{$f(p_T) = 0.147\,p_T \cdot \left(0.354 \left(0.965 - 0.962^{p_T}\right)^{p_T} - 0.000372\right)$} & 17 & $3.46\cdot10^{-5}$ & 0.0292 \\
\makecell{$f(p_T) = 0.140\,p_T \cdot \left(0.342 \left(0.965 - 0.962^{p_T}\right)^{p_T} - 0.000360\right)$\\$+ 0.00141$} & 19 & $3.27\cdot10^{-5}$ & 0.0281 \\
\bottomrule
\end{tabular}
\end{center}
\caption{Same as Table~\ref{table:1d_a1} for the difference $A_0 - A_2$.}
\label{table:a0_minus_a2_1d}
\end{table}

\subsection{Dependence on rapidity}

The SR halls of fame for the 1D $A_4(y)$ and $A_4(m)$ are given respectively in Tables~\ref{table:a4_1d_y} and~\ref{table:a4_1d_m}.

\begin{table}[h]
\footnotesize
\setlength{\tabcolsep}{5pt}
\renewcommand{\arraystretch}{1.5}
\begin{center}
\begin{tabular}{@{}c c c c@{}}
\toprule
Equation & Complexity & Loss & Score \\
\midrule
$f(y) = 0.11277$ & 1 & $0.00689$ & $0.00000$ \\
$f(y) = 0.03670 y^{2}$ & 5 & $0.00042$ & $0.69830$ \\
$f(y) = 0.03176 y^{2} + 0.02567$ & 7 & $0.00011$ & $0.66770$ \\
$f(y) = y \left(0.03176 y - 0.00061\right) + 0.02566$ & 9 & $0.00011$ & $0.00469$ \\
$f(y) = y^{2} \left(0.05396 - 0.00279 y^{2}\right)$ & 11 & $0.00009$ & $0.08918$ \\
\hline
\hline
\rowcolor{lightgray} 
$f(y) = y^{2} \left(0.04498 - 0.00181 y^{2}\right) + 0.01518$ & 13 & $0.00002$ & $0.67610$ \\
$f(y) = y^{2} \left(- y \left(0.00181 y + 0.00012\right) + 0.04498\right) + 0.01518$ & 15 & $0.00002$ & $0.02575$ \\
$f(y) = y^{2} \left(- y^{2} \left(0.00002 y + 0.00181\right) + 0.04498\right) + 0.01518$ & 17 & $0.00002$ & $0.00084$ \\
$f(y) = y^{2} \left(y^{2} \left(0.00030 y^{2} - 0.00511\right) + 0.05344\right) + 0.01221$ & 19 & $0.00002$ & $0.12270$ \\
\bottomrule
\end{tabular}
\end{center}
\caption{Same as Table~\ref{table:1d_a1} for the $A_4$ coefficient in $y$.}
\label{table:a4_1d_y}
\end{table}

\begin{table}[h]
\footnotesize
\setlength{\tabcolsep}{5pt}
\renewcommand{\arraystretch}{1.5}
\begin{center}
\begin{tabular}{@{}c c c c@{}}
\toprule
Equation & Complexity & Loss & Score \\
\midrule
$f(m) = 0.12788$ & 1 & $0.01295$ & $0.00000$ \\
$f(m) = 0.001438 m$ & 3 & $0.01204$ & $0.03648$ \\
\hline
\hline
\rowcolor{lightgray} 
$f(m) = 0.038925 m - 3.4267$ & 5 & $0.00022$ & $1.99600$ \\
$f(m) = 0.03892 m - 3.4266$ & 7 & $0.00022$ & $0.00008$ \\
$f(m) = - m \left(0.0004739 m - 0.12563\right) - 7.3883$ & 9 & $0.00016$ & $0.15940$ \\
$f(m) = - 0.00047148 m^{2} + 0.12517 m - 7.3671$ & 11 & $0.00016$ & $0.00002$ \\
$f(m) = -0.038622 m^{2} \left(e^{-0.070641 m} - 0.0019882\right)$ & 12 & $0.00016$ & $0.03906$ \\
$f(m) = 7.52379 \cdot 10^{-5} m^{2} - 0.04438\, m^{2} e^{-0.07245 m}$ & 16 & $0.00016$ & $0.00052$ \\
\bottomrule
\end{tabular}
\end{center}
\caption{Same as Table~\ref{table:1d_a1} for the $A_4$ coefficient in $m$.}
\label{table:a4_1d_m}
\end{table}

To better understand the limitations and capabilities of symbolic regression in capturing angular structures, we examine how the fitted expressions for $A_0(y)$ 
evolve as we vary the cut on transverse momentum $p_T^{\rm min}$. At very low $p_T$, the coefficient $A_0$ is expected to vanish, and the simulation is dominated 
by statistical fluctuations. In this regime, PySR tends to slightly overfit the noise, returning complex but unphysical expressions.

\begin{figure}
    \centering
    \includegraphics[width=\linewidth]{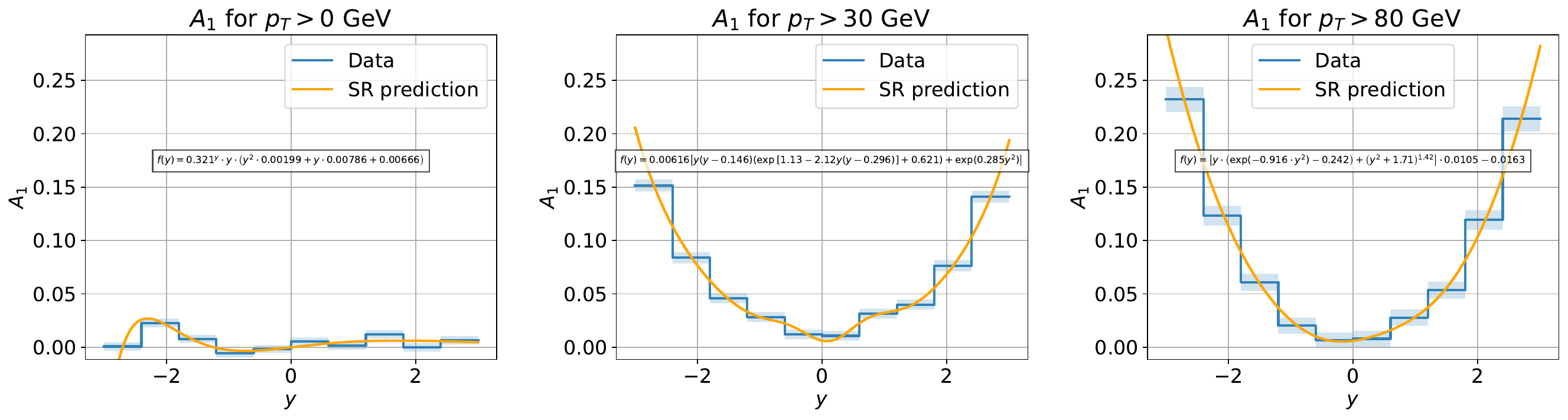}
    \caption{$A_1$ as a function of y for $p_T>p_T^{\rm min}$, with $p_T^{\rm min} = 0$ GeV, 30 GeV and 80 GeV respectively.}
    \label{fig:A1y_pTcuts}
\end{figure}

As we progressively increase the $p_T$ threshold, the true $y$-dependence of $A_1$ becomes more pronounced, and PySR begins to recover meaningful trends. 
Fig.~\ref{fig:A1y_pTcuts} illustrates this progression by showing the fitted $A_1(y)$ for four increasing $p_T$ cuts: $p_T > 0.01$~GeV, 30~GeV, and 80~GeV. 
This exercise highlights the importance of applying suitable phase-space selections when using SR in data-dominated regimes.

\section{2D SR equations}
\label{app:2d_eqs}

In this appendix, we present the central values and uncertainties of the 2D angular coefficients, and their halls of fame for the 2D SR results, following the same structure as in App.~\ref{app:1d_eqs}. Now the the explore the dependence of the angular coefficients in terms of two variables, transverse momentum $p_T$ and the absolute value of the dilepton rapidity $|y|$.

\begin{figure}[ht!]
  \centering

  \begin{minipage}[b]{0.72\linewidth}
    \includegraphics[width=\linewidth]{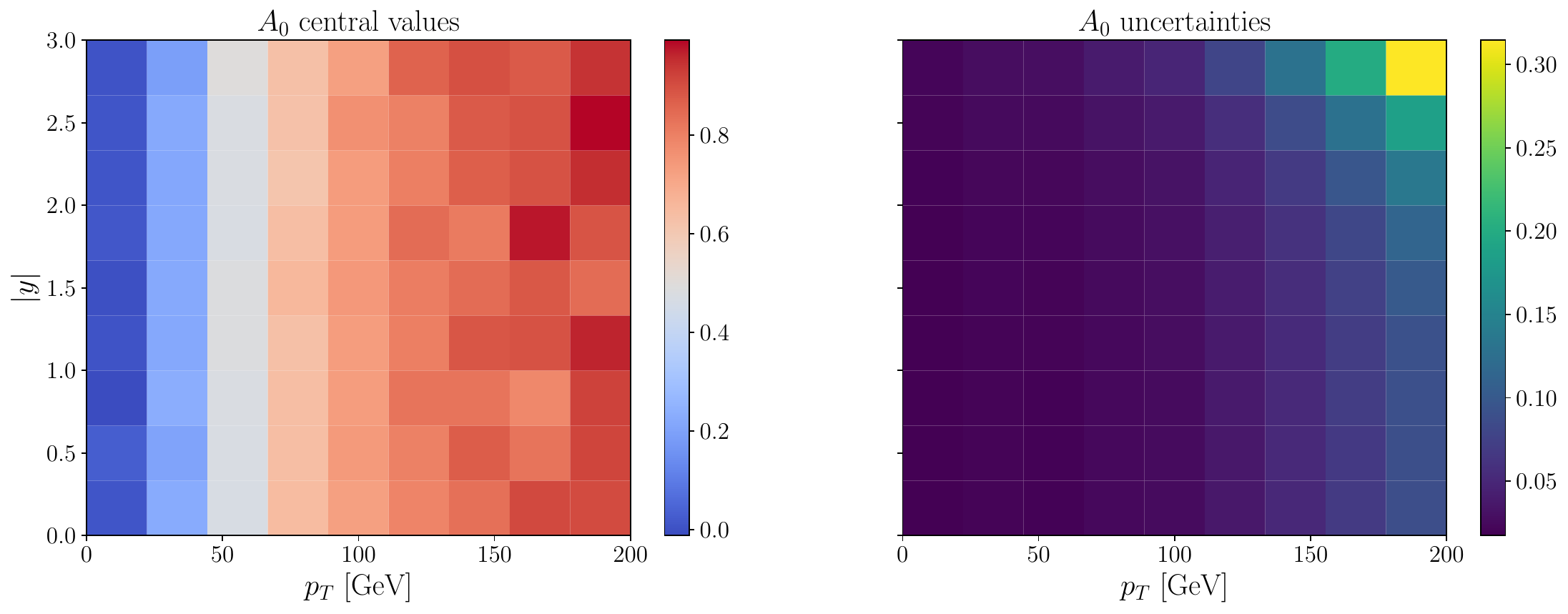}
    \label{fig:a0_2d_cv_unc}
  \end{minipage}

  \begin{minipage}[b]{0.72\linewidth}
    \includegraphics[width=\linewidth]{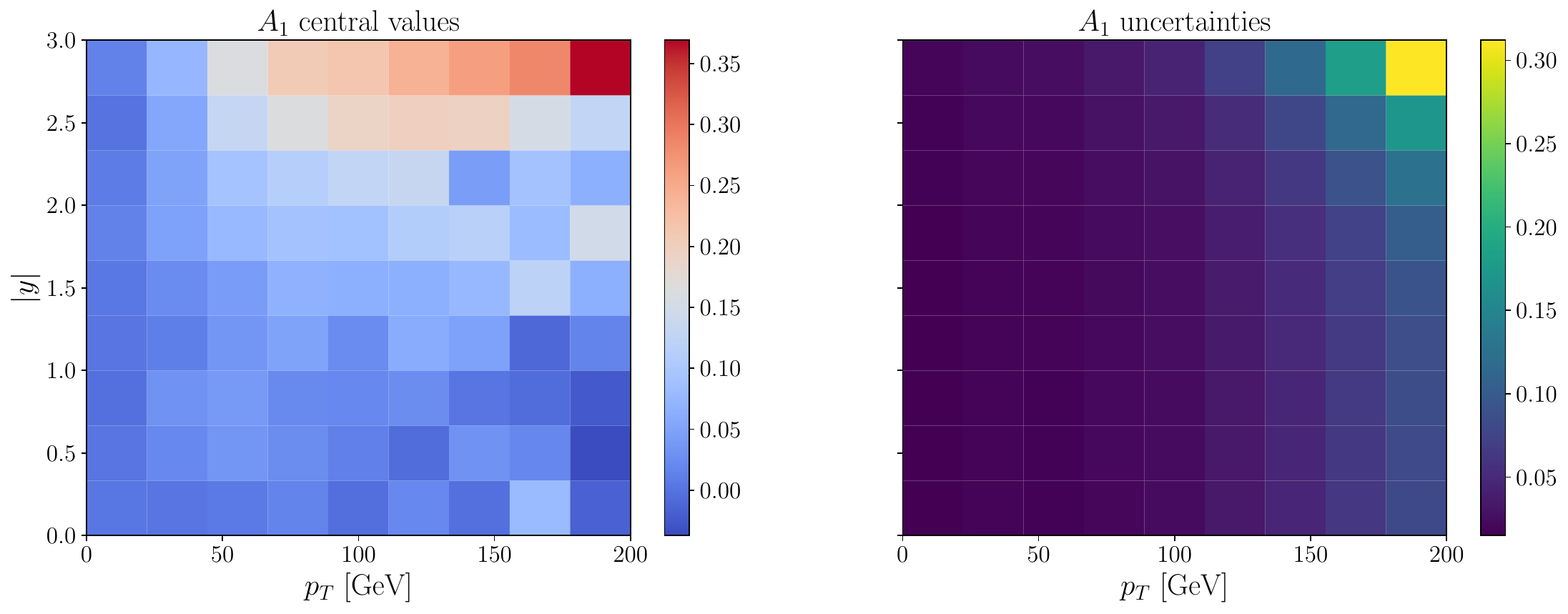}
    \label{fig:a1_2d_cv_unc}
  \end{minipage}

  \begin{minipage}[b]{0.72\linewidth}
    \includegraphics[width=\linewidth]{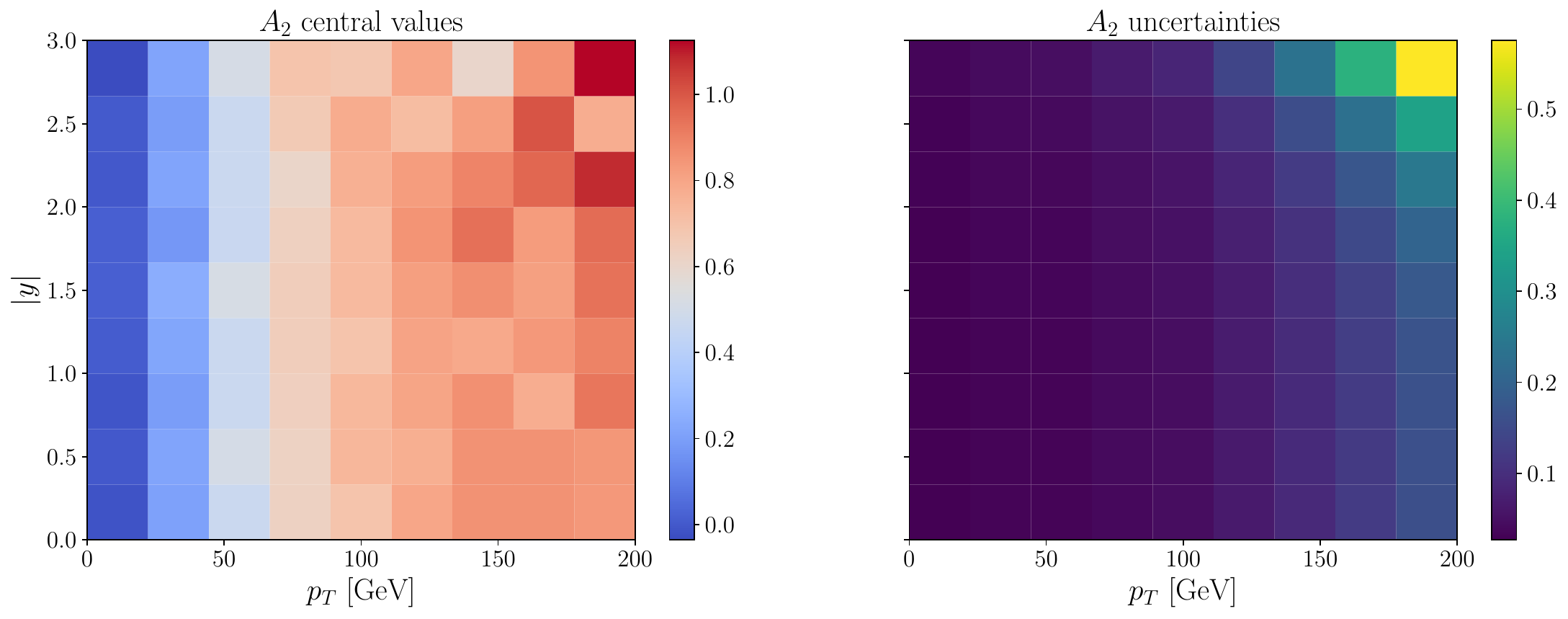}
    \label{fig:a2_2d_cv_unc}
  \end{minipage}

  \caption{2D kinematic distributions for $A_0$ to $A_2$ in $(p_T, |y|)$: central values (left) and uncertainties (right).}
  \label{fig:Ai_2d_cv_unc_1}
\end{figure}

\begin{figure}[ht!]
  \centering

  \begin{minipage}[b]{0.72\linewidth}
    \includegraphics[width=\linewidth]{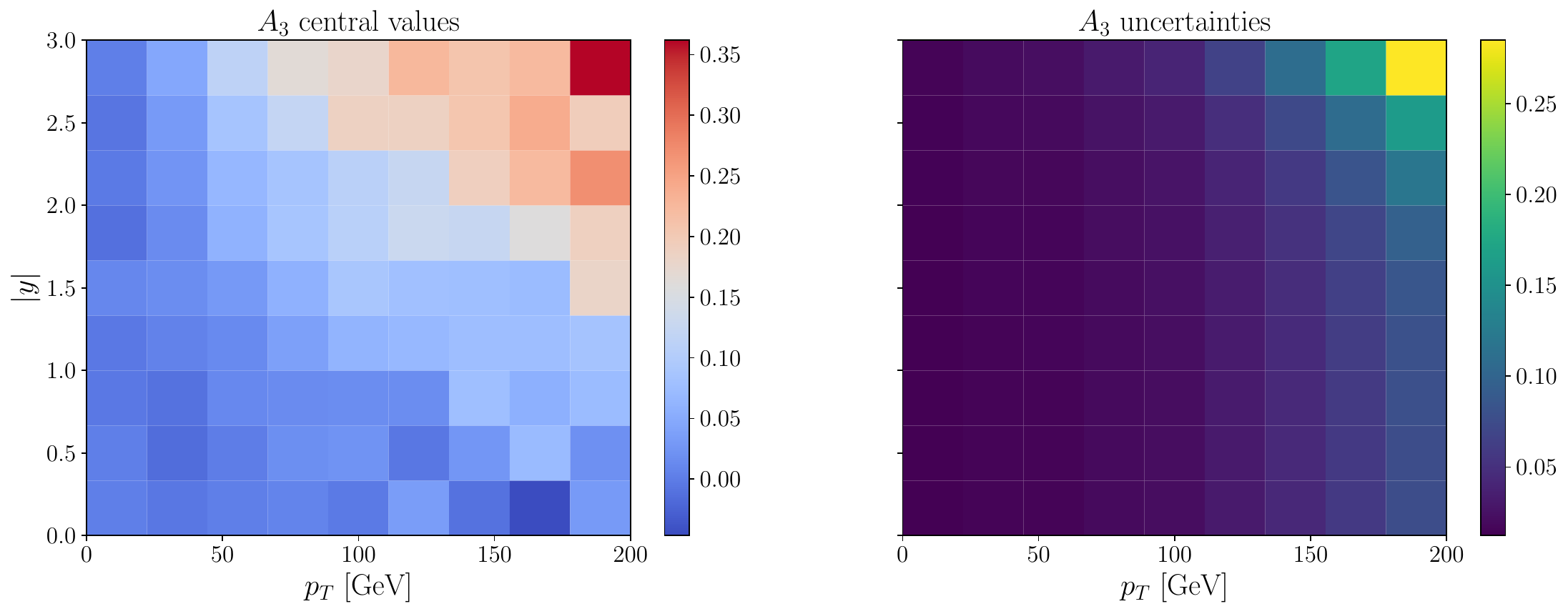}
    \label{fig:a3_2d_cv_unc}
  \end{minipage}

  \begin{minipage}[b]{0.72\linewidth}
    \includegraphics[width=\linewidth]{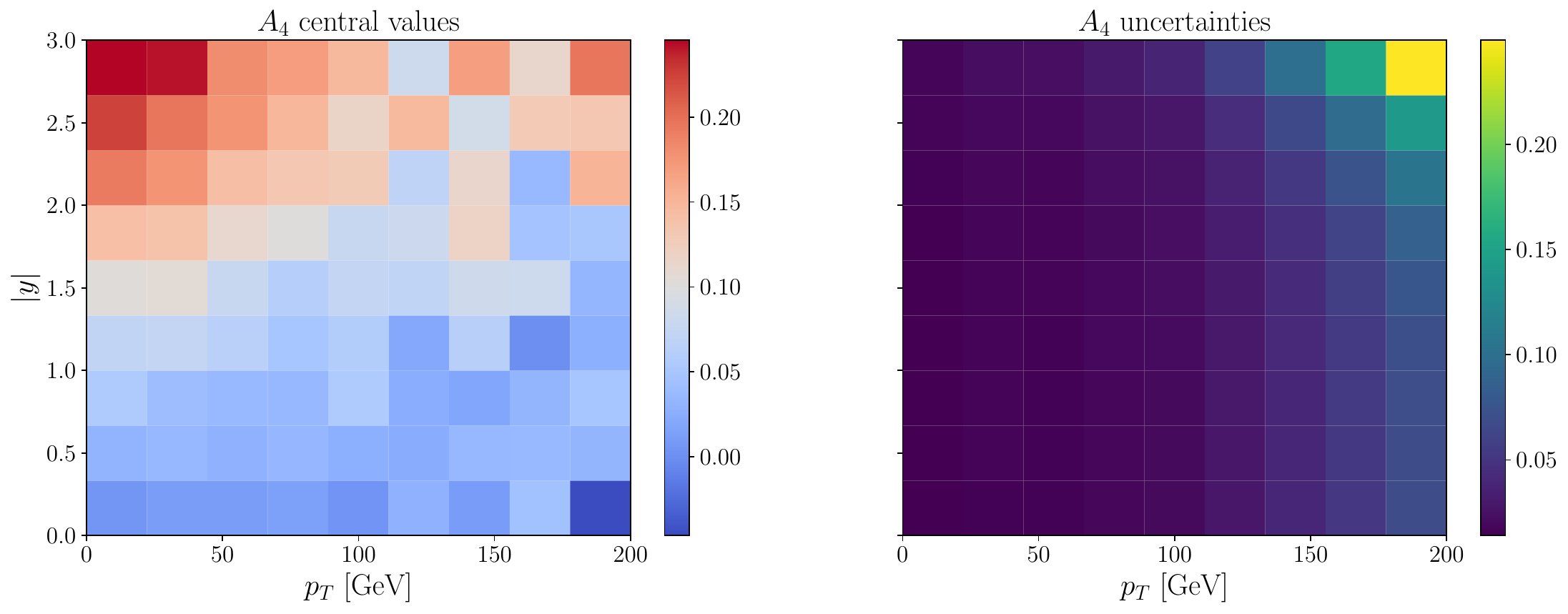}
    \label{fig:a4_2d_cv_unc}
  \end{minipage}

  \caption{Same as Fig.~\ref{fig:Ai_2d_cv_unc_1} for $A_3$ and $A_4$.}
  \label{fig:Ai_2d_cv_unc_2}
\end{figure}

We begin by showing the central values and uncertainties of the $A_0, \dots, A_4$ coefficients in the $(p_T, |y|)$ space in Figs.~\ref{fig:Ai_2d_cv_unc_1} and~\ref{fig:Ai_2d_cv_unc_2}.

Tables \ref{table:a1_2d},  \ref{table:a2_2d},  \ref{table:a3_2d},  \ref{table:a4_2d} show the SR hall of fame in $(p_T, y)$ of the $A_1$, $A_2$, $A_3$, and $A_4$ angular coefficients, respectively.

\begin{table}[h!]
\scriptsize
\setlength{\tabcolsep}{4pt}
\begin{center}
\begin{tabular}{@{}c c c c@{}}
\toprule
Equation & Complexity & Loss & Score \\
\midrule
$f(p_T, y) = 0.0395$ & 1 & 0.00221 & 0.0 \\
$f(p_T, y) = 0.0310 |y|$ & 3 & 0.00144 & 0.215 \\
$f(p_T, y) = 0.000574 p_T |y|$ & 5 & 0.000602 & 0.437 \\
$f(p_T, y) = |y| \left(p_T^{0.0222} - 1.05\right)$ & 7 & 0.000370 & 0.243 \\
\hline
\hline
\rowcolor{lightgray} 
$f(p_T, y) = |y|^{2} \left(p_T^{0.0114} - 1.03\right)$ & 9 & 0.000228 & 0.242 \\
$f(p_T, y) = \left(p_T^{0.0104} - 1.02\right) \left(|y|^{2} + 0.443\right)$ & 11 & 0.000198 & $7.06 \cdot 10^{-2}$ \\
$f(p_T, y) = |y| \left(- 1.02^{|y|} + \left(p_T + p_T^{|y|}\right)^{0.0105}\right)$ & 13 & 0.000194 & $8.87 \cdot 10^{-3}$ \\
$f(p_T, y) = \left(- 1.02^{|y|} + \left(p_T + p_T^{|y|}\right)^{0.0104}\right) \left(|y| + 0.0544\right)$ & 15 & 0.000191 & $7.75 \cdot 10^{-3}$ \\
$f(p_T, y) = \left(- 1.02^{|y|} + 1.01 \left(0.361 p_T + 0.361 p_T^{|y|} - 1\right)^{0.0104}\right) \left(|y| + 0.0614\right)$ & 17 & 0.000190 & $4.21 \cdot 10^{-3}$ \\
$f(p_T, y) = \left(- 1.02^{|y|} + 1.02 \left(0.149 p_T + 0.149 p_T^{|y|} - 1\right)^{0.0104}\right) \left(|y| + 0.0614\right)$ & 19 & 0.000188 & $4.67 \cdot 10^{-3}$ \\
\bottomrule
\end{tabular}
\end{center}
\caption{SR hall of fame for the $A_1$ coefficient in 2D $(p_T, |y|)$.}
\label{table:a1_2d}
\end{table}

\begin{table}[h!]
\scriptsize
\setlength{\tabcolsep}{4pt}
\begin{center}
\begin{tabular}{@{}c c c c@{}}
\toprule
Equation & Complexity & Loss & Score \\
\midrule
$f(p_T, y) = 0.352$ & 1 & 0.0874 & 0.0 \\
$f(p_T, y) = 0.00695 p_T$ & 3 & 0.00824 & 1.18 \\
$f(p_T, y) = p_T^{0.177} - 1.56$ & 5 & 0.00272 & 0.553 \\
$f(p_T, y) = \left(0.948 - 0.972^{p_T}\right)^{2.56}$ & 7 & 0.000810 & 0.607 \\
\hline
\hline
\rowcolor{lightgray} 
$f(p_T, y) = 1.49 \left(0.791 - 0.969^{p_T}\right)^{2.39}$ & 9 & 0.000638 & 0.119 \\
$f(p_T, y) = \left(- 0.953^{p_T} + p_T^{0.0595} - 0.396\right)^{3.42}$ & 11 & 0.000560 & $6.47 \cdot 10^{-2}$ \\
$f(p_T, y) = \left(- 0.952^{p_T} + p_T^{0.0598} - 0.396\right)^{3.42} - 0.00454$ & 13 & 0.000554 & $5.17 \cdot 10^{-3}$ \\
$f(p_T, y) = \left(- 0.952^{p_T} + \left(p_T - |y|^{-1.09}\right)^{0.0598} - 0.396\right)^{3.42}$ & 15 & 0.000547 & $7.08 \cdot 10^{-3}$ \\
$f(p_T, y) = 1.02 |y|^{0.0123} \left(- 0.953^{p_T} + p_T^{0.0603} - 0.404\right)^{3.42}$ & 17 & 0.000536 & $1.02 \cdot 10^{-2}$ \\
$f(p_T, y) = 1.02 |y|^{0.0123} \left(- 0.953^{p_T} + \left(p_T - 0.777\right)^{0.0603} - 0.404\right)^{3.42}$ & 19 & 0.000535 & $1.03 \cdot 10^{-3}$ \\
\bottomrule
\end{tabular}
\end{center}
\caption{SR hall of fame for the $A_2$ coefficient in 2D $(p_T, |y|)$.}
\label{table:a2_2d}
\end{table}

\begin{table}[h!]
\scriptsize
\setlength{\tabcolsep}{4pt}
\begin{center}
\begin{tabular}{@{}c c c c@{}}
\toprule
Equation & Complexity & Loss & Score \\
\midrule
$f(p_T, y) = 0.0265$ & 1 & 0.00196 & 0.0 \\
$f(p_T, y) = 0.000587 p_T$ & 3 & 0.00125 & 0.223 \\
$f(p_T, y) = 0.000521 p_T |y|$ & 5 & 0.000272 & 0.764 \\
\hline
\hline
\rowcolor{lightgray} 
$f(p_T, y) = 0.000623 p_T |y| - 0.0139$ & 7 & 0.000171 & 0.232 \\
$f(p_T, y) = 0.000506 p_T |y|^{1.27} - 0.0109$ & 9 & 0.000143 & $9.09 \cdot 10^{-2}$ \\
$f(p_T, y) = 0.000431 p_T |y|^{1.54} - 0.00855 |y|$ & 11 & 0.000121 & $8.22 \cdot 10^{-2}$ \\
$f(p_T, y) = p_T \left(0.000431 |y|^{1.54} + 1.78 \cdot 10^{-5}\right) - 0.00855 |y|$ & 13 & 0.000120 & $5.04 \cdot 10^{-3}$ \\
$f(p_T, y) = - 0.0107 |y|^{0.819} + 0.000533 \left(p_T |y|^{1.48}\right)^{0.978}$ & 15 & 0.000118 & $6.68 \cdot 10^{-3}$ \\
$f(p_T, y) = - 0.0107 |y| + 0.000533 \left(\left(p_T + |y|\right) \left(|y| - 0.0154\right)^{1.48}\right)^{0.978}$ & 17 & 0.000118 & $1.58 \cdot 10^{-3}$ \\
$f(p_T, y) = - 0.0107 |y| + 0.000533 \left(\left(p_T + |y|\right) \left(|y| - 0.0187\right)^{1.48}\right)^{0.978}$ & 19 & 0.000118 & $1.13 \cdot 10^{-4}$ \\
\bottomrule
\end{tabular}
\end{center}
\caption{SR hall of fame for the $A_3$ coefficient in 2D $(p_T, |y|)$.}
\label{table:a3_2d}
\end{table}

\begin{table}[h!]
\scriptsize
\setlength{\tabcolsep}{4pt}
\begin{center}
\begin{tabular}{@{}c c c c@{}}
\toprule
Equation & Complexity & Loss & Score \\
\midrule
$f(p_T, y) = 0.0849$ & 1 & 0.00426 & 0.0 \\
$f(p_T, y) = 0.0669 |y|$ & 3 & 0.000647 & 0.942 \\
$f(p_T, y) = 0.0533 |y|^{1.32}$ & 5 & 0.000532 & $9.79 \cdot 10^{-2}$ \\
$f(p_T, y) = |y| \left(0.0847 - 0.000336 p_T\right)$ & 7 & 0.000236 & 0.406 \\
\hline
\hline
\rowcolor{lightgray} 
$f(p_T, y) = |y|^{1.14} \left(0.0797 - 0.000336 p_T\right)$ & 9 & 0.000178 & 0.142 \\
$f(p_T, y) = |y| \left(|y| \left(0.0177 - 0.000154 p_T\right) + 0.0464\right)$ & 11 & 0.000142 & 0.114 \\
$f(p_T, y) = |y| \left(|y| \left(- 0.000154 p_T - 0.000154 |y| + 0.0183\right) + 0.0459\right)$ & 13 & 0.000141 & $1.00 \cdot 10^{-3}$ \\
$f(p_T, y) = |y| \left(|y| \left(- 0.000155 p_T - 0.000155 |y|^{|y|} + 0.0218\right) + 0.0406\right)$ & 15 & 0.000133 & $3.07 \cdot 10^{-2}$ \\
$f(p_T, y) = |y| \left(|y| \left(- 0.000156 p_T - 0.000156 |y|^{|y|} + 0.0218\right) + 0.0406\right) + 0.00169$ & 17 & 0.000132 & $4.55 \cdot 10^{-3}$ \\
$f(p_T, y) = \left(|y| \left(|y| \left(- 0.000156 p_T - 0.000156 |y|^{|y|} + 0.0218\right) + 0.0406\right)\right)^{1.00} + 0.00169$ & 19 & 0.000131 & $2.02 \cdot 10^{-3}$ \\
\bottomrule
\end{tabular}
\end{center}
\caption{SR hall of fame for the $A_4$ coefficient in 2D $(p_T, |y|)$.}
\label{table:a4_2d}
\end{table}

\begin{figure}[ht!]
  \centering
  \includegraphics[width=\linewidth]{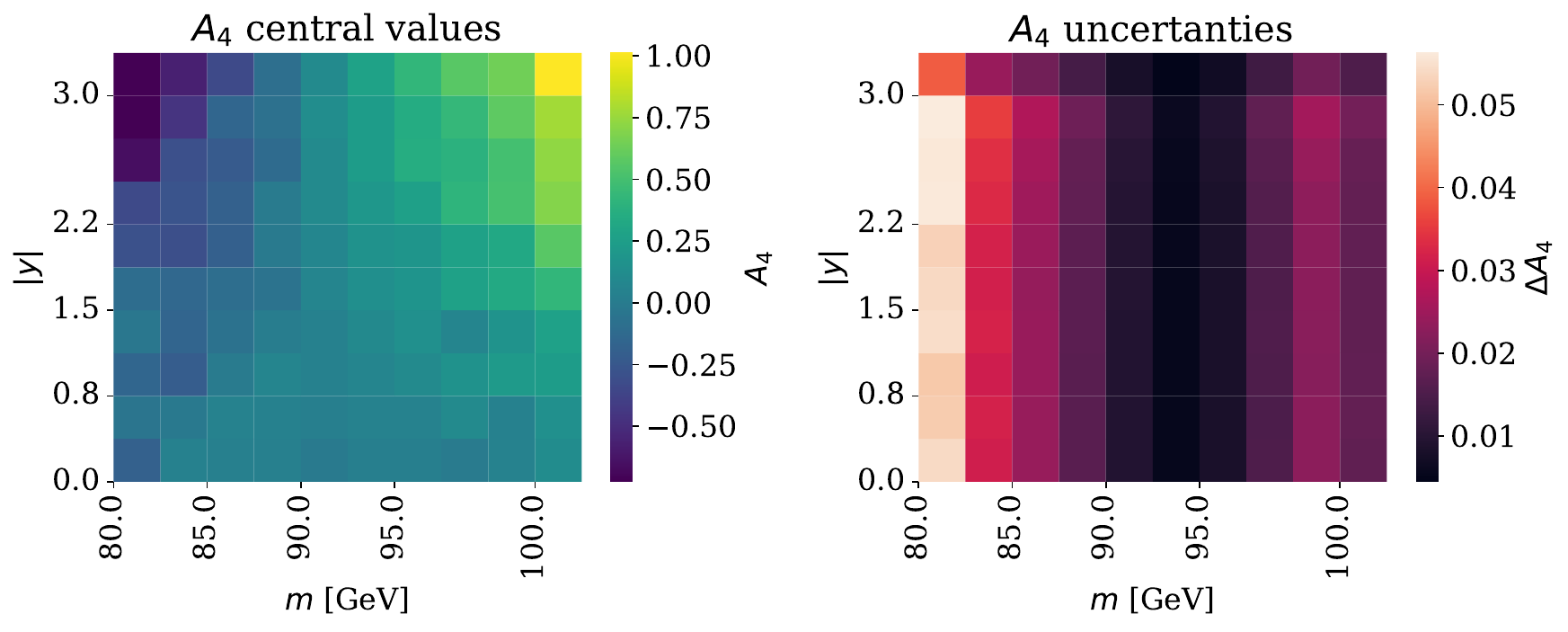}

  \caption{2D kinematic distributions for $A_4$ in $(|y|, m)$: central values (left) and uncertainties (right).}
  \label{fig:A_4_ym_appendix}
\end{figure}

Now, we show the central values and uncertainties for $A_4$ in the $(|y|, m)$ space in Figure \ref{fig:A_4_ym_appendix}.

\bibliographystyle{JHEP}
\bibliography{references}

@inproceedings{Morales-Alvarado:2024jrk,
    author = "Morales-Alvarado, Manuel and Conde, Daniel and Bendavid, Josh and Sanz, Veronica and Ubiali, Maria",
    title = "{Symbolic regression for precision LHC physics}",
    booktitle = "{38th conference on Neural Information Processing Systems}",
    eprint = "2412.07839",
    archivePrefix = "arXiv",
    primaryClass = "hep-ph",
    month = "12",
    year = "2024"
}

@article{Cranmer2023InterpretableML,
  title={Interpretable Machine Learning for Science with PySR and SymbolicRegression.jl},
  author={M. Cranmer},
  journal={ArXiv},
  year={2023},
  volume={abs/2305.01582},
  url={https://api.semanticscholar.org/CorpusID:258436785}
}

@misc{cranmer2019learningsymbolicphysicsgraph,
      title={Learning Symbolic Physics with Graph Networks}, 
      author={Miles D. Cranmer and Rui Xu and Peter Battaglia and Shirley Ho},
      year={2019},
      eprint={1909.05862},
      archivePrefix={arXiv},
      primaryClass={cs.LG},
      url={https://arxiv.org/abs/1909.05862}, 
}

@article{Udrescu:2019mnk,
    author = "Udrescu, Silviu-Marian and Tegmark, Max",
    title = "{AI Feynman: a Physics-Inspired Method for Symbolic Regression}",
    eprint = "1905.11481",
    archivePrefix = "arXiv",
    primaryClass = "physics.comp-ph",
    doi = "10.1126/sciadv.aay2631",
    journal = "Sci. Adv.",
    volume = "6",
    number = "16",
    pages = "eaay2631",
    year = "2020"
}

@article{Cranmer:2020wew,
    author = "Cranmer, Miles and Sanchez-Gonzalez, Alvaro and Battaglia, Peter and Xu, Rui and Cranmer, Kyle and Spergel, David and Ho, Shirley",
    title = "{Discovering Symbolic Models from Deep Learning with Inductive Biases}",
    eprint = "2006.11287",
    archivePrefix = "arXiv",
    primaryClass = "cs.LG",
    month = "6",
    year = "2020"
}

@article{Choi:2010wa,
    author = "Choi, Suyong",
    title = "{Construction of a Kinematic Variable Sensitive to the Mass of the Standard Model Higgs Boson in $H -> WW^* -> l^+ \nu l^- \bar{\nu}$ using Symbolic Regression}",
    eprint = "1006.4998",
    archivePrefix = "arXiv",
    primaryClass = "hep-ph",
    doi = "10.1007/JHEP08(2011)110",
    journal = "JHEP",
    volume = "08",
    pages = "110",
    year = "2011"
}

@article{Butter:2021rvz,
    author = "Butter, Anja and Plehn, Tilman and Soybelman, Nathalie and Brehmer, Johann",
    title = "{Back to the formula - LHC edition}",
    eprint = "2109.10414",
    archivePrefix = "arXiv",
    primaryClass = "hep-ph",
    doi = "10.21468/SciPostPhys.16.1.037",
    journal = "SciPost Phys.",
    volume = "16",
    number = "1",
    pages = "037",
    year = "2024"
}

@article{Dong:2022trn,
    author = "Dong, Zhongtian and Kong, Kyoungchul and Matchev, Konstantin T. and Matcheva, Katia",
    title = "{Is the machine smarter than the theorist: Deriving formulas for particle kinematics with symbolic regression}",
    eprint = "2211.08420",
    archivePrefix = "arXiv",
    primaryClass = "hep-ph",
    doi = "10.1103/PhysRevD.107.055018",
    journal = "Phys. Rev. D",
    volume = "107",
    number = "5",
    pages = "055018",
    year = "2023"
}

@article{Alnuqaydan:2022ncd,
    author = "Alnuqaydan, Abdulhakim and Gleyzer, Sergei and Prosper, Harrison",
    title = "{SYMBA: symbolic computation of squared amplitudes in high energy physics with machine learning}",
    eprint = "2206.08901",
    archivePrefix = "arXiv",
    primaryClass = "hep-ph",
    doi = "10.1088/2632-2153/acb2b2",
    journal = "Mach. Learn. Sci. Tech.",
    volume = "4",
    number = "1",
    pages = "015007",
    year = "2023"
}

@article{Tsoi:2023isc,
    author = "Tsoi, Ho Fung and Pol, Adrian Alan and Loncar, Vladimir and Govorkova, Ekaterina and Cranmer, Miles and Dasu, Sridhara and Elmer, Peter and Harris, Philip and Ojalvo, Isobel and Pierini, Maurizio",
    title = "{Symbolic Regression on FPGAs for Fast Machine Learning Inference}",
    eprint = "2305.04099",
    archivePrefix = "arXiv",
    primaryClass = "cs.LG",
    doi = "10.1051/epjconf/202429509036",
    journal = "EPJ Web Conf.",
    volume = "295",
    pages = "09036",
    year = "2024"
}

@article{AbdusSalam:2024obf,
    author = "AbdusSalam, Shehu and Abel, Steven and Crispim Rom{\~a}o, Miguel",
    title = "{Symbolic regression for beyond the standard model physics}",
    eprint = "2405.18471",
    archivePrefix = "arXiv",
    primaryClass = "hep-ph",
    reportNumber = "IPPP/24/27",
    doi = "10.1103/PhysRevD.111.015022",
    journal = "Phys. Rev. D",
    volume = "111",
    number = "1",
    pages = "015022",
    year = "2025"
}

@article{Soybelman:2024mbv,
    author = "Soybelman, Nathalie and Schiavi, Carlo and Di Bello, Francesco A. and Gross, Eilam",
    title = "{Accelerating graph-based tracking tasks with symbolic regression}",
    eprint = "2406.16752",
    archivePrefix = "arXiv",
    primaryClass = "hep-ex",
    doi = "10.1088/2632-2153/ad8f12",
    journal = "Mach. Learn. Sci. Tech.",
    volume = "5",
    number = "4",
    pages = "045042",
    year = "2024"
}

@article{Tsoi:2024pbn,
    author = "Tsoi, Ho Fung and Rankin, Dylan and Caillol, Cecile and Cranmer, Miles and Dasu, Sridhara and Duarte, Javier and Harris, Philip and Lipeles, Elliot and Loncar, Vladimir",
    title = "{SymbolFit: Automatic Parametric Modeling with Symbolic Regression}",
    eprint = "2411.09851",
    archivePrefix = "arXiv",
    primaryClass = "hep-ex",
    doi = "10.1007/s41781-025-00140-9",
    journal = "Comput. Softw. Big Sci.",
    volume = "9",
    number = "1",
    pages = "12",
    year = "2025"
}

@article{Tsoi:2024ypg,
    author = "Tsoi, Ho Fung and Loncar, Vladimir and Dasu, Sridhara and Harris, Philip",
    title = "{SymbolNet: neural symbolic regression with adaptive dynamic pruning for compression}",
    eprint = "2401.09949",
    archivePrefix = "arXiv",
    primaryClass = "cs.LG",
    doi = "10.1088/2632-2153/adaad8",
    journal = "Mach. Learn. Sci. Tech.",
    volume = "6",
    number = "1",
    pages = "015021",
    year = "2025"
}

@misc{virgolin2022symbolicregressionnphard,
      title={Symbolic Regression is NP-hard}, 
      author={Marco Virgolin and Solon P. Pissis},
      year={2022},
      eprint={2207.01018},
      archivePrefix={arXiv},
      primaryClass={cs.NE},
      url={https://arxiv.org/abs/2207.01018}, 
}

@misc{matchev2021analyticalmodellingexoplanettransit,
      title={Analytical Modelling of Exoplanet Transit Specroscopy with Dimensional Analysis and Symbolic Regression}, 
      author={Konstantin T. Matchev and Katia Matcheva and Alexander Roman},
      year={2021},
      eprint={2112.11600},
      archivePrefix={arXiv},
      primaryClass={astro-ph.EP},
      url={https://arxiv.org/abs/2112.11600}, 
}

@misc{lemos2022rediscoveringorbitalmechanicsmachine,
      title={Rediscovering orbital mechanics with machine learning}, 
      author={Pablo Lemos and Niall Jeffrey and Miles Cranmer and Shirley Ho and Peter Battaglia},
      year={2022},
      eprint={2202.02306},
      archivePrefix={arXiv},
      primaryClass={astro-ph.EP},
      url={https://arxiv.org/abs/2202.02306}, 
}

@article{Dersy:2022bym,
    author = "Dersy, Aur{\'e}lien and Schwartz, Matthew D. and Zhang, Xiaoyuan",
    title = "{Simplifying Polylogarithms with Machine Learning}",
    eprint = "2206.04115",
    archivePrefix = "arXiv",
    primaryClass = "cs.LG",
    doi = "10.1142/S2810939223500028",
    journal = "Int. J. Data Sci. Math. Sci.",
    volume = "1",
    number = "2",
    pages = "135--179",
    year = "2024"
}

@article{ATLAS:2019bsc,
    author = "Aaboud, Morad and others",
    collaboration = "ATLAS",
    title = "{Measurement of $W^{\pm}Z$ production cross sections and gauge boson polarisation in $pp$ collisions at $\sqrt{s} = 13$ TeV with the ATLAS detector}",
    eprint = "1902.05759",
    archivePrefix = "arXiv",
    primaryClass = "hep-ex",
    reportNumber = "CERN-EP-2018-327",
    doi = "10.1140/epjc/s10052-019-7027-6",
    journal = "Eur. Phys. J. C",
    volume = "79",
    number = "6",
    pages = "535",
    year = "2019"
}

@article{CMS:2020etf,
    author = "Sirunyan, Albert M and others",
    collaboration = "CMS",
    title = "{Measurements of production cross sections of polarized same-sign W boson pairs in association with two jets in proton-proton collisions at $\sqrt{s} =$ 13 TeV}",
    eprint = "2009.09429",
    archivePrefix = "arXiv",
    primaryClass = "hep-ex",
    reportNumber = "CMS-SMP-20-006, CERN-EP-2020-168",
    doi = "10.1016/j.physletb.2020.136018",
    journal = "Phys. Lett. B",
    volume = "812",
    pages = "136018",
    year = "2021"
}

@article{CMS:2021icx,
    author = "Tumasyan, Armen and others",
    collaboration = "CMS",
    title = "{Measurement of the inclusive and differential WZ production cross sections, polarization angles, and triple gauge couplings in pp collisions at $ \sqrt{s} $ = 13 TeV}",
    eprint = "2110.11231",
    archivePrefix = "arXiv",
    primaryClass = "hep-ex",
    reportNumber = "CMS-SMP-20-014, CERN-EP-2021-163",
    doi = "10.1007/JHEP07(2022)032",
    journal = "JHEP",
    volume = "07",
    pages = "032",
    year = "2022"
}

@article{ATLAS:2022oge,
    author = "Aad, Georges and others",
    collaboration = "ATLAS",
    title = "{Observation of gauge boson joint-polarisation states in W{\ensuremath{\pm}}Z production from pp collisions at s=13 TeV with the ATLAS detector}",
    eprint = "2211.09435",
    archivePrefix = "arXiv",
    primaryClass = "hep-ex",
    reportNumber = "CERN-EP-2022-202",
    doi = "10.1016/j.physletb.2023.137895",
    journal = "Phys. Lett. B",
    volume = "843",
    pages = "137895",
    year = "2023"
}

@article{ATLAS:2023lsr,
    author = "Aad, Georges and others",
    collaboration = "ATLAS",
    title = "{A precise measurement of the Z-boson double-differential transverse momentum and rapidity distributions in the full phase space of the decay leptons with the ATLAS experiment at $\sqrt s$ = 8 TeV}",
    eprint = "2309.09318",
    archivePrefix = "arXiv",
    primaryClass = "hep-ex",
    reportNumber = "CERN-EP-2023-171",
    month = "9",
    year = "2023"
}

@article{LHCb:2022tbc,
    author = "Aaij, R. and others",
    collaboration = "LHCb",
    title = "{First Measurement of the Z{\textrightarrow}{\ensuremath{\mu}}+{\ensuremath{\mu}}- Angular Coefficients in the Forward Region of pp Collisions at s=13{\,}{\,}TeV}",
    eprint = "2203.01602",
    archivePrefix = "arXiv",
    primaryClass = "hep-ex",
    reportNumber = "LHCb-PAPER-2021-048, CERN-EP-2022-030",
    doi = "10.1103/PhysRevLett.129.091801",
    journal = "Phys. Rev. Lett.",
    volume = "129",
    number = "9",
    pages = "091801",
    year = "2022"
}

@article{Collins:1977iv,
    author = "Collins, John C. and Soper, Davison E.",
    title = "{Angular Distribution of Dileptons in High-Energy Hadron Collisions}",
    reportNumber = "Print-77-0288 (PRINCETON)",
    doi = "10.1103/PhysRevD.16.2219",
    journal = "Phys. Rev. D",
    volume = "16",
    pages = "2219",
    year = "1977"
}

@article{Hagiwara:1984hi,
    author = "Hagiwara, Kaoru and Hikasa, Ken-ichi and Kai, Naoyuki",
    title = "{Parity Odd Asymmetries in $W$ Jet Events at Hadron Colliders}",
    reportNumber = "MAD/PH/141, UT-418-TOKYO",
    doi = "10.1103/PhysRevLett.52.1076",
    journal = "Phys. Rev. Lett.",
    volume = "52",
    pages = "1076",
    year = "1984"
}

@article{Mirkes:1992hu,
    author = "Mirkes, E.",
    title = "{Angular decay distribution of leptons from W bosons at NLO in hadronic collisions}",
    reportNumber = "TTP-92-12",
    doi = "10.1016/0550-3213(92)90046-E",
    journal = "Nucl. Phys. B",
    volume = "387",
    pages = "3--85",
    year = "1992"
}

@article{Mirkes:1994eb,
    author = "Mirkes, E. and Ohnemus, J.",
    title = "{$W$ and $Z$ polarization effects in hadronic collisions}",
    eprint = "hep-ph/9406381",
    archivePrefix = "arXiv",
    reportNumber = "MAD-PH-834, UCD-94-23",
    doi = "10.1103/PhysRevD.50.5692",
    journal = "Phys. Rev. D",
    volume = "50",
    pages = "5692--5703",
    year = "1994"
}

@article{Mirkes:1994dp,
    author = "Mirkes, E. and Ohnemus, J.",
    title = "{Angular distributions of Drell-Yan lepton pairs at the Tevatron: Order $\alpha-s^{2}$ corrections and Monte Carlo studies}",
    eprint = "hep-ph/9412289",
    archivePrefix = "arXiv",
    reportNumber = "MAD-PH-857, UCD-94-39",
    doi = "10.1103/PhysRevD.51.4891",
    journal = "Phys. Rev. D",
    volume = "51",
    pages = "4891--4904",
    year = "1995"
}

@inproceedings{Mirkes:1994nr,
    author = "Mirkes, E. and Ohnemus, J.",
    title = "{Polarization effects in Drell-Yan type processes h1 + h2 ---\ensuremath{>} (W, Z, gamma*, J / psi) + x}",
    booktitle = "{1994 Meeting of the American Physical Society, Division of Particles and Fields (DPF 94)}",
    eprint = "hep-ph/9408402",
    archivePrefix = "arXiv",
    pages = "1721--1723",
    month = "8",
    year = "1994"
}

@article{Bern:2011ie,
    author = "Bern, Z. and others",
    title = "{Left-Handed W Bosons at the LHC}",
    eprint = "1103.5445",
    archivePrefix = "arXiv",
    primaryClass = "hep-ph",
    reportNumber = "SLAC-PUB-14409, CERN-PH-TH-2011-062, UCLA-11-TEP-10, SB-F-386-11, SACLAY-IPHT-T11-040, IPPP-11-15, NIKHEF-2011-006",
    doi = "10.1103/PhysRevD.84.034008",
    journal = "Phys. Rev. D",
    volume = "84",
    pages = "034008",
    year = "2011"
}

@article{Stirling:2012zt,
    author = "Stirling, W. J. and Vryonidou, E.",
    title = "{Electroweak gauge boson polarisation at the LHC}",
    eprint = "1204.6427",
    archivePrefix = "arXiv",
    primaryClass = "hep-ph",
    doi = "10.1007/JHEP07(2012)124",
    journal = "JHEP",
    volume = "07",
    pages = "124",
    year = "2012"
}

@article{Belyaev:2013nla,
    author = "Belyaev, Alexander and Ross, Douglas",
    title = "{What Does the CMS Measurement of W-polarization Tell Us about the Underlying Theory of the Coupling of W-Bosons to Matter?}",
    eprint = "1303.3297",
    archivePrefix = "arXiv",
    primaryClass = "hep-ph",
    reportNumber = "SHEP-13-05",
    doi = "10.1007/JHEP08(2013)120",
    journal = "JHEP",
    volume = "08",
    pages = "120",
    year = "2013"
}

@article{Gauld:2017tww,
    author = "Gauld, R. and Gehrmann-De Ridder, A. and Gehrmann, T. and Glover, E. W. N. and Huss, A.",
    title = "{Precise predictions for the angular coefficients in Z-boson production at the LHC}",
    eprint = "1708.00008",
    archivePrefix = "arXiv",
    primaryClass = "hep-ph",
    reportNumber = "IPPP-17-58, ZU-TH-21-17",
    doi = "10.1007/JHEP11(2017)003",
    journal = "JHEP",
    volume = "11",
    pages = "003",
    year = "2017"
}

@article{Lyubovitskij:2025oig,
    author = "Lyubovitskij, Valery E. and Zhevlakov, Alexey S. and Anikin, Iurii A.",
    title = "{Angular coefficients of the Drell-Yan process across different rapidity and kinematical ranges}",
    eprint = "2503.16008",
    archivePrefix = "arXiv",
    primaryClass = "hep-ph",
    month = "3",
    year = "2025"
}

@article{Li:2025fom,
    author = "Li, Guanghui and Li, Xu and Yan, Bin",
    title = "{Lam-Tung relation breaking effects and weak dipole moments at lepton colliders}",
    eprint = "2503.17663",
    archivePrefix = "arXiv",
    primaryClass = "hep-ph",
    month = "3",
    year = "2025"
}

@article{Gehrmann-DeRidder:2015wbt,
    author = "Gehrmann-De Ridder, A. and Gehrmann, T. and Glover, E. W. N. and Huss, A. and Morgan, T. A.",
    title = "{Precise QCD predictions for the production of a Z boson in association with a hadronic jet}",
    eprint = "1507.02850",
    archivePrefix = "arXiv",
    primaryClass = "hep-ph",
    reportNumber = "IPPP-15-44, ZU-TH-23-15",
    doi = "10.1103/PhysRevLett.117.022001",
    journal = "Phys. Rev. Lett.",
    volume = "117",
    number = "2",
    pages = "022001",
    year = "2016"
}

@article{Gehrmann-DeRidder:2016jns,
    author = "Gehrmann-De Ridder, A. and Gehrmann, T. and Glover, E. W. N. and Huss, A. and Morgan, T. A.",
    title = "{NNLO QCD corrections for Drell-Yan $p_T^Z$ and $\phi^*$ observables at the LHC}",
    eprint = "1610.01843",
    archivePrefix = "arXiv",
    primaryClass = "hep-ph",
    reportNumber = "IPPP-16-74, ZU-TH-36-16, IPPP/16/74, ZU-TH 36/16",
    doi = "10.1007/JHEP11(2016)094",
    journal = "JHEP",
    volume = "11",
    pages = "094",
    year = "2016",
    note = "[Erratum: JHEP 10, 126 (2018)]"
}

@article{Gauld:2024glt,
    author = "Gauld, R. and Haisch, U. and Weiss, J.",
    title = "{A tale of $Z$+jet: SMEFT effects and the Lam-Tung relation}",
    eprint = "2412.13014",
    archivePrefix = "arXiv",
    primaryClass = "hep-ph",
    reportNumber = "MPP-2024-236",
    doi = "10.21468/SciPostPhys.18.5.148",
    journal = "SciPost Phys.",
    volume = "18",
    pages = "148",
    year = "2025"
}

@article{CMS:2011kaj,
    author = "Chatrchyan, Serguei and others",
    collaboration = "CMS",
    title = "{Measurement of the Polarization of W Bosons with Large Transverse Momenta in W+Jets Events at the LHC}",
    eprint = "1104.3829",
    archivePrefix = "arXiv",
    primaryClass = "hep-ex",
    reportNumber = "CERN-PH-EP-2011-043, CMS-EWK-10-014",
    doi = "10.1103/PhysRevLett.107.021802",
    journal = "Phys. Rev. Lett.",
    volume = "107",
    pages = "021802",
    year = "2011"
}

@article{ATLAS:2012au,
    author = "Aad, Georges and others",
    collaboration = "ATLAS",
    title = "{Measurement of the polarisation of $W$ bosons produced with large transverse momentum in $pp$ collisions at $\sqrt{s}=7$ TeV with the ATLAS experiment}",
    eprint = "1203.2165",
    archivePrefix = "arXiv",
    primaryClass = "hep-ex",
    reportNumber = "CERN-PH-EP-2012-016",
    doi = "10.1140/epjc/s10052-012-2001-6",
    journal = "Eur. Phys. J. C",
    volume = "72",
    pages = "2001",
    year = "2012"
}

@article{CMS:2015cyj,
    author = "Khachatryan, Vardan and others",
    collaboration = "CMS",
    title = "{Angular coefficients of Z bosons produced in pp collisions at $\sqrt{s}$ = 8 TeV and decaying to $\mu^+ \mu^-$ as a function of transverse momentum and rapidity}",
    eprint = "1504.03512",
    archivePrefix = "arXiv",
    primaryClass = "hep-ex",
    reportNumber = "CMS-SMP-13-010, CERN-PH-EP-2015-046",
    doi = "10.1016/j.physletb.2015.08.061",
    journal = "Phys. Lett. B",
    volume = "750",
    pages = "154--175",
    year = "2015"
}

@article{ATLAS:2016rnf,
    author = "Aad, Georges and others",
    collaboration = "ATLAS",
    title = "{Measurement of the angular coefficients in $Z$-boson events using electron and muon pairs from data taken at $\sqrt{s}=8$ TeV with the ATLAS detector}",
    eprint = "1606.00689",
    archivePrefix = "arXiv",
    primaryClass = "hep-ex",
    reportNumber = "CERN-EP-2016-087",
    doi = "10.1007/JHEP08(2016)159",
    journal = "JHEP",
    volume = "08",
    pages = "159",
    year = "2016"
}

@article{Grossi:2023fqq,
    author = "Grossi, Michele and Incudini, Massimiliano and Pellen, Mathieu and Pelliccioli, Giovanni",
    title = "{Amplitude-assisted tagging of longitudinally polarised bosons using wide neural networks}",
    eprint = "2306.07726",
    archivePrefix = "arXiv",
    primaryClass = "hep-ph",
    reportNumber = "FR-PHENO-2023-05, MPP-2023-122",
    doi = "10.1140/epjc/s10052-023-11931-y",
    journal = "Eur. Phys. J. C",
    volume = "83",
    number = "8",
    pages = "759",
    year = "2023"
}

@article{Alwall:2014hca,
    author = "Alwall, J. and Frederix, R. and Frixione, S. and Hirschi, V. and Maltoni, F. and Mattelaer, O. and Shao, H. -S. and Stelzer, T. and Torrielli, P. and Zaro, M.",
    title = "{The automated computation of tree-level and next-to-leading order differential cross sections, and their matching to parton shower simulations}",
    eprint = "1405.0301",
    archivePrefix = "arXiv",
    primaryClass = "hep-ph",
    reportNumber = "CERN-PH-TH-2014-064, CP3-14-18, LPN14-066, MCNET-14-09, ZU-TH-14-14",
    doi = "10.1007/JHEP07(2014)079",
    journal = "JHEP",
    volume = "07",
    pages = "079",
    year = "2014"
}

@article{Frederix:2018nkq,
    author = "Frederix, R. and Frixione, S. and Hirschi, V. and Pagani, D. and Shao, H. -S. and Zaro, M.",
    title = "{The automation of next-to-leading order electroweak calculations}",
    eprint = "1804.10017",
    archivePrefix = "arXiv",
    primaryClass = "hep-ph",
    reportNumber = "Nikhef/2018-015, TUM-HEP-1138/18, NIKHEF-2018-015, TUM-HEP-1138-18",
    doi = "10.1007/JHEP11(2021)085",
    journal = "JHEP",
    volume = "07",
    pages = "185",
    year = "2018",
    note = "[Erratum: JHEP 11, 085 (2021)]"
}

@article{Gao:2013xoa,
    author = "Gao, Jun and Guzzi, Marco and Huston, Joey and Lai, Hung-Liang and Li, Zhao and Nadolsky, Pavel and Pumplin, Jon and Stump, Daniel and Yuan, C. -P.",
    title = "{CT10 next-to-next-to-leading order global analysis of QCD}",
    eprint = "1302.6246",
    archivePrefix = "arXiv",
    primaryClass = "hep-ph",
    reportNumber = "SMU-HEP-12-23",
    doi = "10.1103/PhysRevD.89.033009",
    journal = "Phys. Rev. D",
    volume = "89",
    number = "3",
    pages = "033009",
    year = "2014"
}

@book{Ellis:1996mzs,
    author = "Ellis, R. Keith and Stirling, W. James and Webber, B. R.",
    title = "{QCD and collider physics}",
    doi = "10.1017/CBO9780511628788",
    isbn = "978-0-511-82328-2, 978-0-521-54589-1",
    publisher = "Cambridge University Press",
    volume = "8",
    month = "2",
    year = "2011"
}

@book{Peskin:1995ev,
    author = "Peskin, Michael E. and Schroeder, Daniel V.",
    title = "{An Introduction to quantum field theory}",
    doi = "10.1201/9780429503559",
    isbn = "978-0-201-50397-5, 978-0-429-50355-9, 978-0-429-49417-8",
    publisher = "Addison-Wesley",
    address = "Reading, USA",
    year = "1995"
}

@article{Bahl:2025jtk,
    author = "Bahl, Henning and Fuchs, Elina and Menen, Marco and Plehn, Tilman",
    title = "{$\mathcal{CP}$-Analyses with Symbolic Regression}",
    eprint = "2507.05858",
    archivePrefix = "arXiv",
    primaryClass = "hep-ph",
    month = "7",
    year = "2025"
}

@article{Lam:1978zr,
    author = "Lam, C. S. and Tung, Wu-Ki",
    title = "{Structure Function Relations at Large Transverse Momenta in Lepton Pair Production Processes}",
    reportNumber = "RL-78-054",
    doi = "10.1016/0370-2693(79)90204-1",
    journal = "Phys. Lett. B",
    volume = "80",
    pages = "228--231",
    year = "1979"
}

@article{Lam:1978pu,
    author = "Lam, C. S. and Tung, Wu-Ki",
    title = "{A Systematic Approach to Inclusive Lepton Pair Production in Hadronic Collisions}",
    reportNumber = "Print-78-0553 (MCGILL)",
    doi = "10.1103/PhysRevD.18.2447",
    journal = "Phys. Rev. D",
    volume = "18",
    pages = "2447",
    year = "1978"
}

@article{Lam:1980uc,
    author = "Lam, C. S. and Tung, Wu-Ki",
    title = "{A Parton Model Relation Sans {QCD} Modifications in Lepton Pair Productions}",
    reportNumber = "SLAC-PUB-2454",
    doi = "10.1103/PhysRevD.21.2712",
    journal = "Phys. Rev. D",
    volume = "21",
    pages = "2712",
    year = "1980"
}

@article{Piloneta:2024aac,
    author = "Piloneta, Sara and Vladimirov, Alexey",
    title = "{Angular distributions of Drell-Yan leptons in the TMD factorization approach}",
    eprint = "2407.06277",
    archivePrefix = "arXiv",
    primaryClass = "hep-ph",
    reportNumber = "IPARCOS-UCM-24-037",
    doi = "10.1007/JHEP12(2024)059",
    journal = "JHEP",
    volume = "12",
    pages = "059",
    year = "2024"
}

@article{Li:2024iyj,
    author = "Li, Xu and Yan, Bin and Yuan, C. -P.",
    title = "{Lam-Tung relation breaking in Z boson production as a probe of standard model effective field theory effects}",
    eprint = "2405.04069",
    archivePrefix = "arXiv",
    primaryClass = "hep-ph",
    reportNumber = "MSUHEP-24-006",
    doi = "10.1103/PhysRevD.111.073007",
    journal = "Phys. Rev. D",
    volume = "111",
    number = "7",
    pages = "073007",
    year = "2025"
}

@article{Arroyo-Castro:2025slx,
    author = "Arroyo-Castro, Arturo and Scimemi, Ignazio and Vladimirov, Alexey",
    title = "{Leading q$_{T}$/Q correction for Drell-Yan process in TMD factorization}",
    eprint = "2503.24336",
    archivePrefix = "arXiv",
    primaryClass = "hep-ph",
    reportNumber = "IPARCOS-UCM-25-021",
    doi = "10.1007/JHEP06(2025)202",
    journal = "JHEP",
    volume = "06",
    pages = "202",
    year = "2025"
}

@article{Hiller:2025hpf,
    author = "Hiller, Gudrun and Nollen, Lara and Wendler, Daniel",
    title = "{Total Drell{\textendash}Yan in the flavorful SMEFT}",
    eprint = "2502.12250",
    archivePrefix = "arXiv",
    primaryClass = "hep-ph",
    doi = "10.1140/epjc/s10052-025-14349-w",
    journal = "Eur. Phys. J. C",
    volume = "85",
    number = "6",
    pages = "657",
    year = "2025"
}

@inproceedings{Ubiali:2024pyg,
    author = "Ubiali, Maria",
    title = "{Parton Distribution Functions and Their Impact on Precision of the Current Theory Calculations}",
    eprint = "2404.08508",
    archivePrefix = "arXiv",
    primaryClass = "hep-ph",
    month = "4",
    year = "2024"
}

@article{Amoroso:2022eow,
    author = "Amoroso, S. and others",
    title = "{Snowmass 2021 Whitepaper: Proton Structure at the Precision Frontier}",
    eprint = "2203.13923",
    archivePrefix = "arXiv",
    primaryClass = "hep-ph",
    reportNumber = "Edinburgh 2022/08, FERMILAB-PUB-22-222-QIS-SCD-T, MPP-2022-32,
  SLAC-PUB-17652, SMU-HEP-22-02, TIF-UNIMI-2022-6",
    doi = "10.5506/APhysPolB.53.12-A1",
    journal = "Acta Phys. Polon. B",
    volume = "53",
    number = "12",
    pages = "12-A1",
    year = "2022"
}

@article{Buckley:2014ana,
    author = {Buckley, Andy and Ferrando, James and Lloyd, Stephen and Nordstr{\"o}m, Karl and Page, Ben and R{\"u}fenacht, Martin and Sch{\"o}nherr, Marek and Watt, Graeme},
    title = "{LHAPDF6: parton density access in the LHC precision era}",
    eprint = "1412.7420",
    archivePrefix = "arXiv",
    primaryClass = "hep-ph",
    reportNumber = "GLAS-PPE-2014-05, MCNET-14-29, IPPP-14-111, DCPT-14-222",
    doi = "10.1140/epjc/s10052-015-3318-8",
    journal = "Eur. Phys. J. C",
    volume = "75",
    pages = "132",
    year = "2015"
}

@article{Callan:1969uq,
    author = "Callan, Jr., Curtis G. and Gross, David J.",
    title = "{High-energy electroproduction and the constitution of the electric current}",
    doi = "10.1103/PhysRevLett.22.156",
    journal = "Phys. Rev. Lett.",
    volume = "22",
    pages = "156--159",
    year = "1969"
}

@article{Dotson:2025omi,
    author = "Dotson, Andrew and others",
    title = "{Generalized Parton Distributions from Symbolic Regression}",
    eprint = "2504.13289",
    archivePrefix = "arXiv",
    primaryClass = "hep-ph",
    month = "4",
    year = "2025"
}

@article{NNPDF:2021njg,
    author = "Ball, Richard D. and others",
    collaboration = "NNPDF",
    title = "{The path to proton structure at 1{\%} accuracy}",
    eprint = "2109.02653",
    archivePrefix = "arXiv",
    primaryClass = "hep-ph",
    reportNumber = "Edinburgh 2021/12, Nikhef-2021-013, TIF-UNIMI-2021-11",
    doi = "10.1140/epjc/s10052-022-10328-7",
    journal = "Eur. Phys. J. C",
    volume = "82",
    number = "5",
    pages = "428",
    year = "2022"
}

@article{CMS:2020cph,
    author = "Sirunyan, Albert M and others",
    collaboration = "CMS",
    title = "{Measurements of the $W$ boson rapidity, helicity, double-differential cross sections, and charge asymmetry in $pp$ collisions at $\sqrt {s}$ = 13  TeV}",
    eprint = "2008.04174",
    archivePrefix = "arXiv",
    primaryClass = "hep-ex",
    reportNumber = "CMS-SMP-18-012, CERN-EP-2020-116",
    doi = "10.1103/PhysRevD.102.092012",
    journal = "Phys. Rev. D",
    volume = "102",
    number = "9",
    pages = "092012",
    year = "2020"
}

@article{ATLAS:2023lhg,
    author = "Aad, Georges and others",
    collaboration = "ATLAS",
    title = "{A precise determination of the strong-coupling constant from the recoil of $Z$ bosons with the ATLAS experiment at $\sqrt{s} = 8$ TeV}",
    eprint = "2309.12986",
    archivePrefix = "arXiv",
    primaryClass = "hep-ex",
    month = "9",
    year = "2023"
}

@article{ATLAS:2024erm,
    author = "Aad, Georges and others",
    collaboration = "ATLAS",
    title = "{Measurement of the W-boson mass and width with the ATLAS detector using proton{\textendash}proton collisions at $\sqrt{s}=7$ TeV}",
    eprint = "2403.15085",
    archivePrefix = "arXiv",
    primaryClass = "hep-ex",
    reportNumber = "CERN-EP-2024-074",
    doi = "10.1140/epjc/s10052-024-13190-x",
    journal = "Eur. Phys. J. C",
    volume = "84",
    number = "12",
    pages = "1309",
    year = "2024"
}

@article{LHCb:2021bjt,
    author = "Aaij, Roel and others",
    collaboration = "LHCb",
    title = "{Measurement of the W boson mass}",
    eprint = "2109.01113",
    archivePrefix = "arXiv",
    primaryClass = "hep-ex",
    reportNumber = "LHCb-PAPER-2021-024, CERN-EP-2021-170",
    doi = "10.1007/JHEP01(2022)036",
    journal = "JHEP",
    volume = "01",
    pages = "036",
    year = "2022"
}

@article{CMS:2024lrd,
    author = "Chekhovsky, Vladimir and others",
    collaboration = "CMS",
    title = "{High-precision measurement of the W boson mass with the CMS experiment at the LHC}",
    eprint = "2412.13872",
    archivePrefix = "arXiv",
    primaryClass = "hep-ex",
    reportNumber = "CMS-SMP-23-002, CERN-EP-2024-308",
    month = "12",
    year = "2024"
}

@article{CDF:2022hxs,
    author = "Aaltonen, T. and others",
    collaboration = "CDF",
    title = "{High-precision measurement of the $W$          boson mass with the CDF II detector}",
    reportNumber = "FERMILAB-PUB-22-254-PPD",
    doi = "10.1126/science.abk1781",
    journal = "Science",
    volume = "376",
    number = "6589",
    pages = "170--176",
    year = "2022"
}

@article{D0:2013jba,
    author = "Abazov, Victor Mukhamedovich and others",
    collaboration = "D0",
    title = "{Measurement of the $W$ boson mass with the D0 detector}",
    eprint = "1310.8628",
    archivePrefix = "arXiv",
    primaryClass = "hep-ex",
    reportNumber = "FERMILAB-PUB-13-489-E",
    doi = "10.1103/PhysRevD.89.012005",
    journal = "Phys. Rev. D",
    volume = "89",
    number = "1",
    pages = "012005",
    year = "2014"
}

@article{LHC-TeVMWWorkingGroup:2023zkn,
    author = "Amoroso, Simone and others",
    collaboration = "LHC-TeV~MW~Working~Group",
    title = "{Compatibility and combination of world W-boson mass measurements}",
    eprint = "2308.09417",
    archivePrefix = "arXiv",
    primaryClass = "hep-ex",
    doi = "10.1140/epjc/s10052-024-12532-z",
    journal = "Eur. Phys. J. C",
    volume = "84",
    number = "5",
    pages = "451",
    year = "2024"
}

@article{CMS:2024ony,
    author = "Hayrapetyan, Aram and others",
    collaboration = "CMS",
    title = "{Measurement of the Drell{\textendash}Yan forward-backward asymmetry and of the effective leptonic weak mixing angle in proton-proton collisions at s=13TeV}",
    eprint = "2408.07622",
    archivePrefix = "arXiv",
    primaryClass = "hep-ex",
    reportNumber = "CMS-SMP-22-010, CERN-EP-2024-208",
    doi = "10.1016/j.physletb.2025.139526",
    journal = "Phys. Lett. B",
    volume = "866",
    pages = "139526",
    year = "2025"
}

@article{ATLAS:2015ihy,
    author = "Aad, Georges and others",
    collaboration = "ATLAS",
    title = "{Measurement of the forward-backward asymmetry of electron and muon pair-production in $pp$ collisions at $\sqrt{s}$ = 7 TeV with the ATLAS detector}",
    eprint = "1503.03709",
    archivePrefix = "arXiv",
    primaryClass = "hep-ex",
    reportNumber = "CERN-PH-EP-2014-259",
    doi = "10.1007/JHEP09(2015)049",
    journal = "JHEP",
    volume = "09",
    pages = "049",
    year = "2015"
}

@article{LHCb:2024ygc,
    author = "Aaij, R. and others",
    collaboration = "LHCb",
    title = "{Measurement of the effective leptonic weak mixing angle}",
    eprint = "2410.02502",
    archivePrefix = "arXiv",
    primaryClass = "hep-ex",
    reportNumber = "LHCb-PAPER-2024-028, CERN-EP-2024-230",
    doi = "10.1007/JHEP12(2024)026",
    journal = "JHEP",
    volume = "12",
    pages = "026",
    year = "2024"
}

@article{Vent:2025ddm,
    author = "Vent, Sophia and Winterhalder, Ramon and Plehn, Tilman",
    title = "{How to Deep-Learn the Theory behind Quark-Gluon Tagging}",
    eprint = "2507.21214",
    archivePrefix = "arXiv",
    primaryClass = "hep-ph",
    reportNumber = "TIF-UNIMI-2025-16",
    month = "7",
    year = "2025"
}

\end{document}